\PassOptionsToPackage{dvipsnames}{xcolor}
\PassOptionsToPackage{table}{xcolor}

\documentclass[acmsmall,screen]{acmart}
\renewcommand\footnotetextcopyrightpermission[1]{}
\usepackage{etex}
\usepackage{styles/custom}

\AtBeginDocument{%
  }

\acmDOI{10.1145/3776696}
\title{Bayesian Separation Logic}
\subtitle{A Logical Foundation and Axiomatic Semantics for Probabilistic Programming}

\author{Shing Hin Ho}
\orcid{0009-0000-4483-5841}
\affiliation{%
  \institution{Imperial College London}
  \city{London}
  \country{UK}
}
\email{shinghin.ho21@imperial.ac.uk}

\author{Nicolas Wu}
\orcid{0000-0002-4161-985X}
\affiliation{%
  \institution{Imperial College London}
  \city{London}
  \country{UK}
}
\email{n.wu@imperial.ac.uk}

\author{Azalea Raad}
\orcid{0000-0002-2319-3242}
\affiliation{%
  \institution{Imperial College London}
  \city{London}
  \country{UK}
}
\email{azalea.raad@imperial.ac.uk}

\newcommand\appendixRef[1]{\Cref{#1}}
\newcommand\appendixCite[1]{\citet[\cref*{#1}]{ho25}}
\ifx\short\undefined
  \newcommand\appendixref\appendixRef
\else
  \newcommand\appendixref\appendixCite
\fi 

\setcopyright{cc}
\setcctype{by}
\acmDOI{10.1145/3776696}
\acmYear{2026}
\acmJournal{PACMPL}
\acmVolume{10}
\acmNumber{POPL}
\acmArticle{54}
\acmMonth{1}


\begin{abstract}
  Bayesian probabilistic programming languages (BPPLs) let users denote statistical models as code while the interpreter infers the posterior distribution. The semantics of BPPLs are usually mathematically complex and unable to reason about desirable properties such as expected values and independence of random variables. To reason about these properties in a non-Bayesian setting, probabilistic separation logics such as PSL and Lilac interpret separating conjunction as probabilistic independence of random variables. However, no existing separation logic can handle Bayesian updating, which is the key distinguishing feature of BPPLs.
  
  To close this gap, we introduce Bayesian separation logic ($\logic$), a probabilistic separation logic that gives semantics to BPPL. We prove an internal version of Bayes' theorem using a result in measure theory known as the Rokhlin-Simmons disintegration theorem. Consequently, $\logic$ can model probabilistic programming concepts such as Bayesian updating, unnormalised distribution, conditional distribution, soft constraint, conjugate prior and improper prior while maintaining modularity via the frame rule. The model of $\logic$ is based on a novel instantiation of Kripke resource monoid via $\sigma$-finite measure spaces over the Hilbert cube, and the semantics of Hoare triple is compatible with an existing denotational semantics of BPPL based on the category of $\mathit{s}$-finite kernels. Using $\logic$, we then prove properties of statistical models such as the expected value of Bayesian coin flip, correlation of random variables in the collider Bayesian network, the posterior distributions of the burglar alarm model, a parameter estimation algorithm, and the Gaussian mixture model.
\end{abstract}

\ifx\short\undefined
\else
\fi

\begin{document}

\maketitle
\allowdisplaybreaks

\tikzset{
  activecell/.style = {
    rectangle,
    minimum size=6mm,
    thick,
    draw=black, %
    top color=white!60,
    bottom color=white!60,
    font=\itshape
  },
  nonactivecell/.style = {
    rectangle,
    minimum size=6mm,
    thick,
    draw=black, %
    top color=gray!50!black!20, %
    bottom color=gray!50!black!20, %
    font=\itshape\color{gray}
  },
  distcell/.style = {
    rectangle,
    minimum size=6mm,
    thick,
    draw=black, %
    font=\itshape\color{gray}
  },
  distcellempty/.style = {
    rectangle,
    minimum size=6mm,
    thick,
    draw=black, %
    top color=white, %
    bottom color=gray!50!black!20, %
    font=\itshape\color{gray}
  },
  pics/distribution/.style 2 args={
    code={
      \node (mynode) [distcell] {};
      \coordinate (pointA) at (-2mm, -6mm);
      \coordinate (pointB) at (1mm, -6mm);
      \ifdim #1=0mm \else 
        \draw[draw=black, fill=blue] (pointA) rectangle ++(1mm, {#1});
      \fi
      \ifdim #2=0mm \else
      \draw[draw=black, fill=blue] (pointB) rectangle ++(1mm, {#2});
      \fi
      \node[below = 0mm of pointA] {$\color{black}\,\,\false$};
      \node[below = -0mm of pointB] {$\color{black}\,\,\true$};
    }
  },
  pics/distributionwithleftaxis/.style n args= {4}{
    code={
      \node (mynode) [distcell] {};
      \coordinate (pointA) at (-2mm, -6mm);
      \coordinate (pointB) at (1mm, -6mm);
      \coordinate (axleft) at (-3.5mm, {#3});
      \coordinate (axright) at (-2.5mm, {#3});
      \ifdim #1=0mm \else
        \draw[draw=black, fill=blue] (pointA) rectangle ++(1mm, {#1}); 
      \fi
      \ifdim #2=0mm \else
        \draw[draw=black, fill=blue] (pointB) rectangle ++(1mm, {#2});
      \fi
      \node[below = 0mm of pointA] {$\color{black}\,\,\false$};
      \node[below = 0mm of pointB] {$\color{black}\,\,\true$};
      \draw (axleft) to (axright);
      \node[left=of axleft, xshift=11mm] {\tiny{#4}};
    }
  },
  pics/distributionwithrightaxis/.style n args= {4}{
    code={
      \node (mynode) [distcell] {};
      \coordinate (pointA) at (-2mm, -6mm);
      \coordinate (pointB) at (1mm, -6mm);
      \coordinate (axleft) at (3.5mm, {#3});
      \coordinate (axright) at (2.5mm, {#3});
      \ifdim #1=0mm \else
        \draw[draw=black, fill=blue] (pointA) rectangle ++(1mm, {#1});
      \fi
      \ifdim #2=0mm \else
      \draw[draw=black, fill=blue] (pointB) rectangle ++(1mm, {#2});
      \fi
      \node[below = 0mm of pointA] {$\color{black}\,\,\false$};
      \node[below = 0mm of pointB] {$\color{black}\,\,\true$};
      \draw (axleft) to (axright);
      \node[right=of axright, xshift=-10mm] {\tiny{#4}};
    }
  },
  pics/emptydistribution/.style = {
    code={
      \node (mynode) [distcell] {};
      \coordinate (pointA) at (-2mm, -6mm);
      \coordinate (pointB) at (1mm, -6mm);
      \node[below = 0mm of pointA] {$\,\,\false$};
      \node[below = 0mm of pointB] {$\,\,\true$};
      \coordinate (botLeft) at (-3mm, -6mm);
      \coordinate (topLeft) at (3mm, -6mm);
      \coordinate (topRight) at (3mm, -0.1mm);
      \coordinate (botRight) at (-3mm, -0.1mm);
      \draw [thick] (botLeft) to (topRight);
      \draw [thick] (topLeft) to (botRight);
    }
  },
  pics/cemptydistribution/.style = {
    code={
      \node (mynode) [distcell] {};
      \coordinate (pointA) at (-2mm, -6mm);
      \coordinate (pointB) at (1mm, -6mm);
      \node[below = 0mm of pointA] {$\,\,0$};
      \node[below = 0mm of pointB] {$\,\,1$};
      \coordinate (botLeft) at (-3mm, -6mm);
      \coordinate (topLeft) at (3mm, -6mm);
      \coordinate (topRight) at (3mm, 0mm);
      \coordinate (botRight) at (-3mm, 0mm);
      \draw [thick] (botLeft) to (topRight);
      \draw [thick] (topLeft) to (botRight);
    }
  }
}

\newcommand{\bernxgraph}{
  \mathord{%
    \begin{tikzpicture}[baseline={([yshift=0]current bounding box.center)},vertex/.style={anchor=base,
    circle,fill=black!25,minimum size=18pt,inner sep=2pt}]
      \node (mynode) [distcell] {};
      \coordinate (pointA) at (-2mm, -3mm);
      \coordinate (pointB) at (1mm, -3mm);
      \coordinate (axleft) at (3.5mm, 0.5mm);
      \coordinate (axright) at (2.5mm, 0.5mm);
      \coordinate (axleft2) at (-3.5mm, -1.5mm);
      \coordinate (axright2) at (-2.5mm, -1.5mm);
      \draw[draw=black, fill=blue] (pointA) rectangle ++(1mm, 1.5mm);
      \draw[draw=black, fill=blue] (pointB) rectangle ++(1mm, 3.5mm);
      \node[below = 0mm of pointA] {$\color{black}\,\,0$};
      \node[below = 0mm of pointB] {$\color{black}\,\,1$};
      \draw (axleft) to (axright);
      \draw (axleft2) to (axright2);
      \node[right=of axright, xshift=-10mm] {\tiny{$x$}};
      \node[left=of axleft2, xshift=11mm] {\tiny{$1 {-} x$}};
  \end{tikzpicture}}%
}

\newcommand{\bernlxcalcgraph}{
  \mathord{%
    \begin{tikzpicture}[baseline={([yshift=0]current bounding box.center)},vertex/.style={anchor=base,
    circle,fill=black!25,minimum size=18pt,inner sep=2pt}]
      \node (mynode) [distcell] {};
      \coordinate (pointA) at (-2mm, -3mm);
      \coordinate (pointB) at (1mm, -3mm);
      \coordinate (axleft) at (3.5mm, 0.5mm);
      \coordinate (axright) at (2.5mm, 0.5mm);
      \coordinate (axleft2) at (-3.5mm, -3mm);
      \coordinate (axright2) at (-2.5mm, -3mm);
      \draw[draw=black, fill=blue] (pointB) rectangle ++(1mm, 3.5mm);
      \node[below = 0mm of pointA] {$\color{black}\,\,0$};
      \node[below = 0mm of pointB] {$\color{black}\,\,1$};
      \draw (axleft) to (axright);
      \node[right=of axright, xshift=-10mm] {\tiny{$x \cdot \ell_{=1}(1) = x$}};
      \node[left=of axleft2, xshift=11mm] {\tiny{$(1 - x) \cdot \ell_{=1}(0) = 0$}};
  \end{tikzpicture}}%
}

\newcommand{\bernlxgraph}{
  \mathord{%
    \begin{tikzpicture}[baseline={([yshift=0]current bounding box.center)},vertex/.style={anchor=base,
    circle,fill=black!25,minimum size=18pt,inner sep=2pt}]
      \node (mynode) [distcell] {};
      \coordinate (pointA) at (-2mm, -3mm);
      \coordinate (pointB) at (1mm, -3mm);
      \coordinate (axleft) at (3.5mm, 0.5mm);
      \coordinate (axright) at (2.5mm, 0.5mm);
      \coordinate (axleft2) at (-3.5mm, -3mm);
      \coordinate (axright2) at (-2.5mm, -3mm);
      \draw[draw=black, fill=blue] (pointB) rectangle ++(1mm, 3.5mm);
      \node[below = 0mm of pointA] {$\color{black}\,\,0$};
      \node[below = 0mm of pointB] {$\color{black}\,\,1$};
      \draw (axleft) to (axright);
      \node[right=of axright, xshift=-10mm] {\tiny{$x$}};
  \end{tikzpicture}}%
}

\newcommand{\bernlnxgraph}{
  \mathord{%
    \begin{tikzpicture}[baseline={([yshift=0]current bounding box.center)},vertex/.style={anchor=base,
    circle,fill=black!25,minimum size=18pt,inner sep=2pt}]
      \node (mynode) [distcell] {};
      \coordinate (pointA) at (-2mm, -3mm);
      \coordinate (pointB) at (1mm, -3mm);
      \coordinate (axleft) at (3.5mm, -3mm);
      \coordinate (axright) at (2.5mm, -3mm);
      \coordinate (axleft2) at (-3.5mm, -1.5mm);
      \coordinate (axright2) at (-2.5mm, -1.5mm);
      \draw[draw=black, fill=blue] (pointA) rectangle ++(1mm, 1.5mm);
      \node[below = 0mm of pointA] {$\color{black}\,\,0$};
      \node[below = 0mm of pointB] {$\color{black}\,\,1$};
      \draw (axleft2) to (axright2);
      \node[left=of axleft2, xshift=11mm] {\tiny{$1 {-} x$}};
  \end{tikzpicture}}%
}

\section{Introduction}\label{sec:intro}
Statistical techniques have become increasingly prevalent with widespread applications across a multitude of domains from computer vision and data science to computational biology and social science.  As a result, there is a pressing need for developing secure and explainable statistical models. A way of ensuring correctness of such systems is with \emph{formal methods}, a collection of techniques that allows computer scientists to prove properties of algorithms/programming languages by modelling the objects of interest mathematically. In fact, researchers are increasingly applying formal methods to probabilistic/machine learning algorithms in order to provide correctness guarantees on desirable statistical properties \cite{cruz22,slusarz22}. We are interested in formal and compositional reasoning techniques for \emph{Bayesian probabilistic programming languages} -- a class of statistical programming languages that implement algorithms for \emph{Bayesian inference}.

Bayesian inference is a statistical technique that allows users to infer unknown distributions using a statistical model and Bayes' theorem. It has a wide range of applications ranging from medicine \cite{muehlemann23} to computer vision \cite{geman84}. For instance, problems such as clustering and regression can be solved by reframing them as Bayesian inference problems. Computationally, \emph{probabilistic programming} is the programming paradigm that implements Bayesian inference by using general inference algorithms such as Hamiltonian Monte Carlo and Gibbs sampling \cite{brooks11}, while providing a simple interface for users to specify their statistical models \cite{vandemeent21}. \emph{Bayesian probabilistic programming languages} (BPPLs) such as \textsc{Stan} \cite{carpenter17},  \textsc{Anglican} \cite{tolpin16} and \textsc{Gen} \cite{cusumano19} are implementations of the programming paradigm and have specialised constructs for users to easily express statistical models.

\paragraph{Goal}

Our goal is to develop a logical foundation that allows us to reason about statistical properties of BPPL programs (e.g.  independence, expected value, correlation) in a \emph{compositional} manner. To achieve this, we use \emph{separation logic} \cite{reynolds02}, a logical system designed to allow compositional reasoning of computational resources, which in our case, are the random variables produced by the probabilistic program. In particular, we develop \emph{Bayesian separation logic} (\logic, pronounced `basil'), a logical system for Bayesian reasoning based on probabilistic separation logic. From $\logic$, we derive the first Hoare logic for BPPLs, which then allows us to prove the correctness/properties of statistical models such as the \emph{Bayesian coin flip} model, the \emph{collider network}, a \emph{parameter estimation algorithm}, and a \emph{Gaussian-mixture-based clustering model}.

\paragraph{Bayesian Conditioning}

Unlike \emph{randomised} programming languages, which are languages with a sampling construct, $\kp{sample}$, implemented via a pseudo-random number generator (\eg in C and Python), a BPPL has an additional \emph{conditioning} construct, $\kp{observe}$\footnote{For readers familiar with probabilistic programming, $\kp{observe}$ is implemented using the soft-constraint construct $\kp{score}$.}, that allows users to express \emph{conditional probability}. For example, consider an experiment where we toss a fair die $X$, then condition on the event $X > 4$. This can be expressed by the BPPL program in \cref{fig:intro:observe}, which computes the distribution of $X$ given that $X > 4$:

\begin{figure}[h]
  \vspace{-0.75cm}
\[
\hspace{-1cm}
\vcenter{\hbox{
  \begin{minipage}{0.3\linewidth}
    \fbox{$\begin{array}{l}
      \kr{let} X = \kp{sample}(\mathsf{FairDie}) \kl{in} \\
      \kp{observe}(X > 4); \\
      \kpr{return} X
    \end{array}$}
  \end{minipage}
}}
\quad
\vcenter{\hbox{\raisebox{1.2cm}{
  \begin{minipage}{0.25\linewidth}
    \[
      \xrightarrow{\small
        \begin{array}{l}
          \textnormal{`$\kp{observe}$' filters out traces} \\
          \textnormal{that do not satisfy $X > 4$}
        \end{array}
      }
    \]
  \end{minipage}}
}}
\quad
\vcenter{\hbox{
  \raisebox{-1.35cm}{
  \begin{minipage}{0.25\linewidth}
    \input{code/dist.pgf}
  \end{minipage}
}}}
\]
\vspace{-0.5cm}
\caption{Conditioning as the computational effect $\kp{observe}$}
\label{fig:intro:observe}
\end{figure}

\paragraph{Probabilistic Separation Logics}
The reasoning principles for \emph{randomised} languages have been studied in \emph{probabilistic separation logics} \cite{barthe19,tassarotti19,bao21,bao21neg,li23} through a set of axiomatic rules for deriving \emph{Hoare triples}. A Hoare triple, $\{P\}\ M\ \{X. Q\}$, consists of four components: the program $M$, the precondition $P$, the return variable $X$ and the postcondition $Q$. A Hoare triple $\{P\}\ M\ \{X. Q\}$ is \emph{valid} when executing the program $M$ assuming $P$ produces a variable $X$ and the proposition $Q$ holds. For instance, consider the following Hoare triple (using the notation of $\lilac$ \cite{li23}):
\[
  {\color{Purple}\{\top\}}\ \kp{sample}(\msf{FairDie})\ {\color{Purple}\{X. X \sim \msf{Unif}(\{1, ..., 6\})\}}
\]
The triple above describes the pre/postcondition for executing $\kp{sample}(\msf{FairDie})$ for rolling a fair die. Assuming a precondition with no knowledge of variables ($\top$), executing $\kp{sample}(\msf{FairDie})$ yields a random variable $X$ that has a uniform distribution on $\{1, ..., 6\}$. Moreover, since the underlying logic of the Hoare triple is based on probabilistic \emph{separation} logic, we can freely add independent random variables in our pre/postconditions via the \emph{probabilistic separating conjunction} `$*$'. For example, the Hoare triple below states that given a random variable $Y$ with distribution $\mbb P$, executing $\kp{sample}(\msf{FairDie})$ produces an $X$ that is \emph{probabilistically independent} of $Y$:
\begin{align*}
  {\color{Purple}\{Y \sim \mbb P\}}\,\kp{sample}(\msf{FairDie})\,{\color{Purple}\{X.X \sim \msf{Unif}(\{1, ..., 6\}) * Y \sim \mbb P\}}
\end{align*}
This is useful due to the nature of probabilistic reasoning, where a key part is to determine which random variables are independent of each other. Apart from the property of \emph{distribution} (propositions of the form $X \sim \mbb P$), probabilistic separation logic can also reason about different probabilistic properties. For example, the logical entailment
\[
  X \sim \mcal N(0, 1) * Y \sim \mcal{N}(0, 1) \vdash (\mbb E[X] = 0 * \mbb E[Y] = 0) \land \textnormal{Cov}[X, Y] = 0
\]
states that if $X$ and $Y$ are independent, normally-distributed random variables with mean $0$ and standard deviation $1$, then they both have \emph{expected value}\footnote{The expected value $\mathbb{E}[X]$ of a random variable $X$ is the average of the distribution of $X$.}  zero ($\mbb E[X] = 0 * \mbb E[Y] = 0$) and $X, Y$ are uncorrelated (since the covariance is zero $\textnormal{Cov}[X, Y] = 0$). Having statistical propositions and combining them using the separating conjunction `$*$'  allows formal and convenient statistical reasoning in a formal setting.

\paragraph{Limitation of Existing Work: No Support for Bayesian Conditioning}
While probabilistic separation logics are a rich field of study with plenty of variations, existing probabilistic separation logics cannot reason about the $\kp{observe}$ construct. For instance, following the intuition in \cref{fig:intro:observe}, we \emph{should} ideally have a Hoare triple specification as follows:
\[
  \{X \sim \msf{Unif}(\{1, ..., 6\})\}\ \kp{observe}(X > 4)\ \{X \sim \msf{Unif}(\{5, 6\})\}
\]
This technique is called \emph{Bayesian updating}/\emph{Bayesian conditioning}, where we `update' our distributions based on observations. $\logic$ is the \emph{first} probabilistic separation logic that allows reasoning of Bayesian updating, and we do this by drawing an analogy between mutation of memory in standard separation logic. We achieve this by proving an \emph{internal Bayes' theorem} (\cref{thm:basl:bayes}) using a result in measure theory known as the \emph{Rokhlin-Simmons disintegration theorem}.

\paragraph{Limitation of Existing Work: Dependencies of Random Variables}
Existing probabilistic separation logics such as $\lilac$ and $\tsc{BlueBell}$ can reason about dependent random variable via the \emph{conditioning modality}. For instance, the proposition $x \leftarrow X \alt Y \sim \mcal N(x, 1)$ means $Y$ has distribution $\mcal{N}(x, 1)$, conditioning on $X = x$ for some $x \in \R$. $\logic$ extends the conditioning modality to support conditional reasoning in the presence of Bayesian updating. Furthermore, the Hoare logic in $\logic$ supports \emph{conditional sampling}. For example, the following Hoare triple is provable in $\logic$:
\[
  \{X \sim \msf{Unif}(0, 1)\}\ \kp{sample}(\msf{Normal}(X, 1))\ \{Y. \cond{\bc{x}{X}{\msf{Unif}(0, 1)}} Y \sim \mcal{N}(x, 1)\}
\]
It states that assuming $X \sim \msf{Unif}(0, 1)$, sampling a normal with mean $X$ yields a random variable $Y$, which has a \emph{conditional distribution} $Y \sim \mcal{N}(x, 1)$ when $X = x$ for almost all $x \in [0, 1]$.

\paragraph{Proving Properties of the Bayesian Coin Flip Model}

By supporting the above features, together with the compositionality afforded by the frame rule, \logic can serve as a logical foundation for symbolic reasoning probabilistic programming. To demonstrate $\logic$ and its associated Hoare logic, we first consider a simple statistical problem known as \emph{Bayesian coin flip}.
\begin{problem}\label{prob:overview:bayescoin}
  We have a coin and we want to know if it is fair. We toss it once, it comes up heads. We toss it again, it comes up tails. Is it a fair coin?
\end{problem}

\begin{figure}[tbp]
    \centering
    \begin{minipage}[b]{0.42\textwidth}
        \centering
        \begin{align*}
          &X \sim \msf{Unif}(0, 1) \\
          &\mi{Flip}_1 \sim \msf{Bern}(x)\quad \textnormal{when $X = x$} \\
          &\mi{Flip}_2 \sim \msf{Bern}(x)\quad \textnormal{when $X = x$} \\
          &\tb{observation:}\ \mi{Flip}_1 = 1 \\ 
          &\tb{observation:}\ \mi{Flip}_2 = 0
        \end{align*}
        \vspace{0cm}
        \caption{Bayesian solution of \cref{prob:overview:bayescoin}}
        \label{fig:intro:stat-bayescoin}
    \end{minipage}%
    \hfill
    \begin{minipage}[b]{0.58\textwidth}
      \setlength{\arraycolsep}{0pt}
        $\begin{aligned}
          &{\textbf{Precondition:}}\ \textnormal{$\top$} \\
          &{\textbf{Postcondition:}}\ \textnormal{return value $X$ has expected value $\nicefrac{1}{2}$}\!
        \end{aligned}$ \\
      $\begin{array}{ll}
        \kr{let} X = \kp{sample}(\msf{Unif}(0, 1))\kl{in} &\ \annotate{specify prior belief} \\
        \kr{let} \ti{Flip}_1 = \kp{sample}(\msf{Bern}(X))\kl{in} & \ \annotate{toss the coin} \\
        \kp{observe}(\ti{Flip}_1 = 1); &\ \annotate{assert it comes up heads}\, \\
        \kr{let} \ti{Flip}_2 = \kp{sample}(\msf{Bern}(X))\kl{in} & \ \annotate{toss again} \\
        \kp{observe}(\ti{Flip}_2 = 0); &\ \annotate{assert it comes up tails} \\
        \kr{return} X &\ \annotate{what is our belief now?}
      \end{array}$
  \caption{\tsc{BayesCoin}}
  \label{fig:intro:bayescoin}
    \end{minipage}
\end{figure}

\noindent The problem, while simple, encapsulates the three main steps of the Bayesian method: 
\begin{enumerate}
  \item We \emph{assume} a \emph{prior belief} (or \emph{prior} in short) regarding the problem. In our case, we assume a prior regarding whether the coin is fair.
  \item We \emph{observe} real-life data. In our case, we have two observations of coin flip. 
  \item We \emph{update} our belief based on observations. In our case, we update our belief based on \emph{observations} using \emph{Bayes' theorem}.
\end{enumerate}
We defer the explanation of the statistical model to \cref{sec:overview}. For now, we describe the problem using statistical notation in \cref{fig:intro:stat-bayescoin}, and its programmatic counterpart in \cref{fig:intro:bayescoin}. \cref{fig:intro:stat-bayescoin} states that $X$ has (prior) distribution $\msf{Unif}(0, 1)$, $\mi{Flip}_i$ has distribution $\msf{Bern}(x)$ when $X = x$ (the Bernoulli distribution), which is a distribution that returns $1$ with $x$ probability and $0$ with $1 - x$ probability, and we have two observations. Using a BPPL, we can express the same model programmatically as shown in the \rrule{BayesCoin} program in \cref{fig:intro:bayescoin}. Programmatically, the underlying interpreter for the language is performing \emph{rejection sampling}, which, to a first approximation, executes the program many times and filters out traces that do \emph{not} satisfy the required observations, i.e. when $\mi{Flip}_1 \neq 1$ or $\mi{Flip}_2 \neq 0$. Intuitively, since we observed a head and a tail, the expected value (average) of $X$ should remain $\nicefrac{1}{2}$. As we show in \cref{sec:overview}, using $\logic$ we can encode and prove this specification easily (\cref{fig:overview:bayescoin-semantics}), and our proof derivation \emph{formally} describes how the distributions of the random variables evolve. 
\paragraph{Application: Justifying Probabilistic Program Rewrites}
\label{example:rewriting}
It is well-known that probabilistic programming languages are computationally expensive to execute. For example, consider the two programs below (which we will explain in detail later in \cref{subsec:exam:simp}):
\[
  \begin{array}{l}
    \kr{let} X = \kw{sample}(\mathcal{N}(0, 1)) \kl{in} \\
    \kr{let} Y = \kw{sample}(\msf{Beta}(m, n)) \kl{in} \\
    \kr{for} I \kb{in} [1, ..., N] \kl{do} \\
    \quad\kr{observe} x_I \kb{from} \mcal{N}(M, 1); \\
    \quad\kr{observe} c_I \kb{from} \msf{Bern}(W)
  \end{array}
  \qquad
  \begin{array}{l}
    \kr{let} X = \kw{sample}\bigl(\mcal{N}\bigl(\frac{\msf{sum}([x_1, ..., x_N])}{N + 1}, 1\bigr)\bigr) \kl{in} \\
    \kr{let} Y = \kw{sample}(\msf{Beta}(n + \#(c = 1), m + \#(c = 0))) \kl{in} \\
    \kr{return} (X, Y)
  \end{array}
\]
The two programs have \emph{equivalent} distributions and can be considered to be equal. However, the program on the left is computationally costly to sample from due to the existence of two conditioning ($\kp{observe}$-$\kp{from}$) constructs and a loop, while the one on the right is easy to sample from. As we show later in \cref{subsec:exam:simp}, using \logic we can justify \emph{rewriting}  the program on the left to the one on the right by showing that we can prove the \emph{same} Hoare triple (with the same pre- and post-conditions) for both programs, establishing their equivalent distributions.

\newcommand{\unifgraph}{
  \mathord{%
    \begin{tikzpicture}[baseline={([yshift=0]current bounding box.center)},vertex/.style={anchor=base,
    circle,fill=black!25,minimum size=18pt,inner sep=2pt}]
      \coordinate (pointA) at (-2mm, -3mm);
      \coordinate (pointB) at (1mm, -3mm);
      \draw[draw=black, fill=\densitycolor] (pointA) rectangle ++(3.9mm, 5mm);
      \node (mynode) [distcell] {};
      \coordinate (axleft) at (3.5mm, 2mm);
      \coordinate (axright) at (2.5mm, 2mm);
      \node[below = 0mm of pointA] {$\color{black}\,\,0$};
      \node[below = 0mm of pointB] {$\color{black}\,\,1$};
      \draw (axleft) to (axright);
      \node[right=of axright, xshift=-10mm] {\tiny{1}};
  \end{tikzpicture}}%
}

\newcommand{\betatwoonegraph}{
  \mathord{%
    \begin{tikzpicture}[baseline={([yshift=0]current bounding box.center)},vertex/.style={anchor=base,
    circle,fill=black!25,minimum size=18pt,inner sep=2pt}]
      \coordinate (pointA) at (-2mm, -3mm);
      \coordinate (pointB) at (1mm, -3mm);
      \coordinate (pointBtop) at (1.5mm, 2mm);
      \coordinate (pointBbot) at (1.5mm, -3.1mm);
      \draw[draw=black, fill=\densitycolor] (pointA) to (pointBtop) to (pointBbot);
      \node (mynode) [distcell] {};
      \coordinate (axleft) at (3.5mm, 2mm);
      \coordinate (axright) at (2.5mm, 2mm);
      \node[below = 0mm of pointA] {$\color{black}\,\,0$};
      \node[below = -0mm of pointB] {$\color{black}\,\,1$};
      \draw (axleft) to (axright);
      \node[right=of axright, xshift=-10mm] {\tiny{2}};
  \end{tikzpicture}}%
}

\newcommand{\betatwotwograph}{
  \mathord{%
    \begin{tikzpicture}[baseline={([yshift=0]current bounding box.center)},vertex/.style={anchor=base,
    circle,fill=black!25,minimum size=18pt,inner sep=2pt}]
      \coordinate (pointA) at (-3mm, -3mm);
      \coordinate (pointB) at (2mm, -3mm);
      \coordinate (pointBbot) at (3mm, -3.1mm);
      \draw[draw=black, fill=\densitycolor] (pointA) to[out=90,in=90, distance=0.5cm] (pointBbot) to (pointA);
      \node (mynode) [distcell] {};
      \node[below = 0mm of pointA] {$\color{black}\,\,0$};
      \node[below = -0mm of pointB] {$\color{black}\,\,1$};
  \end{tikzpicture}}%
}

\newcommand{\betatwooneunnormgraph}{
  \mathord{%
    \begin{tikzpicture}[baseline={([yshift=0]current bounding box.center)},vertex/.style={anchor=base,
    circle,fill=black!25,minimum size=18pt,inner sep=2pt}]
      \coordinate (pointA) at (-2mm, -3mm);
      \coordinate (pointB) at (1mm, -3mm);
      \coordinate (pointBtop) at (1.5mm, 2mm);
      \coordinate (pointBbot) at (1.5mm, -3.1mm);
      \draw[draw=black, fill=\densitycolor] (pointA) to (pointBtop) to (pointBbot);
      \node (mynode) [distcell] {};
      \coordinate (axleft) at (3.5mm, 2mm);
      \coordinate (axright) at (2.5mm, 2mm);
      \node[below = 0mm of pointA] {$\color{black}\,\,0$};
      \node[below = -0mm of pointB] {$\color{black}\,\,1$};
      \draw (axleft) to (axright);
      \node[right=of axright, xshift=-10mm] {\tiny{1}};
  \end{tikzpicture}}%
}

\newcommand{\diracgraph}[1]{
  \mathord{%
    \begin{tikzpicture}[baseline={([yshift=0]current bounding box.center)},vertex/.style={anchor=base,
    circle,fill=black!25,minimum size=18pt,inner sep=2pt}]
      \coordinate (pointA) at (-0.1mm, -3mm);
      \coordinate (pointB) at (-0.5mm, -3mm);
      \draw[draw=black, fill=blue] (pointA) rectangle ++(0.3mm, 5mm);
      \node (mynode) [distcell] {};
      \coordinate (axleft) at (3.5mm, 2mm);
      \coordinate (axright) at (2.5mm, 2mm);
      \node[below = -0mm of pointB] {$\color{black}\,\,{#1}$};
      \draw (axleft) to (axright);
      \node[right=of axright, xshift=-10mm] {\tiny{100\%}};
  \end{tikzpicture}}%
}

\newcommand{\unifscoregraph}{
  \mathord{%
    \begin{tikzpicture}[baseline={([yshift=0]current bounding box.center)},vertex/.style={anchor=base,
    circle,fill=black!25,minimum size=18pt,inner sep=2pt}]
      \coordinate (pointA) at (-2mm, -3mm);
      \coordinate (pointB) at (1.25mm, -3mm);
      \coordinate (pointBtop) at (2mm, 2mm);
      \coordinate (pointBbot) at (2mm, -3mm);
      \coordinate (pointAtop) at (-2mm, -0.5mm);
      \coordinate (pointABbot) at (-0mm, -0.5mm);
      \coordinate (pointABtop) at (-0mm, 2mm);
      \draw[draw=black, fill=\densitycolor] (pointA) to (pointAtop) to (pointABbot) to (pointABtop) to (pointBtop) to (pointBbot) to (pointA);
      \node (mynode) [distcell] {};
      \coordinate (axleft) at (3.5mm, 2mm);
      \coordinate (axright) at (2.5mm, 2mm);
      \coordinate (axleft2) at (-3.5mm, -0.5mm);
      \coordinate (axright2) at (-2.5mm, -0.5mm);
      \node[below = 0mm of pointA] {$\color{black}\,\,0$};
      \node[below = -0mm of pointB] {$\color{black}\,\,1$};
      \draw (axleft) to (axright);
      \draw (axleft2) to (axright2);
      \node[right=of axright, xshift=-10mm] {\tiny{4}};
      \node[left=of axleft2, xshift=11mm] {\tiny{2}};
  \end{tikzpicture}}%
}

\paragraph{Contributions and Roadmap}

Our contributions are as follows:
\begin{itemize}
  \item We explain how \rrule{BayesCoin} can be specified in $\logic$ (\cref{sec:overview}).
  \item We prove properties of five standard, but non-trivial Bayesian statistical models such as Bayesian networks (\cref{subsec:exam:burglar-alarm}, \cref{subsec:exam:common-effect}), parameter estimation algorithm (\cref{subsec:exam:soft}), and the Gaussian mixture model (\cref{subsec:exam:cluster}). The models have non-trivial features such as conditional dependencies, soft constraint, conjugate prior and improper prior (\cref{subsec:exam:improper}), which no existing probabilistic separation logic can verify (\cref{sec:exam}).
  \item For the semantics of $\logic$, we describe intuitively the existing resource-theoretic model of $\lilac$ for modelling randomness (\cref{subsec:overview:pcm}) and motivate the need for a measure-theoretic semantics to model randomness in a Bayesian setting (\cref{subsec:overview:bayespcm}).
  \item We develop a novel \emph{Kripke resource monoid} for modelling randomness that supports \emph{Bayesian updating} by using $\sigma$-finite measure spaces over the Hilbert cube (\cref{subsec:basl:kripke}).
  \item We show that the Kripke resource model \cite{galmiche05} of $\logic$ is compatible with \emph{partially affine separation logic} \cite{chargueraud20} (\cref{subsec:basl:kripke}).
  \item We generalise $\lilac$'s \emph{disintegration modality} so that conditional reasoning can be performed in the presence of Bayesian updating (\cref{subsec:basl:sem}).
  \item We design new logical propositions to encode the concept of \emph{likelihood} and \emph{normalising constants}, two key concepts in Bayesian probability (\cref{subsec:basl:sem}).
  \item We prove an internal version of Bayes' theorem and show that the concept can be encoded as logical propositions in $\logic$ by combining the disintegration modality and the likelihood proposition (\cref{thm:basl:bayes}).
  \item We develop the \emph{\logic proof system} (a set of Hoare triples) and prove that it is \emph{sound} with respect to our Kripke resource model of $\logic$ (\cref{subsec:basl:sem}).
\end{itemize}
Finally, we discuss related and future work and conclude (\cref{sec:con}).

\section{Overview}\label{sec:overview}

To give an intuition of how $\logic$ can be used to prove properties of statistical models, we demonstrate its proof rules via \rrule{BayesCoin} (\cref{prob:overview:bayescoin}). 
Recall that $\msf{Bern}(0.5)$ represents a fair coin as it returns $1$ and $0$ with equal probability, while $\msf{Bern}(0.9)$ represents a biased coin that comes up heads $90\%$ of the time. 
Hence, we model our belief about whether our coin is fair via a random variable $X$ that takes a value in $[0, 1]$.

Since we do not have additional information about the coin ($X$), we assume $X \sim \msf{Unif}(0, 1)$ as our prior distribution and write the program described in \cref{fig:intro:bayescoin}. To semantically deduce that the return value $X$ has expected value $\nicefrac 1 2$ (see the postcondition of \cref{fig:intro:bayescoin}), we use $\logic$ to describe \rrule{BayesCoin} axiomatically via pre/postcondition style reasoning rules. 
We present the $\logic$ proof sketch of $\rrule{BayesCoin}$ in \cref{fig:overview:bayescoin-semantics}. We proceed with a detailed but informal explanation of our proof.
\newenvironment{leftvruled}[1]{
    \begin{array}{@{}m{10pt}|l@{}}
    \text{\color{BasilGreen}\rotatebox{90}{\,#1\,}}&\begin{array}{@{}l@{}}}
  {\end{array}\end{array}}

\newenvironment{leftvruledline}[1]{
    \begin{array}{l@{}@{\ \ \ \ }m{10pt}|l@{}}
    \begin{array}{@{}l@{}}
      \!\!\!\!\,\linenum{10} \\[2pt]
      \!\!\!\!\,\linenum{11} \\[2pt]
      \!\!\!\!\,\linenum{12} \\[2pt]
      \!\!\!\!\,\linenum{13} \\[2pt]
      \!\!\!\!\,\linenum{14} \\[2pt]
      \!\!\!\!\,\linenum{15}
    \end{array} &
    \text{\color{BasilGreen}\rotatebox{90}{\,#1\,}}
    &\begin{array}{@{}l@{}}}
  {\end{array}\end{array}}

\begin{figure}[t]
\hspace{-1.5cm}
\begin{minipage}[b]{0.5\linewidth}
  $$
\begin{array}{lr}
  \linenum{\ 1} \trp{\{\top_1\}} \\
  \linenum{\ 2}\quad\kr{let} X = \kp{sample}(\msf{Unif}(0, 1))\kl{in} \\
  \linenum{\ 3}\trp{\{X \sim \msf{Unif}(0, 1)\}} \\
  \linenum{\ 4}\quad\kr{let} \ti{Flip}_1 = \kp{sample}(\msf{Bern}(X))\kl{in} \\
  \linenum{\ 5}\trp{\{\cond{\bc{x}{X}{\msf{Unif}(0, 1)}} \mi{Flip}_1 \sim \msf{Bern}(x) \}} \\
  \linenum{\ 6}\quad\kp{observe}(\ti{Flip}_1 = 1); \\
  \linenum{\ 7}\trp{\{\cond{\bc{x}{X}{\msf{Unif}(0, 1)}} \mi{Flip}_1 \sim \ell_{=1} \cdot \msf{Bern}(x)\}} \\
  \linenum{\ 8}\trp{\{\cond{\bc{x}{X}{\msf{Unif}(0, 1)}} \score{x} \}}\ \annotate{likelihood of $X = x$ is $x$}\\
  \linenum{\ 9}\trp{\{X \sim \msf{Beta}(2, 1) * \msf{NormConst}\}} \\
  \begin{leftvruledline}{\rrule{H-Frame}}
    \trp{\{X \sim \msf{Beta}(2, 1)\}} \\
    \quad\kr{let} \ti{Flip}_2 = \kp{sample}(\msf{Bern}(X))\kl{in} \\
    \trp{\{\cond{\bc{x}{X}{\msf{Beta}(2, 1)}} \mi{Flip}_2 \sim \msf{Bern}(x)\}} \\
    \quad\kp{observe}(\ti{Flip}_2 = 0); \\
    \trp{\{\cond{\bc{x}{X}{\msf{Beta}(2, 1)}} \mi{Flip}_2 \sim \ell_{=0} \cdot \msf{Bern}(x)\}} \\
    \trp{\{X \sim \msf{Beta}(2, 2) * \msf{NormConst}\}}
  \end{leftvruledline} \\
  \linenum{16}\trp{\{X \sim \msf{Beta}(2, 2) * \msf{NormConst} * \msf{NormConst}\}} \\
  \linenum{17}\trp{\{X \sim \msf{Beta}(2, 2) * \msf{NormConst}\}} \\
  \linenum{18}\quad\kr{return} X \\
  \linenum{19}\trp{\{X \sim \msf{Beta}(2, 2) * \msf{NormConst}\}} \\
  \linenum{20}\trp{\{\mbb E[X] = \nicefrac 1 2 * \msf{NormConst}\}}
\end{array}
$$
  \vspace{10pt}
  \caption{An axiomatic description of \tsc{BayesCoin}}
  \label{fig:overview:bayescoin-semantics}
  \end{minipage}
\ \ \ \ \ 
\begin{minipage}[b]{0.4\linewidth}
  \scalebox{0.83}{
    $\begin{aligned}
      \input{code/beta11.pgf} & \input{code/bernx.pgf} \\
      \input{code/beta21.pgf} & \input{code/bernx0.pgf} \\
      \input{code/beta22.pgf} & \input{code/bernx1.pgf}
    \end{aligned}$}
  \caption{Visualisation of distributions}
  \label{fig:overview:bayescoin-vis}
\end{minipage}

\vspace{-10pt}
\end{figure}

\paragraph{Lines 1-3}
Initially, we have no random variables, as captured by the \emph{trivial} precondition $\top_1$, where $1$ in $\top_1$ is the current \emph{normalising constant} (we explain this on Page \pageref{page:normconst-exp}). 
After executing line 2 (of \cref{fig:overview:bayescoin-semantics}), the interpreter produces a random variable $X$ with the desired distribution. 
To reason about this, we apply the \rrule{H-Sample} rule below at line 2,  which states that sampling from a distribution $\mbb P$ of type $\R$ returns a random variable $X$ of type $\R$ distributed according to $\mbb P$ (for line 2, $\mbb P$ is instantiated to $\msf{Unif}(0, 1)$):
\[
  \inferrule
    {}
    {\vdash \{\top_1\}\ \kp{sample}(\mbb P)\ \{X : \R.\ X \sim \mbb P\}}\ \ \rrule{H-Sample}
\]
To `chain' the postcondition to the rest of the program, we apply the sequencing rule \rrule{H-Let} below,  chaining the postcondition of the first program to the precondition of the second program,  thus obtaining $X \sim \msf{Unif}(0, 1)$ on line 3.
\[
  \inferrule
    {\vdash \{P\}\ M\ \{X : \mcal A. Q\} \\ X : \mcal A \vdash \{Q\}\ N\ \{Y : \mcal B. R\}}
    {\vdash \{P\}\ \kr{let} X = M \kb{in} N\ \{Y : \mcal B. R\}}\ \rrule{H-Let}
\]

\paragraph{Lines 4-5}
At line 4 we sample a Bernoulli variable according to our sampled $X$. 
While this line certainly typechecks, it is more challenging to verify: it is ambiguous statistically, as it semantically describes a \emph{conditional distribution}. 
Given $X \sim \msf{Unif}(0, 1)$, we write $\mi{Flip}_1|X = x \sim \msf{Bern}(x)$ to mean that conditioning on $X = x$ for almost all $x \in [0, 1]$, the random variable $\mi{Flip}_1$ has distribution $\msf{Bern}(x)$. 
To reason about this in $\logic$, we introduce our conditional sampling axiom:
\[
  \inferrule
    {}
    {\vdash \{X \sim \mbb P\}\ \kp{sample}(p(X))\ \{Y : \R. \cond{\bc{x}{X}{\mbb P}} Y \sim p(x)\}}\ {\rrule{H-CondSample}\vphantom{\cond{\bc{x}{X}{\mbb P}}}}
\]

\vspace{-3pt}
The arrow/bar notation $\cond{\bc{x}{X}{\mbb P}}{P}$ is our novel \emph{conditioning modality}, which assumes $X$ has distribution $\mbb P$ and binds it to a deterministic name $x$, while the proposition $P$ on the right is a proposition on the conditioned space after conditioning $X = x$. Specifically, $\cond{\bc{x}{X}{\mbb P}} Y \sim p(x)$ in the postcondition is read as follows: assuming $X \sim \mbb P$, if we condition on $X = x$ for some deterministic $x$, then the sampled $Y$ has distribution $p(x)$. Instantiating the axiom above, line 5 says $\mi{Flip}_1$ has distribution $\msf{Bern}(x)$ whenever $X = x$.

\begin{wrapfigure}{r}{0.54\textwidth}
  \vspace{-14pt}
  $\bernxgraph \xmapsto{\ (\ell_{=1}\,\cdot\,{-})\ } \bernlxcalcgraph$
  \vspace{-12pt}
  \caption{Applying likelihood function}
  \label{fig:overview:likelihood}
   \vspace{-5pt}
\end{wrapfigure}

\paragraph{Lines 6-8} Line 6 asserts $\mi{Flip}_1 = 1$ and rejects all program traces where $\mi{Flip}_1 = 0$. 
Semantically, this means we apply a \emph{likelihood function} $\ell_{=1} : \{0, 1\} \to \Rnn$ to $\msf{Bern}(x)$, where $\ell_{=1}(1) \deq 1$ and $\ell_{=1}(0) \deq 0$. Specifically, since $\mi{Flip}_1 \sim \msf{Bern}(x)$ (as depicted on the left side of \cref{fig:overview:likelihood}), applying the likelihood function $\ell_{=1}$ to $\msf{Bern}(x)$ \emph{updates} the distribution to the one on the right side of \cref{fig:overview:likelihood}. Notice that this operation is done in the conditioned space conditioning on $X = x$. In light of this, we introduce the $\rrule{H-CondObserve}$ axiom below for \emph{conditional observation}:
\[
  \inferrule
    {}
    {\vdash \{\cond{\bc{x}{\!X}{\pi}} Y\! \sim p(x)\}\,\kp{observe}(P(Y))\,\{\cond{\bc{x}{\!X}{\pi}} Y\! \sim \ell_{P} \cdot p(x)\}}\ \rrule{H-CondObserve}
\]
where $P$ is a Boolean predicate on $Y$. The axiom states that assuming $Y \sim p(x)$ when conditioning on $X = x$, where $X \sim \pi$ in the unconditioned space, then observing $P(Y)$ updates the distribution from $p(x)$ to $\ell_{P} \cdot p(x)$, where $\ell_P$ is the likelihood function with $\ell_P(x) \deq 1$ if $P(x)$ holds, and $0$ otherwise. In our case, $Y = \mi{Flip}_1$ and $P(\mi{Flip}_1)$ is defined to be $(\mi{Flip}_1 = 1)$, which yields the postcondition on line 7. For readers familiar with probabilistic programming, $\kp{observe}$ is implemented using the \emph{soft constraint} construct $\kp{score}$, and there is a more general rule $\rrule{H-CondScore}$ as we explain in \cref{sec:basl}.

Since $\kp{observe}(\mi{Flip}_1 = 1)$ multiplies the likelihood of $\mi{Flip}_1 = 1$ by $1$, the likelihood of $\mi{Flip}_1 = 1$ is $x \cdot 1$ = $x$ when $X = x$. 
Similarly, $\kp{observe}(\mi{Flip}_1 = 1)$ multiplies the likelihood of $\mi{Flip}_1 = 0$ by $0$;
\ie $\mi{Flip}_1 = 0$ has likelihood $(1 - x) \cdot 0 = 0$ when $X = x$. In other words, this describes traces where $\mi{Flip}_1 \sim \msf{Bern}(x)$ has likelihood $x$. Intuitively this makes sense: when $X = 0.1$, the likelihood of $\mi{Flip}_1 = 1$ is $0.1$, which is lower than the likelihood when $X = 0.99$. 
To reason about this, we introduce a \emph{likelihood proposition} $\score{x}$, which denotes that the likelihood of our current state is $x$. Since $\mi{Flip}_1 \sim \msf{Bern}(x)$ has likelihood $x$, the entailment $\mi{Flip}_1 \sim \ell_{=1} \cdot \msf{Bern}(x)\ \vdash\ \score{x}$ holds, which results in the postcondition on line 8 using the standard rule of consequence \rrule{H-Cons} (see \cref{fig:basl:proofrule}).

\paragraph{Line 9} In light of our observation regarding how likely $X = x$ is, we can now \emph{update} our belief. To achieve this, we apply an internal, logical version of \emph{Bayes' theorem}. Recall that Bayes' theorem, in the context of Bayesian statistics, states the following: 
\[
  \msf{posterior} \propto \msf{prior} \cdot \msf{likelihood}
\]
Or, equivalently, suppose $\nicefrac 1 Z$ is the \emph{normalising constant} for some $Z > 0$, then Bayes' theorem can be stated as $\msf{unnormalised\ posterior} \cdot Z = \msf{prior} \cdot \msf{likelihood}$. Notice that line 8 has a similar structure to above -- we have the prior $X \sim \msf{Unif}(0, 1)$ and the likelihood $\score{x}$. The question is: can we find a suitable proposition that represents the (unnormalised) posterior? That is, finding a suitable proposition $?P$ such that the following entailment holds:
\[
  \cond{\overbrace{\bc{x}{X}{\msf{Unif}(0, 1)}}^{\tx{prior}}} \overbrace{\score{x}\vphantom{\bc{x}{X}{\msf{Unif}(0, 1)}}}^{\tx{likelihood}} \vdash \overbrace{?P\vphantom{\bc{x}{X}{\msf{Unif}(0, 1)}}}^{\tx{posterior}}
\]
Before explaining what the proposition looks like, let us first intuitively visualise how our belief regarding the coin has changed.

\begin{wrapfigure}{r}{0.45\textwidth}
  $\biggl(X \sim \!\!\unifgraph\!\biggr) \xrightarrow{\tx{observing $\mi{Flip}_1 = 1$}} \biggl(X \sim \!\!\betatwooneunnormgraph\!\biggr)$
  \vspace{-10pt}
  \caption{Bayesian updating of $\msf{Unif}(0, 1)$}
  \label{fig:overview:first-update}
  \vspace{-8pt}
\end{wrapfigure}

After observing the coin landing on heads, we shift towards believing $X$ being more likely to be larger. For instance, it is less likely for $X$ to be close to zero, as this would make landing on heads less likely. Without doing the calculations (which can be found in a standard text on Bayesian methods, \eg by 
\citet[\S2.2]{mcelreath20}), the original distribution $\msf{Unif}(0, 1)$ (left of \cref{fig:overview:first-update}) is updated to $\nicefrac 1 2 \cdot \msf{Beta}(2, 1)$ (right of \cref{fig:overview:first-update}). There is, however, a caveat: since we applied Bayes' theorem, the Beta distribution is \emph{unnormalised}: the area under the triangle of the graph is $\nicefrac{1}{2} \neq 1$, as denoted by $\nicefrac 1 2$ in $\nicefrac 1 2 \cdot \msf{Beta}(2, 1)$. To remedy this, we use the separating conjunction `$*$' to factor out the normalising constant.\label{page:normconst-exp} Specifically, for all $Z > 0$ (e.g. $Z = \nicefrac 1 2$ here), the entailment $X \sim Z \cdot \msf{Beta}(2, 1) \vdash X \sim \msf{Beta}(2, 1) * \msf{NormConst}$ holds. Intuitively, $\msf{NormConst}$ factors out and hides the normalising factor and asserts the existence of a non-zero normalising constant, i.e. $\msf{NormConst} \deq \exists k : (0, \infty).\ \score{k}$. This answers our question: the `posterior proposition' $?P$ is of the form:
\begin{align*}
  \cond{\overbrace{\bc{x}{X}{\msf{Unif}(0, 1)}}^{\tx{prior}}} \overbrace{\score{x}\vphantom{\bc{x}{X}{\msf{Unif}(0, 1)}}}^{\tx{likelihood}}
  \vdash\ \overbrace{X \sim \msf{Beta}(2, 1)\vphantom{\bc{x}{X}{\msf{Unif}(0, 1)}}}^{\tx{posterior}}\, *\ \msf{NormConst}
\end{align*}
As we will see in \cref{sec:basl} (\cref{thm:basl:bayes}), the entailment is sound in $\logic$ via properties of \emph{disintegration} (\cref{lem:basl:density}). Next, by applying the rule of consequence, we obtain the postcondition on line $9$.

\begin{wrapfigure}{r}{0.45\textwidth}
  $\biggl(X \sim \!\!\betatwoonegraph\!\biggr) \xrightarrow{\tx{observing $\mi{Flip}_2 = 0$}} \biggl(X \sim \!\!\betatwotwograph\!\biggr)$
  \vspace{-10pt}
  \caption{Bayesian updating of $\msf{Beta}(2, 1)$}
  \label{fig:overview:second-update}
\end{wrapfigure}

\paragraph{Lines 10-16} Note that lines 10-15 are similar to line 3-7: we perform another coin toss, observe it comes up tails, and return our updated belief. Intuitively, after observing the coin landing on tails, we `shift back' our belief about $X$ being likely to take on higher values -- $X$ is more likely to take on values in the middle (e.g. $0.4 \le X \le 0.6$ has higher probability than $X \ge 0.8$ or $X \le 0.2$), i.e. we update the distribution to the one on the right of \cref{fig:overview:second-update}. This distribution is known as the $\msf{Beta}(2, 2)$ distribution. To prove this in $\logic$, we first assume $X \sim \msf{Beta}(2, 1)$ as our precondition of line 10, repeat the steps above and obtain the postcondition on line 15.

We now enter another key step of our proof: using the \emph{frame rule} of separation logic:
\[
  \inferrule
    {\vdash \{P\}\ M\ \{X. Q\}}
    {\vdash \{P * F\}\ M\ \{X. Q * F\}}\ (X \notin \msf{fv}(F))\ \rrule{H-Frame}
\]
The \rrule{H-Frame} rule allows us to \emph{frame off} (factor out) propositions that are probabilistically independent of our current resources. It states that in order to prove $\{P * F\}\ M\ \{X. Q * F\}$, where $F$ is a proposition probabilistically independent of $P$ and $Q$, it suffices to prove $\{P\}\ M\ \{X. Q\}$. 

As the normalising constant proposition $\msf{NormConst}$ is `separated' from $X \sim \msf{Beta}(2, 1)$ on line 9, we can `frame off' the normalising constant $\msf{NormConst}$ prior to line 10 and frame it back on after line 15 and obtain the postcondition on line 16.

\paragraph{Lines 17-20}
At line 16 we have two normalising constants $\msf{NormConst}$ and $\msf{NormConst}$, each created from an update of $X$. We now combine them: intuitively, if two independent unnormalised random variables have normalising constants $Z$ and $Z'$ respectively, the overall distribution has normalising constant $Z \cdot Z'$. This justifies the entailment $\msf{NormConst} * \msf{NormConst} \vdash \msf{NormConst}$. This means we can obtain the postcondition on line 17. Returning $X$ yields the same postcondition (see \rrule{H-Ret} in \cref{fig:basl:proofrule}), which leads to our desired postcondition on line 19. 
Since from line 19 we know $X \sim \msf{Beta}(2, 2)$ and a $\msf{Beta}(2, 2)$-distributed random variable has expected value (average) $\nicefrac 1 2$, the entailment $X \sim \msf{Beta}(2, 2) * \msf{NormConst} \vdash \mbb E[X] = \nicefrac 1 2 * \msf{NormConst}$ holds, and we obtain the postcondition on line 20 using the standard rule of consequence $\rrule{H-Cons}$ (see \cref{fig:basl:proofrule}).

\section{Verifying Statistical Models with \logic}\label{sec:exam}
We demonstrate the expressivity and verification capability of $\logic$ by verifying five programs described below, each with distinct features. 
To this end, in \cref{subsec:exam:bppl} we first present the \logic programming language, $\lang$, a standard Bayesian probabilistic programming language, and then present the \logic proof system as a set of Hoare triples (most of which we have described in \cref{sec:overview}). Specifically, we verify and prove probabilistic properties of six programs listed in \cref{tbl:exam:stat}.
\begin{table}[t]
  \centering
  \begin{tabular}{>{\centering\arraybackslash} c c c } \toprule
    \S & \tb{Statistical model} & \tb{Distinct feature(s)} \\ \midrule
    \cref{subsec:exam:soft} & Parameter estimation algorithm & Soft constraint; conjugate priors \\
    \cref{subsec:exam:burglar-alarm} & The burglar alarm Bayesian network & Joint conditioning; Bayes' theorem \\
    \cref{subsec:exam:common-effect} & The common effect Bayesian network & Conditional dependence and correlation \\
    \cref{subsec:exam:improper} & The semantic Lebesgue measure & Improper prior; handling $\sigma$-finite measures \\ 
    \cref{subsec:exam:cluster} & Gaussian-mixture-based clustering & Intractable posterior; bounded loops \\
    \cref{subsec:exam:simp} & Models with equivalent posterior & Verifying program rewriting \\ \bottomrule
  \end{tabular}
  \caption{Statistical models considered in this section and their distinct features}
  \label{tbl:exam:stat}
  \vspace{-10pt}
\end{table}

\subsection{\logic Programming Language and Proof System}\label{subsec:exam:bppl}

The \logic programming language, $\lang$, is a typed, first-order language equipped with two effects: the \emph{probabilistic sampling} effect $\kp{sample}(\mbb P)$, which samples from a distribution $\mbb P$, and the \emph{soft conditioning} effect $\kp{score}(\ell)$, which \emph{scales} the current distribution according to a non-negative number $\ell$. The $\lang$ \emph{terms and types} are defined by the following grammar, where $X$ ranges over a countably infinite set of names, $n \in \N$, $r \in \R$, and $f$ ranges over (measurable) functions.
\[
	 \begin{aligned}
	    \Terms \ni M &\ddeq
	    () \alt \pv{x} \alt n \alt r \alt f(M) \alt (M, M) \alt M.\kw{1} \alt M.\kw{2} \alt
	    \kw{true} \alt \kw{false} \alt \kr{if} M \kb{then} M \kb{else} M \\
	    &\quad|\ \,\kp{sample}(M) \alt \kp{score}(M) \alt \kp{return}(M) 
	    \alt \kr{let} \pv{x} = M \kb{in} M  \\
	    \Types \ni \tau &\ddeq 
	    \tunit \alt \tnat \alt \treal \alt \tbool \alt \tau \times \tau \alt \tprob(\tau) \tagi{types}
	  \end{aligned}
\]

\paragraph{Additional Encodings} We encode the following syntactic shorthands: 
\[
\begin{array}{c}
	M; N \deq \kr{let} \_ = M \kb{in} N
	\qquad
	\kp{observe}(M) \deq \kp{score}(\kr{if} M \kb{then} 1 \kb{else} 0) \\
	\kpr{observe} M \kw{\ from\ } \mbb P \deq \kp{score}(\msf{density}_{\mbb P}(M))
\end{array}	
\]
The sequential composition ($;$) shorthand is standard; we describe the $\kpr{observe}$ construct shortly below, and elaborate on the \emph{soft constraint} construct `$\kpr{observe} \kw{from}$' in \cref{subsec:exam:soft}.

\paragraph{$\lang$ Typing Judgements}
We present the $\lang$ typing judgements in the technical appendix (\appendixref{append:ppl}), where a term $M$ is typed via a judgement $\vdashp$. 
Semantically, every open term $\Delta \vdashp M : \tau$ denotes an \emph{$s$-finite kernel} $\semo{M} : \semo{\Delta} \rightsquigarrow \semo{\tau}$. 
We refer the reader to the work of \citet{staton17} for a detailed explanation of the semantics as this is not essential for understanding \logic.

\begin{figure}[t]
  \begin{mdframed}
    \small
    \begin{mathpar}
      \inferrule[H-Sample]
        {}
        {\{\topp\}\,\kp{sample}(\mbb P)\,\{X : \R. X \sim \mbb P\}} \and
      \inferrule[H-CondSample]
        {}
        {\{X \sim \mbb P\}\,\kp{sample}(p(X))\,\{Y : \R. \cond{\bc{x}{X}{\mbb P}} Y \sim p(x)\}} \and
      \inferrule[H-Score]
        {}
        {\{X \sim \pi\}\,\kp{score}(f(X))\,\{X \sim f \cdot \pi\}} \and
      \inferrule[H-CondScore]
        {}
        {\{\cond{\bc{x}{X}{\pi}} Y \sim p(x)\}\,\kp{score}(f(Y)) \{\cond{\bc{x}{X}{\pi}} Y \sim f \cdot p(x)\}} \and
      \inferrule[H-Observe]
        {}
        {\{X \sim \pi\}\,\kp{observe}(P(X))\,\{X \sim \ell_P \cdot \pi\}} \and
      \inferrule[H-CondObserve]
        {}
        {\{\cond{\bc{x}{X}{\pi}} Y \sim p(x)\}\,\kp{observe}(P(Y)) \,\{\cond{\bc{x}{X}{\pi}} Y \sim \ell_P \cdot p(x)\}} \and
      \inferrule[H-Return]
        {}
        {\{Q[\semo{M}/X]\}\,\kpr{return} M\,\{X : \mcal A. Q\}} \and
      \inferrule[H-Frame]
        {\vdash \set{P} M \set{X : \mcal A. Q}}
        {\vdash \set{P * F} M \set{X : \mcal A. Q * F}}\ (X \notin \msf{fv}(F)) \and
      \inferrule[H-Let]
        {\vdash \{P\}\, M \,\{X : \mcal A. Q\} \\
        \vdash \forallrv X : \mcal A. \set{Q} N \set{Y : \mcal B. R}}
        {\vdash \set{P} \kr{let} X = M \kb{in} N \set{Y : \mcal B. R}} \and
      \inferrule[H-Cons]
        {P' \vdash P \\
          \vdash \set{P} M \set{X : \mcal A. Q} \\
          Q \vdash Q'}
        {\vdash \{P'\}\,M\,\{X : \mcal A. Q'\}}
    \end{mathpar}
  \end{mdframed} \phantom{a}\vspace{-15pt}\\
  \caption{$\logic$ proof rules}
  \label{fig:basl:proofrule}
  \vspace{-10pt}
\end{figure}

\paragraph{\logic Assertions}%
The \logic \emph{assertions} are defined by the following grammar:
\begin{align*}
  P &\ddeq \top \alt \bot \alt P \land P \alt P \lor P \alt P \Rightarrow P \alt \forall x : A. P \alt \exists x : A. P \alt P * P \alt P \wand P \\
  &\quad|\,\ E \sim \pi \alt \mbb E[E] = e \alt \own{E} \alt (\cond{\bc{x}{E}{\pi}} P) \alt \score{e} \alt \forallrv X : \mcal A. P \alt \existsrv X : \mcal A. P \alt \set{P} M \set{X : \mcal A. P}
\end{align*}
where $\pi$, $E$, $e$, $M$ range over maps defined later in \cref{fig:basl:typing}, and $\top_1$, as seen in \cref{sec:overview}, represents states with normalising constant $1$ and is defined to be $\score{1}$. Intuitively, $E$ is a random expression, $e$ is a deterministic expression, $M$ is a $\lang$ term, and $\pi$ is a measure (distribution). The first-order and separation logic assertions (first line of the grammar) are standard. For probabilistic assertions (second line),  we have described most of them intuitively in \cref{sec:overview}, but for $\forallrv, \existsrv$ quantifiers, which quantify over \emph{random variables},  and $\msf{own}$, which asserts ownership of random expression $E$.

\paragraph{\logic Proof System}
We present the \emph{\logic axiomatic proof system} through a set of rules in \cref{fig:basl:proofrule}.
We have intuitively described most of the axioms with the coin flip example in \cref{sec:overview} 
except for axioms related to the $\kw{score}$ construct. To understand what $\kp{score}$ does intuitively, suppose we have a random variable $X \sim \msf{Unif}(0, 1)$. 
Executing $\kp{score}(\kr{if} X < \nicefrac 1 2 \kb{then} 2 \kb{else} 4)$ then increases the likelihood of values $< \nicefrac 1 2$ being drawn by a factor of $2$, and that of values $\geq \nicefrac 1 2$ by a factor of  $4$:
\[
  \Biggl\{X \sim \!\!\unifgraph\!\Biggr\}\ 
  \kp{score}(\kr{if} X < \nicefrac{1}{2} \kb{then} 2 \kb{else} 4)\ \Biggl\{X \sim \!\!\unifscoregraph\!\Biggr\}
\]
Note that the distribution of $X$ in the postcondition is \emph{unnormalised}, \ie the shaded region has area (normalising constant) $2 \cdot \nicefrac 1 2 + 4 \cdot \nicefrac 1 2 = 3$. 
Moreover, when the distribution of $X$ is normalised, we can calculate that $\Pr[X \ge \nicefrac 1 2] = 2 \cdot \Pr[X < \nicefrac 1 2]$. 
Statistically speaking, it is useful to think of $\kp{score}$ as a way for users to specify the likelihood of an observation, allowing the interpreter to `mutate' the current distribution.
Indeed, as we described above, we encode the hard conditioning $\kp{observe}(M)$ construct as $\kp{score}(\kr{if} M \kb{then} 1 \kb{else} 0)$, setting to $0$ the likelihood of observations where $M$ does not hold, rendering them impossible.

\subsection{Conjugate Priors as Hoare Triples: Verifying a Parameter Estimation Algorithm}\label{subsec:exam:soft}
We begin our journey of verifying statistical models by considering a useful language feature known as \emph{soft constraint} in languages such as \textsc{Stan} and \textsc{Anglican}, which allows users to specify observations even when they are drawn from \emph{continuous} distributions. 
$\logic$ is the \emph{first} probabilistic separation logic that can reason about soft constraints and Bayesian updating.

\begin{wrapfigure}{r}{0.38\textwidth}
$\begin{array}{l}
	\kr{let} \Theta = \kp{sample}(\msf{Normal}(\theta_0, \sigma_0)) \kb{in} \\
	\kpr{observe} x_1 \kpb{from} \msf{Normal}(\Theta, \sigma); \\ 
	\kpr{observe} x_2 \kpb{from} \msf{Normal}(\Theta, \sigma); \\ 
	\kpr{return} \Theta
	\vspace{-10pt}
\end{array}$
\caption{The \tsc{GaussParam} program}\label{fig:exam:gaussparam}
\end{wrapfigure}
As described above,  we encode the \emph{soft constraint} construct as $\kpr{observe} x \kb{from} \mbb P \deq \kp{score}(\msf{density}_{\mbb P}(x))$. 
The $\kpr{observe} x \kb{from} \mbb P$ denotes a distribution $\mbb P$ that has density with respect to either the Lebesgue measure $\lambda_{\R}$ or the counting measure $\#_{\N}$ (which includes `common' distributions such as normal, binomial, gamma, \etc). 
We write $\msf{density}_{\mbb P}$ to denote the corresponding density function of $\mbb P$
(see works of \citet[\S7]{vakar18} and \citet[\S4]{staton20}). 
Intuitively, the score is higher if the observed $x$ is more likely to be generated from $\mbb P$, and the score is $0$ if the observation is not possible, \eg $\kpr{observe} 2 \kb{from} \msf{Unif}(0, 1)$ is tantamount to $\kp{score}(0)$.

We next consider the \tsc{GaussParam} example by \citet[\S2]{lee12} in \cref{fig:exam:gaussparam}, where we have a normally distributed population with a known standard deviation $\sigma$, and we would like to estimate its mean $\Theta$.
To do this, we assume a normal prior $\Theta \sim \mcal N(\theta_0, \sigma_0)$, where $\theta_0$, $\sigma_0$ are constants. 
Suppose we draw two samples $\{x_1, x_2\}$ from the dataset; we can then write $\kpr{observe} x_i \kpb{from} \msf{Normal}(\Theta, \sigma)$ for $i \in \{1, 2\}$ to compute our updated belief of $\Theta$, as shown in \cref{fig:exam:gaussparam}.
Intuitively, the mean of $\Theta$ shifts \emph{up} when $x_i > \Theta$ and shifts down otherwise. 
Using Bayes' rule, we can compute the updated mean and standard deviation as follows \cite[\S2.2.1]{lee12}:
  \begin{align*}
    \bi{\theta}^{\theta_0, \sigma_0}_{\msf{new}, x} \deq (\bi{\sigma}_{\msf{new}}^{\theta_0, \sigma_0})^2\biggl(\frac{\theta_0}{\sigma_0^2} + \frac{x}{\sigma^2}\biggr)
    \qquad\qquad
    \bi{\sigma}_{\msf{new}}^{\theta_0, \sigma_0} \deq \sqrt{\frac{1}{\sigma_0^{-2} + \sigma^{-2}}}
  \end{align*}

  \begin{figure}[t]
	\hrule
    \begin{mathpar}
      \inferrule[H-Norm-Conj-Norm]
        {}
        {\{\Theta \sim \mcal N(\theta_0, \sigma_0)\}
        \ \kpr{observe} x \kpb{from} \msf{Normal}(\Theta, \sigma)\ 
        \{\Theta \sim \mcal N(\bi{\theta}_{\msf{new}, x}^{\theta_0, \sigma_0}, \bi{\sigma}_{\msf{new}}^{\theta_0, \sigma_0}) * \msf{NormConst}\}} \and
      \inferrule[H-Beta-Conj-Bern]
        {}
        {\{\Theta \sim \msf{Beta}(m, n)\}\ \kr{observe} 1 \kb{from} \msf{Bern}(\Theta)\ \{\Theta \sim \msf{Beta}(m + 1, n) * \msf{NormConst}\}} \and
    \inferrule[H-Gamma-Conj-Poisson]
        {}
        {\{\lambda \sim \Gamma(k, \theta)\}\ \kr{observe} x \kb{from} \msf{Poisson}(\lambda)\ \{\lambda \sim \Gamma(k + x, \nicefrac \theta {\theta + 1}) * \msf{NormConst}\}}
    \end{mathpar} 
	\rule{0.95\textwidth}{.4pt}
    $\begin{array}{l l}
    		\vspace{-5pt}\\
        \trp{\{\Theta \sim \mcal N(\theta_0, \sigma_0)\}} 
        & \annotate{\rrule{H-Sample}} \\
        \quad \kw{observe}\ x\ \kw{from}\ \msf{Normal}(\Theta, \sigma) 
 	   & \annotate{desugars to $\kfont{score}(\msf{normal\tx{-}pdf}(x_1 \alt \Theta, \sigma))$} \\
        \trp{\{\Theta \sim \msf{normal\tx{-}pdf}(x \alt {-}, \sigma) \cdot \mcal N(\theta_0, \sigma_0)\}} 
        & \annotate{\rrule{H-Score}} \\
        \trp{\{\Theta \sim \mcal N(\bi{\theta}_{\msf{new},x}^{\theta_0, \sigma_0}, \bi{\sigma}_{\msf{new}}^{\theta_0, \sigma_0}) * \msf{NormConst}\}} 
        & \annotate{\rrule{H-Cons}}
      \end{array}
      $ \vspace{5pt}\\
      \hrule\vspace{-5pt}
	\caption{Examples of derived conjugate priors in \logic (above); a \logic derivation of \textsc{H-Norm-Conj-Norm} (below), where the \annotate{\!annotation} denotes the \logic rule(s) applied to obtain the associated postcondition.}
    \label{fig:exam:conj}
  \end{figure}

  \begin{figure}[t]
  \hrule\vspace{-5pt}
  \[
    \begin{array}{l l}
      \trp{\{\top_1\}} \\
      \quad \kr{let} \Theta = \kp{sample}(\msf{Normal}(\theta_0, \sigma_0)) \kb{in} \\
      \trp{\{\Theta \sim \mcal N(\theta_0, \sigma_0)\}} 
      & \annotate{\rrule{H-Sample}} \\
      \quad\kpr{observe} x_1 \kpb{from} \msf{Normal}(\Theta, \sigma) 
      & \annotate{let $\theta_{i + 1} \deq \bi{\theta}_{\msf{new}, x_{i + 1}}^{\theta_i, \sigma_i}$; $\sigma_{i + 1} \deq \bi{\sigma}_{\msf{new}}^{\theta_i, \sigma_i}$ for $i \geq 0$} \\ 
      \trp{\{\Theta \sim \mcal N(\theta_1, \sigma_1) * \msf{NormConst}\}} 
      & \annotate{\rrule{H-Norm-Conj-Norm}} \\
      \begin{leftvruled}{\rrule{H-Frame}}
        \trp{\{\Theta \sim \mcal N(\theta_1, \sigma_1)\}} \\
        \quad \kpr{observe} x_2 \kpb{from} \msf{Normal}(\Theta, \sigma) \\
        \trp{\{\Theta \sim \mcal N(\theta_2, \sigma_2) * \msf{NormConst}\}} \\
      \end{leftvruled} 
      & \annotate{\rrule{H-Frame} and \rrule{H-Norm-Conj-Norm}} \\
      \trp{\{\Theta \sim \mcal{N}(\theta_2, \sigma_2) * \msf{NormConst} * \msf{NormConst}\}} \\
      \trp{\{\Theta \sim \mcal{N}(\theta_2, \sigma_2) * \msf{NormConst} \}} 
      & \annotate{\rrule{H-Cons}} \\
      \quad \kpr{return} \Theta \\
      \trp{\{\Theta \sim \mcal{N}(\theta_2, \sigma_2) * \msf{NormConst} \}} 
      & \annotate{\rrule{H-Ret}}
    \end{array}
  \]
  \hrule\vspace{-5pt}
    \caption{A \logic proof sketch of $\tsc{GaussParam}$.
    }
    \label{fig:exam:gauss-posterior}
  \end{figure}

Note that the updated distribution has a closed-form solution (which is not usually the case). 
Indeed, the $\msf{Normal}$-prior/$\msf{Normal}$-likelihood pair is an instance of a \emph{conjugate distribution}, which is a prior/likelihood pair that leads to a closed-form posterior distribution via Bayes' rule. 
  
In \logic we can obtain this fact using the \emph{derived} $\rrule{H-Norm-Conj-Norm}$ rule in \cref{fig:exam:conj} (above), with its \logic derivation given at the bottom of \cref{fig:exam:conj}. 
In fact, the ability to derive conjugate priors in \logic allows us to perform symbolic reasoning. To this end,  in \cref{fig:exam:conj} we list a few common conjugate priors represented as Hoare triples. 
The ability to derive conjugate priors in \logic, along with the frame rule, gives us a foundation for \emph{modular and symbolic reasoning} of probabilistic programs: for $\kpr{observe} x_1 \kb{from} \msf{Normal}(\Theta, \sigma)$, we apply the frame rule to `frame out' the normalising constant, then apply the conjugate distribution triple, which then allows us to symbolically derive the posterior distribution for $\Theta$ in \cref{fig:exam:gauss-posterior}.

\subsection{Verifying the `Hello World' of Probabilistic Programming: Burglar Alarm}\label{subsec:exam:burglar-alarm}

With soft constraint conditioning defined, we now consider a classic example in Bayesian statistics, the \tsc{BurglarAlarm} program at the top of \cref{fig:exam:burglar-alarm}. The problem is as follows: your home has a burglar alarm that activates when there is a burglar, but it also accidentally activates when there is an earthquake. There is a $1\%$ probability that there is a burglar, and a $10\%$ probability that there is an earthquake. When the alarm activates, there is a $99\%$ probability of the phone ringing. Assuming that your phone is ringing now, what is the probability of there being an earthquake? %

We use \logic in \cref{fig:exam:burglar-alarm-post} to prove that the probability of having an earthquake is roughly $84.7\%$ by using a technique called \emph{joint conditioning}. Note that random variable $A$ takes a value in $\{0, 1\}$ depending on whether there is a burglar (captured by $B$) or an earthquake ($E$). By conditioning on the \emph{joint} random variable $(B, E) = (b, e)$ for some $(b, e) \in \{0, 1\}^2$, we know $A$ has distribution $\delta_{b \lor e}$. 
We then apply the rule of consequence \rrule{H-Cons} with our internal notion of Bayes' theorem $\rrule{E-Bayes}$ (explained later in \cref{thm:basl:bayes}), we obtain a posterior distribution stating that there is an $84.7\%$ probability of there being an earthquake, as shown below, with $f$ defined in \cref{fig:exam:burglar-alarm-post}:
\begin{align*}
  \overbrace{{\bc{(b, e)}{(B, E)}{\msf{Bern}(0.01) \otimes \msf{Bern}(0.1)}}}^{\textnormal{prior}}\ \alt\  \overbrace{\score{f(b, e)}\vphantom{\bc{(b, e)}{(B, E)}{\msf{Bern}(0.01) \otimes \msf{Bern}(0.1)}}}^{\textnormal{likelihood}}
  &\vdash \overbrace{(B, E) \sim f \cdot (\msf{Bern}(0.01) \otimes \msf{Bern}(0.1))\vphantom{\bc{(b, e)}{(B, E)}{\msf{Bern}(0.01) \otimes \msf{Bern}(0.1)}}}^{\textnormal{posterior}} \tag{\cref{thm:basl:bayes}} \\
  &\vdash E \sim \msf{Bern}(\nicefrac {0.099}{0.11682}) \tag{calculation}
\end{align*}
\newcommand{\fff}{
  f
  }
\begin{figure}[t]
  $\begin{array}{l l}
    \kr{let} B = \kp{sample}(\msf{Bern}(0.01))\kl{in} & \annotate{probability of burglary is $1\%$} \\
    \kr{let} E = \kp{sample}(\msf{Bern}(0.1))\kl{in} & \annotate{probability of earthquake is $10\%$} \\
    \kr{let} A = \kp{return}(B \lor E) \kl{in} & \annotate{alarm activates in case of burglary or earthquake}\\
    \kp{observe}\ 1 \kb{from} \msf{Bern}(\kr{if}\!A \kb{then} 0.99 \kb{else} 0.01); & \annotate{observe the phone is ringing} \\
    \kr{return} E & \annotate{is there an earthquake?}
  \end{array}$ \\ \vspace{2pt}
	\rule{\textwidth}{.4pt}
\[
  \begin{array}{l l}
    \trp{\{\top_1\}} \\
    \quad\kr{let} B = \kp{sample}(\msf{Bern}(0.01))\kl{in} \\
    \trp{\{B \sim \msf{Bern}(0.01)\}} & \annotate{\rrule{H-Sample}} \\
    \quad\kr{let} E = \kp{sample}(\msf{Bern}(0.1))\kl{in} \\
    \trp{\{B \sim \msf{Bern}(0.01) * E \sim \msf{Bern}(0.1)\}} & \annotate{\rrule{H-Sample}, \rrule{H-Frame}} \\
    \trp{\{(B, E) \sim \msf{Bern}(0.01) \otimes \msf{Bern}(0.1)\}} & \annotate{\rrule{H-Cons}} \\
    \quad\kr{let} A = \kp{return}(B \lor E) \kl{in}\\
    \trp{\{\cond{\bc{(b, e)}{(B, E)}{\msf{Bern}(0.01) \otimes \msf{Bern}(0.1)}} A \sim  \delta_{b \lor e} \}} & \annotate{\rrule{H-CondSample}} \\
    \quad\kp{observe}\ 1 \kb{from} \msf{Bern}(\kr{if}\!A \kb{then} 0.99 \kb{else} 0.01); & \annotate{with $f(b, e) \deq \textnormal{if}\ b \lor e\ \textnormal{then}\ 0.99\ \textnormal{else}\ 0.1 $} \\
    \trp{\left\{\bc{(b, e)}{(B, E)}{\msf{Bern}(0.01) \otimes \msf{Bern}(0.1)}\ \middle|\ A \sim \fff \cdot \delta_{b \lor e} \right\}} & \annotate{\rrule{H-CondScore}} \\
    \trp{\left\{\bc{(b, e)}{(B, E)}{\msf{Bern}(0.01) \otimes \msf{Bern}(0.1)}\ \middle|\ \scorelr{\fff\!(b, e)} \right\}} & \annotate{\rrule{H-Cons}} \\
    \trp{\left\{(B, E) \sim \fff \cdot (\msf{Bern}(0.01) \otimes \msf{Bern}(0.1)) \right\}} & \annotate{\rrule{H-Cons} with \rrule{E-Bayes}} \\
    \trp{\{E \sim \msf{Bern}(\nicefrac{0.099}{0.11682}) * \msf{NormConst} \}} & \annotate{\rrule{H-Cons}} \\
    \trp{\{\mbb{E}[E] = \nicefrac{0.099}{0.11682} * \msf{NormConst} \}} & \annotate{\rrule{H-Frame}, \rrule{H-Cons}} \\
    \quad\kr{return} E &  \\
    \trp{\{\mbb{E}[E] = \nicefrac{0.099}{0.11682} * \msf{NormConst} \}} & \annotate{\rrule{H-Return}}\\
    \trp{\{\Pr[E = 1] \approx 0.847 * \msf{NormConst} \}} & \annotate{\rrule{H-Cons}}
    \vspace{-10pt}
  \end{array}
\]
\caption{\tsc{BurglarAlarm} (above); \logic proof sketch for deriving the posterior probability of earthquake (below)
}
\vspace{-10pt}
\label{fig:exam:burglar-alarm-post}
\label{fig:exam:burglar-alarm}
\end{figure}

\subsection{Reasoning about Independence and Correlations: The Collider Bayesian Network}\label{subsec:exam:common-effect}

We now consider a \emph{Bayesian network} in which two random variables $X$ and $Y$ that are initially independent become negatively correlated after performing Bayesian conditioning.
\logic is the \emph{first} logic that can reason about such conditional dependence brought about by Bayesian updating. 

\begin{wrapfigure}{r}{0.5\textwidth}
  \vspace{-\baselineskip}
  \centering
  \resizebox{0.5\textwidth}{!}{%
    $\displaystyle
    \begin{array}{@{}c@{\quad}c@{}}
      \begin{array}{l}
        \kr{let} X=\kp{sample}(\msf{Bern}(\nicefrac 1 2)) \kb{in}\\
        \kr{let} Y=\kp{sample}(\msf{Bern}(\nicefrac 1 2)) \kb{in}\\
        \kr{let} Z=\kp{return}(X\lor Y)\;\kb{in}\\
        \kp{observe}(Z = 1);\\
        \kr{return}(X,Y)
      \end{array}
      &
      \vcenter{\hbox{%
        \begin{tikzpicture}[>=latex,node distance=1.2cm,baseline=(Z.center)]
          \node[draw,circle]            (X) {$X$};
          \node[draw,circle,right of=X] (Y) {$Y$};
          \node[draw,circle,fill=gray!20,below=of $(X)!0.5!(Y)$] (Z) {$Z$};
          \draw[->] (X) -- (Z);
          \draw[->] (Y) -- (Z);
        \end{tikzpicture}%
      }}
    \end{array}
    $
  }
  \caption{\tsc{Collider} and its Bayesian network}
  \label{fig:exam:common-effect}
  \vspace{-10pt}
\end{wrapfigure}

A Bayesian network is a directed acyclic graph modelling relationship of random variables. A common structure in Bayesian networks is known as \emph{colliders} (or \emph{common effects}), where the distribution of a random variable $Z$ is conditionally independent upon $X$ and $Y$. 
For our example in \cref{fig:exam:common-effect}, we flip two coins $X$ and $Y$ (taking values in $\{0, 1\}$) and let $Z$ be the maximum of $X$ and $Y$. Before observing $Z$, $X$ and $Y$ are independent.
That is, we can use \logic to prove $X \sim \msf{Bern}(\nicefrac 1 2) * Y \sim \msf{Bern}(\nicefrac 1 2)$ prior to observing $Z$. 
However, once we perform Bayesian conditioning by observing $Z = 1$, $X$ and $Y$ are no longer independent: knowing the value of $Z$ gives us information about $X$ and $Y$. For example, knowing $Z \!=\! 1$ and $X \!=\! 0$ gives us extra information $Y = 1$. In fact, not only $X$ and $Y$ are now dependent, they are also \emph{negatively correlated}: when $X$ has a higher value, $Y$ is likely to be lower, and vice versa.

We use \logic to prove this negative correlation as shown in \cref{fig:exam:corr-proof}. 
We first apply $\rrule{H-Sample}$ to sample $X$ with distribution $\msf{Bern}(\nicefrac{1}{2})$, then apply $\rrule{H-Frame}$ and $\rrule{H-Sample}$ to obtain $Y \sim \msf{Bern}(\nicefrac{1}{2})$ and `frame' $Y$ onto $X$ to obtain $X \sim \msf{Bern}(\nicefrac{1}{2}) * Y \sim \msf{Bern}(\nicefrac 1 2)$.
As we show shortly in \cref{sec:semantics}, the (bi)entailment $X \sim \mu * Y \sim \nu \vdashv (X, Y) \sim \mu \otimes \nu$ holds and we thus use $\rrule{H-Cons}$ to obtain $(X, Y) \sim  \msf{Bern}(\nicefrac 1 2) \otimes  \msf{Bern}(\nicefrac 1 2)$.

\begin{wrapfigure}{r}{0.52\textwidth}
  \vspace{-10pt}
  $\begin{array}{l}
    \trp{\{\top_1\}} \\
    \quad\kr{let} X = \kp{sample}(\msf{Bern}(\nicefrac 1 2)) \kb{in} \\
    \trp{\{X \sim \msf{Bern}(\nicefrac 1 2)\}} \\
    \quad\kr{let} Y = \kp{sample}(\msf{Bern}(\nicefrac 1 2)) \kb{in} \\
    \trp{\{X \sim \msf{Bern}(\nicefrac 1 2) * Y \sim \msf{Bern}(\nicefrac 1 2)\}} \\
    \trp{\{(X, Y) \sim \msf{Bern}(\nicefrac 1 2) \otimes \msf{Bern}(\nicefrac 1 2)\}} \\
    \quad\kr{let} Z = \kpr{return} X \lor Y \kb{in} \\
    \trp{\{\cond{\bc{(x, y)}{(X, Y)}{\msf{Bern}(\nicefrac 1 2) \otimes \msf{Bern}(\nicefrac 1 2)}} Z \sim \delta_{x \lor y}\}} \\
    \quad\kp{observe}(Z = 1); \\
    \trp{\{\cond{\bc{(x, y)}{(X, Y)}{\msf{Bern}(\nicefrac 1 2) \otimes \msf{Bern}(\nicefrac 1 2)}} Z \sim \ell_{=1} \cdot \delta_{x \lor y}\}} \\
    \trp{\{\cond{\bc{(x, y)}{(X, Y)}{\msf{Bern}(\nicefrac 1 2) \otimes \msf{Bern}(\nicefrac 1 2)}} \score{1 - [x = y = 0]}\}} \\
    \quad\kr{return} (X, Y) \\
    \trp{\{(X, Y) \sim (1 - [x = y = 0]) \cdot \msf{Bern}^2(\nicefrac 1 2)\}} \\
    \trp{\{(X, Y) \sim \msf{Unif}\{(0, 1), (1, 0), (1, 1)\} * \msf{NormConst}\}} \\
    \trp{\{\mbb E[XY] = \nicefrac 1 3 \land \mbb E[X] = \mbb E[Y] = \nicefrac 2 3 * \msf{NormConst}\}} \\
    \trp{\{\msf{Cov}[X, Y] < 0 * \msf{NormConst}\}}
  \end{array}$
  \vspace{-10pt}
  \caption{A \logic proof sketch of \tsc{Collider} showing that return values $(X, Y)$ are negatively correlated}
  \label{fig:exam:corr-proof}
  \vspace{-10pt}
\end{wrapfigure}
We next condition on $(X, Y)$ and sample $Z$ using the conditional sampling axiom $\rrule{H-CondSample}$ to show $Z$ has a conditional distribution $\delta_{x \lor y}$, (jointly) conditioning on the random variable $(X, Y) \!=\! (x, y)$ for $(x, y) \!\in\! \{0, 1\}^2$. Upon subsequently observing $Z \!=\! 1$,  we apply the conditional observation axiom $\rrule{H-CondObserve}$ to prove the updated distribution `ignores' the case when $X \!=\! 0$ and $Y \!=\! 0$. 
Let us consider the updated likelihood of all four cases of $(x, y) \in \{0, 1\}^2$: when $x \!=\! y \!=\! 0$, the likelihood is $0$, otherwise it is $1$. That is, the likelihood proposition $\score{\msf{if}\ x \!=\! y \!=\! 0\ \msf{then}\ 0\ \msf{else}\ 1}$ holds, or equivalently, $\score{1 - [x \!=\! y \!=\! 0]}$ holds. 
Using the \emph{internal Bayes' theorem} (formulated later in \cref{thm:basl:bayes}), we thus know the updated $(X, Y)$ has distribution $(X, Y) \sim (1 - [x \!=\! y \!=\! 0]) \cdot \msf{Bern}(\nicefrac 1 2)^2$. 
We compute the probability of $(X, Y)$ being $(0, 1)$, $(1, 0)$ or $(1, 1)$ is $ \nicefrac 1 3$ in each case,  and thus calculate the covariance $\msf{Cov}[X, Y] \!=\! \mbb E[XY] - \mbb E[X]\mbb E[Y] \!=\! \nicefrac 1 3 - \nicefrac 2 3 \cdot \nicefrac 2 3 = -\nicefrac 1 9$. 
Using $\rrule{H-Cons}$, we then formally prove $\msf{Cov}[X, Y] < 0$, stating that $X$ and $Y$ are negatively correlated.

\subsection{Modelling Improper Prior: Correctness of the Semantic Lebesgue Measure}\label{subsec:exam:improper}

We now consider a common technique in Bayesian statistical modelling known as \emph{improper priors}. An improper prior is a distribution with an infinite normalising constant. One canonical such example \cite{staton20,narayanan16} is the \emph{Lebesgue measure}, $\msf{Leb}_{[0, \infty)}$, which assigns length $b - a$ to any interval $(a, b)$ with $0 \le a < b$. 
As shown by \citet{staton20},  a $\lang$ program can simulate the Lebesgue measure  as the computation $\kr{let} X = \kp{sample}(\msf{Exp}(1)) \kb{in} \kw{score}(e^X)$ since the \emph{measure} (as opposed to probability) of $X \in (a, b)$ is $\int_{(a, b)} e^x\,e^{-x}\dd{x} = b - a$. In fact, $X$ has an \emph{improper prior distribution} -- the normalising constant is currently $\int_0^\infty e^x \cdot e^{-x}\,\dd{x} = \int_{0}^\infty 1\,\dd{x} = \infty$.

\logic is the \emph{first} logic that can reason about improper priors because its semantic domain include \emph{$\sigma$-finite measure spaces}. 
We present a \logic proof sketch of the computation in \cref{fig:exam:leb}, proving that $X$ is distributed according to the Lebesgue measure: $X \sim \msf{Leb}_{[0, \infty)}$. 
In fact, since $X \sim \msf{Leb}_{[0, \infty)}$, we know that for any measure $\mu$ with a density with respect to $\msf{Leb}_{[0, \infty)}$ (\eg $\mu = \msf{Unif}(0, 1)$), observing $X$ to have distribution $\mu$ results in $X \sim \mu$. We can derive this in \logic as shown  in \cref{fig:exam:lebcond}.
\begin{figure}[t]
  \begin{minipage}[b]{0.45\textwidth}
    $\begin{array}{l l}
    \trp{\{\top_1\}} \\
    \quad\kr{let} X = \kp{sample}(\msf{Exp}(1)) \kl{in} \\
    \trp{\{X \sim \msf{Exp}(1)\}} & \annotate{\rrule{H-Sample}} \\
    \quad\kw{score}(e^X) \\
    \trp{\{X \sim \msf{exp} \cdot \msf{Exp}(1)\}} & \annotate{\rrule{H-Score}} \\
    \trp{\{X \sim \msf{Leb}_{[0, \infty)}\}} & \annotate{\rrule{H-Cons}}  
    \vspace{-10pt}
  \end{array}$
  \caption{\tsc{Lebesgue}}
  \vspace{-10pt}
  \label{fig:exam:leb}
  \end{minipage}
  \hspace{35pt}
  \begin{minipage}[b]{0.45\textwidth}
    $\begin{array}{l l}
    \quad\trp{\{X \sim \msf{Leb}_{[0, \infty)}\}} \\
    \quad\quad \kpr{observe} X \kpb{from} \mu \\
    \quad\trp{\{X \sim \msf{pdf}_{\mu} \cdot \msf{Leb}_{[0, \infty)}\}} & \annotate{\rrule{H-Score}} \\
    \quad\trp{\{X \sim \mu\}} & \annotate{\rrule{H-Cons}}
    \vspace{-10pt}
  \end{array}$
  \caption{Conditioning of improper Lebesgue prior}
  \vspace{-10pt}
  \label{fig:exam:lebcond}
  \end{minipage}
\end{figure}

\subsection{Representing the Posterior of a Bayesian Clustering Algorithm}\label{subsec:exam:cluster}

For our next example, we use $\logic$ to calculate the posterior of a clustering algorithm that implements the Gaussian mixture model (GMM). The problem is as follows: suppose we have a dataset $\{x_i \in \R\}_{i = 1}^n$ and we want to cluster them into two groups. 
The model works by assuming there are two normally-distributed latent variables $\mu_0, \mu_1 \in \R$ that represent the mean of the two groups and another latent variable $\pi \in [0, 1]$ representing the proportion of samples belonging to the two groups. For each $x_i$, we sample a Bernoulli random variable $N$ with success probability $\pi$ and we assume $x_i$ is a sample drawn from $\mcal N(\mu_N, 1)$ by performing soft conditioning. The diagram at the top of \cref{fig:exam:gmm} illustrates the main idea of GMM: given a dataset $\{x_i\}_{i = 1}^n$, we infer the mean of clusters $\mu_0$ and $\mu_1$ and fit the data via the mixture of two Gaussian distributions.

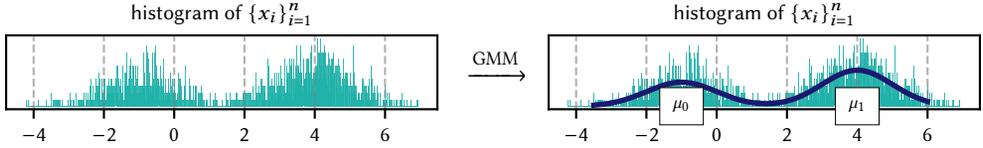
\begin{wrapfigure}{r}{0.38\textwidth}
  $\inferrule[H-BoundedFor]
    { \textnormal{for $i = 1, ..., n$, } \{P_{i - 1}\}\ M[\nicefrac {x_i} X]\ \{P_{i}\}}
    {\{P_0\} \kb{for} X \kb{in} [x_1, ..., x_n] \kb{do} M\ \{P_n\} }$
  \vspace{-5pt}
\end{wrapfigure}
Unlike the previous examples, GMM does not have a closed-form solution -- we must rely on inference algorithms such as MCMC to approximate the posterior distribution of $(\mu, \pi)$. However, with $\logic$, we can symbolically represent the formula for the posterior distribution. Similar to $\lilac$, we extend $\lang$ with a syntactic construct that encodes bounded loop over a literal list $[x_1, ..., x_n]$ by defining $\kr{for} X \kb{in} [x_1, ..., x_n] \kb{do} M \deq M[x_1/X]; ...; M[x_n/X]$.
By applying the sequencing rule $\rrule{H-Let}$ inductively, we obtain the rule $\rrule{H-BoundedFor}$, which allows us to express GMM and represent the posterior of $(\mu, \pi)$ in \cref{fig:exam:gmm}. Even though the posterior distribution cannot be explicitly simplified, we can still represent it as the postcondition $(\mu, \pi) \sim \mbb P_n$.
\begin{figure}[t]
  \[
    \mathord{\input{code/samples.pgf}}
  \vcenter{\hbox{\raisebox{2.7cm}{
    \begin{minipage}{0.05\linewidth}
      $\xrightarrow{\small\textnormal{GMM}}$
    \end{minipage}}
  }}
  \mathord{\input{code/mixture.pgf}}\vspace{-50pt}
  \]
  \hrule
\[
\begin{array}{ll}
  \trp{\{\top_1\}} \\
  \quad\kr{let} \pi = \kp{sample}(\msf{Unif}(0, 1)) \kl{in} & \annotate{sample the proportion of each cluster} \\
  \trp{\{\pi \sim \msf{Unif}(0, 1)\}} & \annotate{\rrule{H-Sample}, \rrule{H-Frame}} \\
  \quad\kr{let} \mu = \kp{sample}(\mcal N(0, 10) \otimes \mcal N(0, 10)) \kl{in} & \annotate{sample from two normal distributions} \\
  \trp{\{\mu \sim \mcal{N}^2(0, 10) * \pi \sim \msf{Unif}(0, 10)\}} & \annotate{\rrule{H-Sample}} \\
  \quad\kr{for} X \kb{in} [x_1, ..., x_n] \kl{do} & \annotate{let $\mathbb{P}_0 \deq \mcal N^2(0, 1) \otimes \msf{Unif}(0, 1)$} \\
  \quad\quad\trp{\{(\mu, \pi) \sim \mbb P_i\}} & \annotate{\rrule{H-Cons}} \\
  \quad\quad\quad \kr{let} I = \kp{sample}(\msf{Bern}(\pi)) \kl{in} & \annotate{let $f_i(m, p) \deq \mcal (1 - p)\mcal N(x_i \alt m_0, 1) + p\mcal N(x_i \alt m_1, 1)$} \\
  \quad\quad\trp{\{\cond{\bc{(m, p)}{(\mu, \pi)}{\mbb P_i}} I \sim \msf{Bern}(p)\}} & \annotate{\rrule{H-CondSample}} \\
  \quad\quad\quad \kr{observe} X \kb{from} \msf{Normal}(\mu_I, 1) & \annotate{let $\mbb P_{i + 1} \deq f_i \cdot \mbb P_i$} \\
  \quad\quad\trp{\{\cond{\bc{(m, p)}{(\mu, \pi)}{\mbb P_i}} I \sim f_i \cdot \msf{Bern}(p)\}} & \annotate{\rrule{H-CondScore}} \\
  \quad\quad\trp{\{(\mu, \pi) \sim \mbb P_{i} \}} & \annotate{\rrule{H-Cons}, \rrule{E-Bayes}} \\
  \trp{\{(\mu, \pi) \sim \mbb P_n\}} & \annotate{\rrule{H-BoundedFor}}
\end{array}
\]
\vspace{-5pt}
\caption{An illustration Gaussian mixture model (above); an implementation of the Gaussian mixture model and a \logic proof sketch deriving its posterior (below)}
\label{fig:exam:gmm}
\vspace{-5pt}
\end{figure}

\subsection{Rewriting Probabilistic Programs}\label{subsec:exam:simp}

As we briefly discussed in \cref{sec:intro}, we can use \logic to justify \emph{rewriting} probabilistic program by proving that they satisfy equivalent specifications (Hoare triples).
Let us revisit the example in \cref{sec:intro} (\p\pageref{example:rewriting}); 
as sown in \cref{fig:exam:simp}, we can combine the techniques introduced thus far (in \cref{subsec:exam:soft} to \cref{subsec:exam:cluster}) to show that the two programs satisfy equivalent specifications. 
The program on the left contains normally-distributed and beta-distributed random variables $X$ and $Y$, respectively, along with conditioning constructs on two lists of observed data $\{x_i\}_{i = 1}^N$ and $\{c_i\}_{i = 1}^N$. 
Note that two kinds of conjugacy exist within the program: the normal-normal conjugate and the beta-Bernoulli conjugate. 
As such, we can simplify the program on the left by \emph{pre-computing} the posterior and sampling from them directly, as in the program on the right.  
However, how do we prove that the simplification is sound, \ie the two programs are equivalent (up to a normalising constant)? 
We can do this in \logic by proving the \emph{same specification} (Hoare triple) for both programs, as shown in \cref{fig:exam:simp}.
\begin{figure}[t]
\[
\begin{array}{@{}l@{\hspace{-15pt}}l@{}}
  \begin{array}{l}
    \trp{ \{\top_1\} } \\
    \quad\kr{let} X = \kw{sample}(\mathcal{N}(0, 1)) \kl{in} \\
    \trp{ \{X \sim \mcal{N}(0, 1)\} } \;\annotate{\rrule{H-Sample}}\\
    \quad\kr{let} Y = \kw{sample}(\msf{Beta}(m, n)) \kl{in} \\
    \trp{ \{X \sim \mcal{N}(0, 1) * Y \sim \msf{Beta}(m, n)\} } \;\annotate{\rrule{H-Sample}}\\
    \trp{ \{X \sim \mcal{N}(0, 1) * Y \sim \msf{Beta}(m, n) * \msf{NC}\} } \;\annotate{\rrule{H-Cons}}\\
    \annotate{define $a_i \deq \sum_{j = 1}^{i} [c_j = 1]$, $b_i \deq \sum_{j = 1}^{i} [c_j = 0]$} \\
    \annotate{define $c_i \deq \sum_{j = 1}^{i} \frac{\sum^{i}_{j = 1} x_j}{i}$} \\
    \quad\kr{for} I \kb{in} [1, ..., N] \kl{do} \\
    \postc{X \sim \mcal{N}(c_i, 1) \ *\\ Y \sim \msf{Beta}(m + a_i, n + b_i) * \msf{NC} } \;\annotate{loop invariant}\\
    \quad\quad\kr{observe} x_I \kb{from} \mcal{N}(M, 1); \\
    \begin{array}{@{} l @{}  l @{}}
	    \postc{X \sim \mcal{N}(c_{i + 1}, 1) * \msf{NC} \ * \\
	     Y \sim \msf{Beta}(m + a_i, n + b_i) } 
	     & 
	     \begin{array}{@{} l @{}} 
	     	\;\annotate{\rrule{H-Norm-Conj-Norm}}\\
	     	\;\annotate{\rrule{H-Cons}}
	      \end{array}	
	 \end{array}      \\
	\quad\quad\kr{observe} c_I \kb{from} \msf{Bern}(W) \\ 
    \postc{
    	X \sim \mcal{N}(c_{i + 1}, 1) * \msf{NC}\ * \\
    	Y \sim \msf{Beta}(m + a_{i + 1}, n + b_{i + 1})
    } \;\annotate{\rrule{H-Beta-Conj-Bern}}\\
    \postc{X \sim \mcal{N}(c_N, 1) \ * \\Y \sim \msf{Beta}(m + a_{N}, n + b_{N}) * \msf{NC}} 
    \;\annotate{\rrule{H-BoundedFor}}\\
  \end{array}
&
  \begin{array}{l}
    \trp{ \{\top_1\} } \\
    \quad\kr{let} X = \kw{sample}(\mcal{N}(c_N, 1)) \kl{in} \\
    \trp{ \{X \sim \mcal{N}(c_N, 1)\} } \;\annotate{\rrule{H-Sample}} \\
    \quad\kr{let} Y = \kw{sample}(\msf{Beta}(n + a_N, m + b_N)) \kl{in} \\
    \postc{X \sim \mcal{N}(c_N, 1) \ * \\ Y \sim \msf{Beta}(m + a_N, n + b_N) } \;\annotate{\rrule{H-Sample}} \\
    \quad\kr{return} (X, Y) \\
    \postc{X \sim \mcal{N}(c_N, 1) \ * \\ Y \sim \msf{Beta}(m + a_N, n + b_N) } \;\annotate{\rrule{H-Ret}}
  \end{array}
\end{array} 
\]
\vspace{-5pt}
\caption{Using \logic to justify that the program on the left can be \emph{soundly rewritten} to that on the right by proving that they both satisfy the \emph{same \logic specification} (\logic triple).}
\label{fig:exam:simp}
\vspace{-5pt}
\end{figure}

Specifically, using \rrule{H-Norm-Conj-Norm} and \rrule{H-Beta-Conj-Bern}, we establish a loop invariant and reason about how the distributions of $X$ and $Y$ change at every iteration $I$. 
We then apply the Hoare triple for bounded for-loops \rrule{H-BoundedFor}. 
As such, we can show that the random variables in both programs have the same distribution (up to a normalising constant $\msf{NC}$, which can be safely ignored since inference algorithms return the same samples, regardless of the normalising constant), and therefore we can justify rewriting the program on the left to that on the right.

\smallskip
Through numerous examples showcasing the novel and hitherto-unsupported features of \logic, we have demonstrated the expressivity of \logic, and we believe this can serve as a logical foundation for static analysis/symbolic execution tools for probabilistic programs.

\section{The Semantics of $\logic$}\label{sec:basl}
\label{sec:semantics}

Now that we have demonstrated how $\logic$ can function as a logical framework for proving properties in statistical models, we explain its resource-theoretic semantics in \cref{subsec:basl:sem}. But before this, we review the Kripke resource model of the $\lilac$ separation logic \cite{li23} in \cref{subsec:overview:pcm}, which the model of $\logic$ is based on, then we motivate the need for a generalised model of randomness for Bayesian updating in \cref{subsec:overview:bayespcm} and prove that it is indeed a resource model in \cref{subsec:basl:kripke}.

\subsection{Background: A Resource Monoid for Randomisation}\label{subsec:overview:pcm}

In separation logic, \emph{computational resources} such as heaps are modelled by \emph{partial commutative monoids} $(\mcal M, \bullet, \mb 1)$ \cite{calcagno07}. The set $\mcal M$ represents  states of the resource and the partial function $(\bullet) : \mcal M \times \mcal M \pto \mcal M$ combines two states $m_1, m_2 \in \mcal M$ if they are \emph{compatible} (\eg $m_1$ and $m_2$ describe different parts of the resource). For example, the heap is modelled by $\mcal M_{\msf{heap}} \deq \msf{Loc} \pto_{\tx{fin}} \msf{Val}$, where $\msf {Loc}$ is the set of memory addresses and $\msf{Val}$ is the set of values. For instance, $\{42 \mapsto \msf{``a"}\} \in \mcal M_{\msf{heap}}$ represents a heap where address $42$ stores value $\msf{``a"}$. Moreover, given two heaps $m_1, m_2$, $(\bullet)$ combines them if they contain separate addresses. For example, $m_1 \deq \set{21 \mapsto 2025, 42 \mapsto \msf{``a"}}$ and $m_2 \deq \set{52 \mapsto 123}$ can be combined to $m_1 \bullet m_2 = \set{21 \mapsto 2025, 42 \mapsto \msf{``a"}, 52 \mapsto 123}$ since the addresses in $m_1$ ($21$ and $42$) do not overlap with $52$ in $m_2$. There is also an identity element $\mb 1 \in \mcal M$ that represents the empty resource. For heaps, $\mb 1$ is defined to be the empty heap $\set{}$.

Applying the same intuition to probability, one reasonable model of computational resource is that of a \emph{random number generator}. 

The $\lilac$ separation logic defined a partial commutative monoid that models a random generator by encoding the following two properties:
\begin{enumerate}
  \item the \emph{distribution} of the generated numbers, \ie a random number generator should have information about the distribution of the numbers generated; and
  \item the \emph{usage} of the generator, which has information regarding whether the $i$-th number $\omega_i$ has been generated.
\end{enumerate}

To model (1) and (2), $\lilac$ uses a \emph{measure space}. In fact, the distribution and usage property can be modelled via the \emph{measures} and \emph{$\sigma$-algebras}, respectively. For readers unfamiliar with these concepts, a \emph{measurable space} is a pair $(\Omega, \mcal F)$, where $\Omega$ is a set and $\mcal F \subseteq \mcal P(\Omega)$ is a set of subsets such that $\varnothing, \Omega \in \mcal F$ and $\mcal F$ is closed under complements and countable unions. We call $\Omega$ the \emph{sample space} and $\mcal F$ a \emph{$\sigma$-algebra} of $\Omega$. A \emph{measure} of $(\Omega, \mcal F)$ is a function $\mu : \mcal F \to [0, \infty]$ satisfying $\mu(\varnothing) = 0$ and $\mu(\biguplus_{i \in \N} U_i) = \sum_{i \in \N} \mu(U_i)$. The triple $(\Omega, \mcal F, \mu)$ forms a \emph{measure space}. If $\mu(\Omega) = 1$, we additionally call $(\Omega, \mcal F, \mu)$ a \emph{probability space}.  For example, to model a fair die, we set $\Omega \deq \set{1, 2, 3, 4, 5, 6}$, $\mcal F \deq \mcal P(\Omega)$, and $\mu(U) \deq \sum_{i = 1}^6 \ch{U}(i) \cdot \nicefrac{1}{6}$, where $\ch{U}(i)$ is 1 when $i \in U$ and $0$ when $i \notin U$. We can then compute that the probability of the die having a value greater than four is $\mu(\set{5, 6}) = \nicefrac{1}{3}$.

We now consider a simplified model of $\lilac$'s random number generator -- we define our sample space to be $\Omega \deq \set{\true, \false}^2$ (as opposed to $[0, 1]^\N$ in the full model of $\lilac$), which, intuitively, can be thought of as the computer having access to a source of randomness that can generate two independent booleans. 
The states of the random generator is then modelled by the following set:
\[
  \mcal M \deq \{(\mcal F, \mu) \alt
    (\Omega, \mcal F, \mu) \tx{ is a probability space}
  \}
\]
To explain how $\mcal M$ models random generators, we construct the empty generator $\mb 1 \deq (\mcal F_{\mb 1}, \mu_{\mb 1})$ -- an element of $\mcal M$ that represents a state where nothing has been generated. This means:
\begin{enumerate}
  \item We \emph{do not} know the probability of the first boolean being $\false$ and second boolean being $\true$. Hence, $\set{(\false, \true)} \notin \mcal F_{\mb 1}$ and we cannot apply $\mu_{\mb 1} : \mcal F_{\mb 1} \to [0, \infty]$ to $\set{(\false, \true)}$ to get the probability. Similarly, $\set{(a, b)} \notin \mcal F_{\mb 1}$ for $a, b \in \set{\true, \false}$.
  \item We \emph{do not} know the probability of the first boolean being $\true$, which is equivalent to saying the first boolean being $\true$ and the second boolean being $\true$ \emph{or} $\false$. Hence, $\set{(\true, \true), (\true, \false)} \notin \mcal F_{\mb 1}$. Similarly, $\set{(a,b), (c, d)} \notin \mcal F_1$ for $a, b, c, d \in \set{\true, \false}$.
  \item Even though the first number has not been generated, we know that the first and/or second boolean will either be $\true$ or $\false$. Hence, $\set{(\true, \false), (\true, \true), (\false, \true), (\false, \false)} = \Omega \in \mcal F_{\mb 1}$.
\end{enumerate}
For $\mcal F_{\mb 1}$ to be a $\sigma$-algebra, we need $\varnothing \in \mcal F_{\mb 1}$; hence, $\mcal F_{\mb 1} = \set{\varnothing, \Omega}$. For the probability measure $\mu_{\mb 1}: \mcal F_{\mb 1} \to [0, \infty]$, we know that the first and second boolean have a $100\%$ probability of being $\true$ or $\false$. Hence, $\mu_{\mb 1}(\set{(\true, \false), (\true, \true), (\false, \true), (\false, \false)}) = 1$. See \cref{fig:overview:rng} for an illustration.

Next, we construct $m_2 \deq (\mcal F_2, \mu_2) \in \mcal M$ where the first boolean has a $42\%$ probability of being $\true$ and $58\%$ probability of being $\false$, and only the first boolean has been generated. This means:
\begin{enumerate}
  \item We know the probability of the first boolean being $\true$ is $42\%$, which is equivalent to saying the first boolean being $\true$ and the second boolean being $\true$ or $\false$ is $42\%$. Hence, $\set{(\true, \false), (\true, \true)} \in \mcal F_2$ and $\mu_2(\set{(\true, \false), (\true, \true)}) \deq 0.42$. Similarly, $\set{(\false, \false), (\false, \true)} \in \mcal F_2$ and $\mu_2(\set{(\false, \false), (\false, \true)}) \deq 0.58$.
  \item We \emph{do not} know the probability of the second boolean being $\false$, which is equivalent to saying the second boolean being $\false$ and the first boolean being $\true$ or $\false$. Hence, $\set{(\true, \false), (\false, \false)} \notin \mcal F_2$. Similarly, the singletons $\set{(a, b)}$ with $a, b \in \set{\true, \false}$ and $\set{(\true, \true), (\false, \true)}$ should not be in $\mcal F_2$ since they contain information about the second boolean.
\end{enumerate}
Since $\set{(\true, \true), (\false, \true)}$ and $\set{(\false, \false), (\false, \true)}$ are in $\mcal F_2$, their union $\Omega$ is also in $\mcal F_2$ and $\mu_2(\Omega) = 1$, which makes sense since the probability of either events happening is $42\% + 58\% = 100\%$.

Similarly, we can construct $m_3 = (\mcal F_3, \mu_3) \in \mcal M$ where only the second boolean has been generated with $30\%$ probability of being $\true$ and $70\%$ probability of being $\false$. Now, since $m_2$ only has information about the first boolean and $m_3$ only has information about the second boolean, they can be combined to form $m_2 \bullet m_3 = (\mcal F_{23}, \mu_{23})$ such that we can refer to \eg the probability of the first boolean being $\true$ and the second boolean being $\false$, \ie$\set{(\true, \false)} \in \mcal F_{23}$ and $\mu_{23}(\set{
\true, \false}) \deq \mu_2(\set{(\true, \false), (\true, \true)}) \cdot \mu_3(\set{(\false, \false), (\false, \true)}) = 42\% \cdot 70\% = 29.4\%$ (see \cref{fig:overview:rng}). 
As shown by \citet{li23}, $(\mcal M, \bullet, \mb 1)$ forms a partial commutative monoid (PCM). 

\begin{figure}
  \[\begin{array}{@{}c@{\ }c@{\ }c@{\ }c@{\ }c@{\ }c@{}}
  \begin{tikzpicture}
    \node (m1) {
      \begin{tikzpicture}
        \node (r1) [nonactivecell] {$\scriptstyle B_1$};
        \node (r2) [nonactivecell, right = -1pt of r1] {$\scriptstyle B_2$};
        \pic [below = -1pt of r1, xshift=-0.1pt] {emptydistribution};
        \pic [below = -1pt of r2, xshift=0.1pt] {emptydistribution};
      \end{tikzpicture}
    };
    \node[left=of m1, xshift=33pt, yshift=5pt] {$\mb 1 =$};
  \end{tikzpicture}
  &
  \begin{tikzpicture}
    \node (m2) {
      \begin{tikzpicture}
        \node (r1) [activecell] {$\scriptstyle B_1$};
        \node (r2) [nonactivecell, right = -1pt of r1] {$\scriptstyle B_2$};
        \pic [below = -1pt of r1] {distribution={2.1mm}{2.9mm}};
        \pic [below = -1pt of r2] {emptydistribution};
      \end{tikzpicture}
    };
    \node[left=of m2, xshift=33pt, yshift=5pt] {$m_2 =$};
  \end{tikzpicture}
  &
  \begin{tikzpicture}
    \node (m3) {
      \begin{tikzpicture}
        \node (r1) [nonactivecell] {$\scriptstyle B_1$};
        \node (r2) [activecell, right = -1pt of r1] {$\scriptstyle B_2$};
        \pic [below = -1pt of r1] {emptydistribution};
        \pic [below = -1pt of r2] {distribution={4mm}{1mm}};
      \end{tikzpicture}
    };
    \node[left=of m3, xshift=33pt, yshift=5pt] {$m_3 =$};
  \end{tikzpicture}
  &
    \begin{tikzpicture}
      \node (m12) {
        \begin{tikzpicture}
          \node (r1) [activecell] {$\scriptstyle B_1$};
          \node (r2) [nonactivecell, right = -1pt of r1] {$\scriptstyle B_2$};
          \pic [below = -1pt of r1] {distribution={2.1mm}{2.9mm}};
          \pic [below = -1pt of r2] {emptydistribution};
        \end{tikzpicture}
      };
      \node[left=of m12, xshift=33pt, yshift=5pt] {$\mb 1 \bullet m_2 = $};
    \end{tikzpicture}
    &
    \begin{tikzpicture}
      \node (m23) {
        \begin{tikzpicture}
          \node (r1) [activecell] {$\scriptstyle B_1$};
          \node (r2) [activecell, right = -1pt of r1] {$\scriptstyle B_2$};
          \pic [below = -1pt of r1] {distribution={2.1mm}{2.9mm}};
          \pic [below = -1pt of r2] {distribution={4mm}{1mm}};
        \end{tikzpicture}
      };
      \node[left=of m23, xshift=33pt, yshift=5pt] {$m_2 \bullet m_3 = $};
    \end{tikzpicture}
  \end{array}
\]
  \caption{Illustrations of random number generators}
  \label{fig:overview:rng}
\end{figure}

\subsection{The Need for an Extended Resource Model of Randomness}\label{subsec:overview:bayespcm}

We now motivate the semantics of $\logic$ by considering a problem we would like to solve; namely, how a random generator, such as the ones presented above, can be \emph{updated}. In the heap model of separation logic (see first paragraph of \cref{subsec:overview:pcm}), an update on address $l$ with value $v$ is modelled as an \emph{update function} $(-)[l \mapsto v] : \mcal M_{\msf{heap}} \to \mcal M_{\msf{heap}}$ defined by $m[l \mapsto v] \deq m \cup \{l \mapsto v\}$. 
We develop an analogous operation on random number generators, for which we must model the $\kp{observe}$/$\kp{score}$ construct. But first, let us establish the motivation: why would we want to update a distribution? And why is it such a notoriously hard problem? The answer is twofold: we need to handle both \emph{unnormalised measures} and the update's effect on \emph{dependent random variables}.

\paragraph{Unnormalised Measures}
Consider an experiment where we flip two fair coins $B_1$ and $B_2$ (modelled as booleans). 
The resulting random generator can be visualised as $m_\tx{before}$ in \cref{fig:overview:update}. Suppose that, based on data, we know $B_2$ must be $\true$; we can then update our belief about $B_2$ via a \emph{likelihood function}, which is a function of type $\ell : \Omega \to \Rnn$, defined by $\ell(b_1, \true) \deq 1$ and $\ell(b_1, \false) \deq 0$. For any likelihood function $\ell$ and $m = (\mcal P(\Omega), \mu) \in \mcal M$, there is a natural update operation $(\cdot)$ such that $(\ell \cdot \mu) : \mcal P(\Omega) \to \W$ is a measure on $(\Omega, \mcal P(\Omega))$ defined as follows:
\[
  (\ell \cdot \mu)(E) \deq \sum_{(b_1, b_2) \in E} \ell(b_1, b_2) \cdot \mu(\{(b_1, b_2)\})
\]
This update scales the measure $\mu$ according to $\ell$. 
Let us write $(\ell \cdot m) \deq (\mcal F, \ell \cdot \mu)$ when $m = (\mcal F, \mu) \in \mathcal M$ as a shorthand. 
Then the update $m_\tx{after} \deq \ell \cdot m_{\tx{before}}$ reflects our belief: the probability that $B_2 = \false$ is now zero since $\ell(b_1, \false) = 0$ and $m_{\tx{after}}(\{B_2 = \false\}) = m_{\tx{after}}(\{(\true, \false), (\false, \false)\}) = 0$. However, there is a problem: $m_\tx{after}$ is no longer a probability measure because it does not add up to $100\%$:
\[
  \mu_\tx{after}(\Omega) = \mu_{\tx{after}}(\{B_2 = \true\}) + \mu_{\tx{after}}(\{B_2 = \false\}) = 0 + \nicefrac 1 2 = \nicefrac 1 2
\]
This means the current distribution is \emph{unnormalised}, with the \emph{normalising constant} being $\nicefrac{1}{\nicefrac 1 2} = 2$. This also means $m_{\tx{after}} \notin \mcal M$ as $\mcal M$ only includes probability spaces. To solve this problem, we extend $\mcal M$ to include non-probability measures as well.

\begin{figure}
  \small
\begin{minipage}[b]{0.4\textwidth}
  $\begin{array}{@{}c@{\quad}c@{\quad}c@{}}
    \begin{tikzpicture}
      \node (mbefore) {
        \begin{tikzpicture}
          \node (r1) [activecell] {$\scriptstyle B_1$};
          \node (r2) [activecell, right = -1pt of r1] {$\scriptstyle B_2$};
          \pic [below = -1pt of r1] {distributionwithleftaxis={2.5mm}{2.5mm}{-3.5mm}{50\%}};
          \pic [below = -1pt of r2] {distribution={2.5mm}{2.5mm}};
          \node at ($(r1)!0.5!(r2) + (0, 0.5)$) {$m_{\tx{before}}$};
        \end{tikzpicture}
      };
      \node[right=of mbefore, xshift=-10mm, yshift=1mm] (mapsto) {$\xmapsto{\ell\,\cdot\, {-}}$};
    \end{tikzpicture}
    \begin{tikzpicture}
      \node[right=of mbefore] (mafter) {
        \begin{tikzpicture}
          \node (r1) [activecell] {$\scriptstyle {B_1}$};
          \node (r2) [activecell, right = -1pt of r1] {$\scriptstyle B_2$};
          \pic [below = -1pt of r1] {distribution={2.5mm}{2.5mm}};
          \pic [below = -1pt of r2] {distributionwithrightaxis={0mm}{2.5mm}{-3.5mm}{50\%}};
          \node at ($(r1)!0.5!(r2) + (-0.5, 0.5)$) {$m_{\tx{after}}$};
        \end{tikzpicture}
      };
    \end{tikzpicture}
  \end{array}$
  \caption{Updating via likelihood function}
  \label{fig:overview:update}
\end{minipage}
\begin{minipage}[b]{0.45\textwidth}
  \small
  $\begin{tikzpicture}
    \node (core) {
      \begin{tikzpicture}
        \node (r1) [activecell] {$\scriptstyle B_1$};
        \node (r2) [activecell, right = -1pt of r1] {$\scriptstyle B_2$};
        \pic [below = -1pt of r2] {distributionwithrightaxis={1.25mm}{3.75mm}{-2.25mm}{75\%}};
        \pic [below = -1pt of r1] {distributionwithleftaxis={2.5mm}{2.5mm}{-3.5mm}{50\%}};
        \node at ($(r1)!0.5!(r2) + (0, 0.5)$) {$m$};
      \end{tikzpicture}
    };
    \node[above right= -1cm and 2cm of core] (condtrue) {
      \begin{tikzpicture}
        \node (r1) [activecell] {$\scriptstyle B_1$};
        \node (r2) [activecell, right = -1pt of r1] {$\scriptstyle B_2$};
        \pic [below = -1pt of r1] {distribution={0mm}{5mm}};
        \pic [below = -1pt of r2] {distributionwithrightaxis={0mm}{5mm}{-1mm}{100\%}};
        \node at ($(r1)!0.5!(r2) + (-0.5, 0.25)$) {$m_{B_1 = \false}$};
      \end{tikzpicture}
    };
    \node[below=0mm of condtrue] (condfalse) {
      \begin{tikzpicture}
        \node (r1) [activecell] {$\scriptstyle B_1$};
        \node (r2) [activecell, right = -1pt of r1] {$\scriptstyle B_2$};
        \pic [below = -1pt of r1] {distribution={5mm}{0mm}};
        \pic [below = -1pt of r2] {distributionwithrightaxis={2.5mm}{2.5mm}{-3.5mm}{50\%}};
        \node at ($(r1)!0.5!(r2) + (0, 0.7)$) {$m_{B_1 = \true}$};
      \end{tikzpicture}
    };
    \draw[arrows = {-Latex}] (core) to node[midway,above,sloped] { \tiny\!\!\!\!\!\!conditioning $\scriptstyle B_1 = \true$} (condtrue);
    \draw[arrows = {-Latex}] (core) to node[midway,below,sloped] {\tiny \!\!\!\!\!\!conditioning $\scriptstyle B_1 = \false$} (condfalse);
  \end{tikzpicture}$
  \vspace{-20pt}
  \caption{A dependent random generator}
  \label{fig:overview:depdist}
\end{minipage}
\end{figure}

\paragraph{Dependent Random Variables}

A far more challenging problem is to handle dependency between random variables. 
Consider an experiment where we toss a fair coin $B_1$, and depending on the result of $B_1$, we obtain $B_2$ by sampling from two different distributions:
  \begin{align*}
    B_2 \sim \begin{cases}
      \tx{toss a biased coin that is always $\true$} & \tx{if $B_1 = \true$} \\
      \tx{toss a fair coin} & \tx{if $B_1 = \false$}
    \end{cases}
  \end{align*}
Assuming that $B_2 = \true$, what is the probability that $B_1 = \true$?

Notice that $B_1$ and $B_2$ are not probabilistically independent: knowing $B_1$ is $\true$ means $B_2$ is $\true$ $100\%$ of the time, and knowing $B_1$ is $\false$ means $B_2$ is $\true$ $50\%$ of the time (which means overall, $B_2$ is $\true$ $75\%$ of the time). Suppose $m \in \mcal M$ represents this distribution; we represent the dependency between $B_1$ and $B_2$ in $m$ pictorially in \cref{fig:overview:depdist}. To see the problem, note that when we mutate the distribution of $B_2$ via $\ell$, the distribution of $B_1$ changes as well: indeed, knowing $B_2 = \true$ makes $B_1 = \true$ more likely, as $B_2 = \true$ is more likely to be caused by flipping the biased coin that is always $\true$, rather than the fair coin. To demonstrate how updating $B_2$ causes the update of $B_1$, we perform a `case analysis' and apply $(\ell \cdot {-})$ to the conditioned space $m_{B_1 = \true}$ and $m_{B_1 = \false}$:  
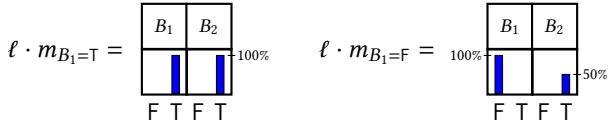
\[
  \begin{tikzpicture}
    \node (condtrue) {
      \begin{tikzpicture}
        \node (r1) [activecell] {$\scriptstyle B_1$};
        \node (r2) [activecell, right = -1pt of r1] {$\scriptstyle B_2$};
        \pic [below = -1pt of r1] {distribution={0mm}{5mm}};
        \pic [below = -1pt of r2] {distributionwithrightaxis={0mm}{5mm}{-1mm}{100\%}};
      \end{tikzpicture}
    };
    \node[right=of condtrue, xshift=-20pt, yshift=5pt] (cflabel) {$\ell \cdot m_{\tiny B_1 = \false} = $};
    \node[right= of cflabel, xshift=-33pt, yshift=-5pt] (condfalse) {
      \begin{tikzpicture}
        \node (r1) [activecell] {$\scriptstyle B_1$};
        \node (r2) [activecell, right = -1pt of r1] {$\scriptstyle B_2$};
        \pic [below = -1pt of r1] {distributionwithleftaxis={5mm}{0mm}{-1mm}{100\%}};
        \pic [below = -1pt of r2] {distributionwithrightaxis={0mm}{2.5mm}{-3.5mm}{50\%}};
      \end{tikzpicture}
    };
    \node[left=of condtrue, xshift=33pt, yshift=5pt] {$\ell \cdot m_{\tiny B_1 = \true} = $};
  \end{tikzpicture}
\]
Since we now have information regarding the original distribution of $B_1$ (the \emph{prior} distribution) as well as the likelihood, we can apply Bayes' theorem. Specifically, for all $b \in \{\true, \false\}$:
\begin{align*}
  \overbrace{(\ell \cdot m)(\{B_1 = b\})}^{\tx{unnormalised posterior}}
  &= \overbrace{\mu(\{B_1 = b\})}^{\tx{prior}} \cdot \overbrace{(\ell \cdot m_{B_1 = b})(\{B_2 = \true\})}^{\tx{likelihood}}
  = \begin{cases}
    \nicefrac 1 2 & \textnormal{if}\ b = \true \\
    \nicefrac 1 4 & \textnormal{if}\ b = \false
  \end{cases}
\end{align*}
This means $(\ell \cdot m)(\{B_1 = \true\}) = \nicefrac{1}{2}$ and $(\ell \cdot m)(\{B_1 = \false\}) = \nicefrac{1}{4}$ and $\ell \cdot m$ is an unnormalised measure ($\nicefrac{1}{2} + \nicefrac{1}{4} \neq 1$). 
Normalising $\ell \cdot m$ yields the desired distribution -- assuming $B_2 = \true$, $B_1$ has $\nicefrac{1}{3}$ probability of being $\false$, and $\nicefrac{2}{3}$ probability of being $\true$, as shown below:
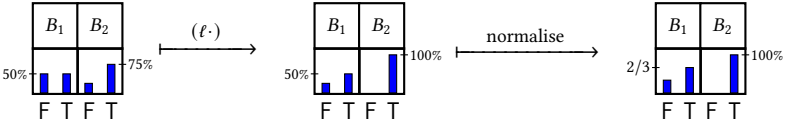
\[
  \begin{array}{@{}c@{\quad}c@{\quad}c@{}}
    \begin{tikzpicture}
      \node (mbefore) {
        \begin{tikzpicture}
          \node (r1) [activecell] {$\scriptstyle B_1$};
          \node (r2) [activecell, right = -1pt of r1] {$\scriptstyle B_2$};
          \pic [below = -1pt of r1] {distributionwithleftaxis={2.5mm}{2.5mm}{-3.5mm}{50\%}};
          \pic [below = -1pt of r2] {distributionwithrightaxis={1.25mm}{3.75mm}{-2.25mm}{75\%}};
        \end{tikzpicture}
      };
      \node[right=of mbefore, xshift=-13mm, yshift=3.5mm] (mapsto) {$\xmapsto{\quad(\ell \cdot)\quad}$};
    \end{tikzpicture}
    \begin{tikzpicture}
      \node[right=of mbefore] (mafter) {
        \begin{tikzpicture}
          \node (r1) [activecell] {$\scriptstyle B_1$};
          \node (r2) [activecell, right = -1pt of r1] {$\scriptstyle B_2$};
          \pic [below = -1pt of r1] {distributionwithleftaxis={1.25mm}{2.5mm}{-3.5mm}{50\%}};
          \pic [below = -1pt of r2] {distributionwithrightaxis={0mm}{5mm}{-1mm}{100\%}};
        \end{tikzpicture}
      };
      \node[right=of mafter, xshift=-12mm, yshift=3mm] (mapsto) {$\xmapsto{\quad\tx{normalise}\quad}$};
    \end{tikzpicture}
    \begin{tikzpicture}
      \node[right=of mafter] (mnorm) {
        \begin{tikzpicture}
          \node (r1) [activecell] {$\scriptstyle B_1$};
          \node (r2) [activecell, right = -1pt of r1] {$\scriptstyle B_2$};
          \pic [below = -1pt of r1] {distributionwithleftaxis={1.67mm}{3.33mm}{-2.67mm}{$2/3$}};
          \pic [below = -1pt of r2] {distributionwithrightaxis={0mm}{5mm}{-1mm}{100\%}};
        \end{tikzpicture}
      };
    \end{tikzpicture}
  \end{array}
\]

To model these features in $\logic$, we develop a novel resource model for randomisation as follows:
\begin{enumerate}
  \item To handle unnormalised measures, we extend $\lilac$'s partial commutative monoid by allowing non-probability measure spaces and impose several \emph{finiteness} restrictions (\cref{defn:basl:rng}) and show that the resulting structure remains a partial commutative monoid (\cref{thm:basl:pcm}), \ie a model of separation logic. 
  In fact (as in $\lilac$), it forms an even richer structure known as a \emph{Kripke resource monoid} (\cref{cor:basl:krm}).
  \item Given an unnormalised random number generator $m \in \mcal M$ and a random variable $X$, if the underlying space of $X$ is a \emph{standard Borel space} (described later in \cref{fig:basl:typing}), then the family of generators $m_{X = x}$ conditioning on $X = x$ (formally, the \emph{disintegration} of $m$ over $X$) exists (\cref{prop:basl:cond}). 
  This allows us to perform conditional reasoning: in order to reason about the dependencies between random variables, we need to reason about conditioned spaces.
  \item Random variables arising from random generators in $\logic$ can be \emph{updated} via a logical version of Bayes' theorem (\cref{thm:basl:bayes}). 
  This result relies on $\logic$'s \emph{partially affine} structure (\cref{prop:basl:affine}) and a theorem in disintegration theory (\cref{lem:basl:density}).
\end{enumerate}

With the above theorems and guarantees, $\logic$ admits a standard resource-theoretic semantics via a construction known as a \emph{Kripke resource model} \cite{galmiche05}, which we formally develop in the rest of this section.

\subsection{The Kripke Resource Model of $\logic$}\label{subsec:basl:kripke}

Recall from \cref{subsec:overview:pcm} that our sample space was set to the booleans $\{\true, \false\}$. However, in \logic (as in $\lilac$) we fix the Hilbert cube $(\Omega, \BO) \deq ([0, 1]^\N, \mcal B[0, 1]^\N)$ as our underlying sample space, 
which can be thought of as the random number generator having the ability to independently generate a stream of numbers between $0$ and $1$. 
We next define the Kripke resource monoid of $\logic$, which intuitively comprises the unnormalised random number generators described in \cref{subsec:overview:bayespcm}.
\begin{definition}\label[definition]{defn:basl:rng}
  Let $\mcal F$ be a $\sigma$-algebra of $\Omega$ and $\mu : \mcal F \to \W$ a measure. The pair $(\mcal F, \mu)$ is a \emph{random generator} if the following conditions hold:
  \begin{enumerate}
    \item $\mu$ is a $\sigma$-finite measure: there exists a countable sequence $\{U_i \in \mcal F\}_{i \in \N}$ such that $\{U_i\}_{i \in \N}$ covers $\Omega$ (\ie $\bigcup_{i \in \N} U_i = \Omega$) and $\mu(U_i) < \infty$, for all $i \in \N$.
    \item $\mu$ has non-zero total measure: $\mu(\Omega) > 0$.
    \item $\mcal F$ is a sub-$\sigma$-algebra of $\BO$: $\mcal F \subseteq \BO$.
    \item $\mcal F$ is countably generated: there exists a countable set of subsets $\{F_i \subseteq \Omega\}_{i \in \N}$ such that $\mcal F$ is the least $\sigma$-algebra containing $\{F_i\}_{i \in \N}$.
    \item $\mcal F$ has a \emph{finite footprint}: there exists an  $n \in \N$ such that:
    $\fora{F \!\in\! \mcal F} \exsts{F'\! \subseteq [0, 1]^n} F \!=\! F'\! \times \Omega$.
  \end{enumerate}
\end{definition}
\noindent We write $\mcal M$ for the set of random generators of the Hilbert cube $(\Omega, \Sigma_\Omega)$.
Condition (1) describes the space of interest -- we are not only interested in finite measures, but also measures with infinite normalising constants so that we can model \emph{improper priors},  as demonstrated in \cref{subsec:exam:improper}. 
Condition (2) is needed so that $\mcal M$ forms a PCM, and it affects the way we interpret Hoare triples (\cref{subsec:basl:kripke}). 
Conditions (3) and (4) are needed so that we can consider \emph{disintegrations} of $\mu$ with respect to its random variables (\cref{prop:basl:cond}). 
Condition (5) is needed because we want to ensure there is enough space to generate new random variables (as in $\lilac$ \cite[\S2.5]{li23}).

\begin{definition}[\text{\cite[Definition 2.2]{li23}}]
  Let $(\mcal F, \mu), (\mcal G, \nu) \in \mcal M$. Then
  $(\mcal H, \rho)$ is the \emph{independent combination} of $(\mcal F, \mu)$ and $(\mcal G, \nu)$ if $\mcal H$ is the smallest $\sigma$-algebra containing $\mcal F$ and $\mcal G$, and
  for all $F \in \mcal F, G \in \mcal G$, $\rho(F \cap G) = \mu(F) \cdot \nu(G)$.
\end{definition}

\begin{theorem}[PCM]\label{thm:basl:pcm}
  Let $m$ be the independent combination of $m_1, m_2 \in \mcal M$. Then $m \in \mcal M$ and it is unique. 
  Let $(\bullet) : \mcal M \times \mcal M \pto \mcal M$, mapping $m_1, m_2 \in \mcal M$ to their independent combination if it exists, and $\mb 1$ for the trivial probability space over $\Omega$. Then $(\mcal M, \bullet, \mb 1)$ is a partial commutative monoid.
\end{theorem}
\begin{proof}[Proof notes]
  The proof of uniqueness, similar to $\lilac$, relies on the uniqueness of measures theorem, but for $\sigma$-finite measures instead. Identity and commutativity of $(\bullet)$ follow from properties of $\sigma$-algebras. For associativity, suppose $m_1, m_2, m_3 \in \mcal M$ and $m_{(12)3}$ is defined; we define $m_{23} = (\mcal F_{23}, \mu_{23})$ by choosing a set $V \in {\mcal F}_{1}$ satisfying $0 < \mu_{1}(V) < \infty$ and define $\mu_{23}(U) \deq \frac{\mu_{(12)3}(V \cap U)}{\mu_{1}(V)}$.
  We then construct a $\lambda$-system $\Lambda \subseteq \mcal F_{23}$ of $\mcal F_{23}$-measurable sets such that $F_{23} \in \Lambda$ satisfies the property $\mu_{(12)3}(F_1 \cap F_{23}) = \mu_1(F_1)\mu_{23}(F_{23})$ whenever $F_1 \in \mcal F_1$, $\mu_1(F_1) < \infty$ and $\mu_{23}(F_{23}) < \infty$. Finally, we apply the $\pi$-$\lambda$ theorem and show that all measurable sets in $\mcal F_{23}$ satisfy the property and therefore establish associativity (see \appendixref{thm:append-pcm:pcm} in the technical appendix for the full proof).
\renewcommand{\qed}{}
\end{proof}

In a logic of bunched implications (which includes separation logics), a \emph{Kripke resource monoid} (KRM) is the basis for providing a satisfiability relation by defining a \emph{Kripke resource model} \cite{galmiche05}. We now show that our PCM can be extended to a KRM.

\begin{definition}[\text{\cite[Definition 1]{barthe19}}]
  A \emph{(partial) Kripke resource monoid} is a tuple $(\mcal M, \bullet, \mb 1, \sqsubseteq)$ such that $(\mcal M, \bullet, \mb 1)$ is a partial commutative monoid and $({\sqsubseteq}) \subseteq \mcal M \times \mcal M$ is a preorder such that $(\bullet)$ is bifunctorial over $(\sqsubseteq)$; \ie for all $m, n, m', n' \in \mcal{M}$, if $m \sqsubseteq n$, $m' \sqsubseteq n'$ and $m \bullet m'$, $n \bullet n'$ are defined, then $m \bullet m' \sqsubseteq n \bullet n'$.
\end{definition}

\begin{corollary}[KRM]\label[corollary]{cor:basl:krm}
  Let $(\sqsubseteq) \subseteq \mcal M \times \mcal M$ be a preorder defined by $(\mcal F_1, \mu_1) \sqsubseteq (\mcal F_2, \mu_2)$ if $\mcal F_1 \subseteq \mcal F_2$ and $\mu_2|_{\mcal F_1} = \mu_1$ (\citet[Theorem 2.4]{li23}). Then $(\mcal M, \bullet, \mb 1, \sqsubseteq)$ is a Kripke resource monoid.
\end{corollary}

A desirable property of $\mcal M$ is that it is closed under conditioning of random variables.
That is, suppose $(\mcal F, \mu) \in \mcal M$, $X : \Omega \to \mcal A$ is a measurable function and $\pi$ is a distribution of $X$ generated by mutating the likelihood of its current distribution; then the space $(\mcal F, \mu^+_x|_{\mcal F})$ conditioning on $X = x$ (almost) always exists (\cref{prop:basl:cond}). To prove this, we apply the following \emph{Rokhlin-Simmons disintegration theorem} \cite{simmons12}, which is a variant of the disintegration theorem that holds for $\sigma$-finite measures in standard Borel spaces.
\begin{lemma}[disintegration]\label[lemma]{lem:basl:disintegration}
  Let $\mu : \Sigma_{\Omega} \to \W$ be a $\sigma$-finite Borel measure, $X : \Omega \to \mcal A$ a $(\Sigma_\Omega, \Sigma_A)$-measurable function and $\pi : \Sigma_A \to \W$ a $\sigma$-finite measure that dominates the pushforward $X_*\mu$. Then there exists a $\pi$-almost-surely-unique $(X, \pi)$-disintegration $\{\mu_x\}_{x \in A}$.
\end{lemma}
\begin{theorem}[conditional]\label{prop:basl:cond}
  Let $(\mcal F, \mu) \in \mcal M$, $X : \Omega \to \mcal A$ be a $(\BO, \Sigma_A)$-measurable function and $\pi : \Sigma_A \to \W$ a $\sigma$-finite measure that dominates $X_*\mu$. Then there exists a measure $\mu^+ : \BO \to \W$ satisfying $\mu^+|_{\mcal F} = \mu$. Further, the $(X, \pi)$-disintegration $\{\mu^+_{x}\}_{x \in A}$ of $\mu^+$ exists and $(\mcal F, \mu^+_x|_{\mcal F}) \in \mcal M$ for $\pi$-almost-all $x \in A$.
\end{theorem}
\begin{proof}
  Readers familiar with disintegration will note that this is a disintegration theorem in disguise. However, there are several non-trivialities. For existence of a Borel measure, since $\mu : \mcal F \to \W$ is assumed to be a countably-generated $\sigma$-finite measure on a sub-Borel $\sigma$-algebra $\mcal F$, the extension $\mu^+$ exists following from the result of \citet[Proposition 433K]{fremlin11}.
  We then apply the Rokhlin-Simmons disintegration theorem (\cref{lem:basl:disintegration}) and obtain the conditional measure $\{\mu^+_x\}_{x \in A}$ and restrict them to the $\sigma$-algebra $\mcal F$.
\end{proof}

\subsection{Semantics of \logic Assertions}\label{subsec:basl:sem}

With the Kripke resource monoid defined, we now formulate a \emph{satisfiability relation} for $\logic$ assertions. In particular, we give semantics to \emph{well-typed assertions}, where an assertion $P$ is well-typed under context $\Gamma; \Delta$ if $\Gamma; \Delta \vdash P$, as defined in \cref{fig:basl:typing}. 
For instance, the assertion $P \deq (X \sim \msf{Unif}(0, a) \Rightarrow \mbb E[X] = \nicefrac{a}{2})$ is well-typed under the context $\Gamma \deq \Gamma', a : \R$ and $\Delta \deq \Delta', X : (\R, \mcal B(\R))$. 
Notice that the map $(\gamma, a) \mapsto \msf{Unif}(0, a)$ is a function $\sem{\Gamma} \to \mcal G(\R, \mcal B(\R))$, which means we have $\Gamma \vdash_{\msf{meas}} \msf{Unif}(0, a) : (\R, \mcal B(\R))$ by $\rrule{T-ProbMeas}$. Then, notice that $X$ is a name of $\Delta$, which means $\Gamma ; \Delta \vdash X \sim \msf{Unif}(0, 1)$ by $\rrule{T-Dist}$. Similarly, $\Gamma ; \Delta \vdash \mbb E[X] = x$ by $\rrule{T-Expectation}$. Combining both assertions with $\rrule{T-BinModality}$ yields the correct conclusion $\Gamma ; \Delta \vdash P$.

\begin{figure}[t]
  \begin{mdframed}
    \small
    \tb{Deterministic context:} $\Gamma = [x_1 : A_1, ..., x_n : A_n]$ \\
    \tb{Probabilistic context:} $\Delta = [X_1 : \mcal A_1, ..., X_n : \mcal A_n]$ \\
  \begin{mathpar}
    \inferrule[T-RandExpr]
      {E \in \sem{\Gamma} \to \mfn{\sem{\Delta}} {\mcal A}}
      {\Gamma ; \Delta \vdash_{\msf{re}} E : \mcal A} \and
    \inferrule[T-DetExpr]
      {e \in \sem{\Gamma} \to A}
      {\Gamma \vdash_{\msf{de}} e : A} \and
        \inferrule[T-Meas]
      {\pi \in \sem{\Gamma} \to M_\sigma(\mcal A)}
      {\Gamma \vdash_{\msf{meas}} \pi : \mcal A} \and
    \inferrule[T-True]
      {\ }
      {\Gamma ; \Delta \vdash \top} \and
    \inferrule[T-False]
      {\ }
      {\Gamma ; \Delta \vdash \bot} \and
    \inferrule[T-BinModality]
      {\Gamma ; \Delta \vdash P \\
       \Gamma ; \Delta \vdash Q}
      {\Gamma ; \Delta \vdash P \odot Q} \and
    \inferrule[T-DetQuant]
      {\Gamma, x : A ; \Delta \vdash P}
      {\Gamma ; \Delta \vdash \mcal Q x : A. P} \and
    \inferrule[T-RandQuant]
      {\Gamma ; \Delta, X : \mcal A \vdash P}
      {\Gamma ; \Delta \vdash \mcal Q_{\msf{rv}} X : \mcal A. P} \and
    \inferrule[T-Expectation]
      {\Gamma ; \Delta \vdash_{\msf{re}} E : (\R, \BR) \\\\
       \Gamma ; \Delta \vdash_{\msf{de}} e : \R}
      {\Gamma ; \Delta \vdash \mbb E[E] = e} \and
    \inferrule[T-Dist]
      {\Gamma ; \Delta \vdash_{\msf{re}} E : \mcal A \\
       \Gamma \vdash_{\msf{meas}} \pi : \mcal A}
      {\Gamma ; \Delta \vdash E \sim \pi}
    \and
    \inferrule[T-Ownership]
      {\Gamma ; \Delta \vdash_{\msf{re}} E : \mcal A}
      {\Gamma ; \Delta \vdash \own E} \and
    \inferrule[T-Conditioning]
      {\Gamma ; \Delta \vdash_{\msf{re}} E : \mcal A \\\\
       \Gamma \vdash_{\msf{meas}} \pi : \mcal A \\
       \Gamma, x : A ; \Delta \vdash P}
      {\Gamma ; \Delta \vdash \cond{\bc{x}{E}{\pi}}\, P} \and
    \inferrule[T-Likelihood]
      {\Gamma \vdash_{\msf{de}} e : \Rnn}
      {\Gamma ; \Delta \vdash \score{e}} \and
    \inferrule[T-Hoare]
      {\Gamma ; \Delta \vdash P \\
       \Gamma ; \Delta \vdash_{\msf{prog}} M : \mcal A \\
       \Gamma ; \Delta, X : \mcal A \vdash Q}
      {\Gamma ; \Delta \vdash \set{P} M \set{X : \mcal A. Q}} \and
  \end{mathpar}
  where $A$ ranges over sets, $\mcal A$ ranges over standard Borel spaces, ${\odot} \in \set{\land, \lor, \Rightarrow, *, \wand}$, $\mcal Q \in \set{\forall, \exists}$ and $M_\sigma(\mcal A)$ is the set of $\sigma$-finite measures over $\mcal A$.
  \end{mdframed}
  \caption{Typing judgements for $\logic$ assertions}\label{fig:basl:typing}
\end{figure}

\begin{figure}[t]
  \small
  \begin{tabular}{m{3.7cm}lm{8.3cm}}
    $(\gamma, D, m) \vDash \top$ & always & \\
    $(\gamma, D, m) \vDash \bot$ & never & \\
    $(\gamma, D, m) \vDash P \land Q$ &\iffdef
    & $(\gamma, D, m) \vDash P$ and $(\gamma, D, m) \vDash Q$ \\
    $(\gamma, D, m) \vDash P \lor Q$ &\iffdef &
    $(\gamma, D, m) \vDash P$ or $(\gamma, D, m) \vDash Q$ \\
    $(\gamma, D, m) \vDash P \Rightarrow Q$ &\iffdef & for all $m' \sqsupseteq m$, $(\gamma, D, m) \vDash P$ implies $(\gamma, D, m') \vDash Q$ \\
    $(\gamma, D, m) \vDash P * Q$ &\iffdef & there exists $m_1 \bullet m_2 \sqsubseteq m$ such that $(\gamma, D, m_1) \vDash P$, $(\gamma, D, m_2) \vDash Q$\\
    $(\gamma, D, m) \vDash P \wand Q$
        & \iffdef
        & if $m' \bullet m$ is defined then 
          $(\gamma, D, m') \vDash P$  implies $(\gamma, D, m' \bullet m) \vDash Q$ \\
    $(\gamma, D, m) \vDash \forall x : A. P$ 
      & \iffdef
      & for all $x \in A$, $((\gamma, x), D, m) \vDash P$
    \\
    $(\gamma, D, m) \vDash \exists x : A. P$ 
      & \iffdef
      & for some $x \in A$,  $((\gamma, x), D, m) \vDash P$
    \\
    $(\gamma, D, m) \vDash \forallrv X : \mcal A. P$
      & \iffdef
      & for all $X \in \msf{RV}(\mcal A)$, $(\gamma, (D, X), m) \vDash P$
    \\
    $(\gamma, D, m) \vDash \existsrv X : \mcal A. P$
      & \iffdef
      & for some $X \in \msf{RV}(\mcal A)$, $(\gamma, (D, X), m) \vDash P$\\
  $(\gamma, D, m) \vDash \own{E}$ & \iffdef & $E(\gamma) \circ D$ is $\mcal F$-measurable
\\
    $(\gamma, D, m) \vDash E \sim \pi$
    & \iffdef
    & $E(\gamma) \circ D$ is $\mcal F$-measurable and
    $\pi(\gamma) = (E(\gamma) \circ D)_*\mu$ \\
  $(\gamma, D, m) \vDash \mbb E[E] = e$ & \iffdef & $(\gamma, D, (\mcal F, \mu)) \vDash \score{1}$ and $\int_{\Omega} E(\gamma) \circ D \,\dd{\mu} = e(\gamma)$ \\
  $(\gamma, D, m) \vDash \score{e}$ 
    & \iffdef
    & $\mu(\Omega) = e(\gamma)$ \\
  \multicolumn{1}{m{3.7cm}}{
  $(\gamma, D, (\mcal F, \mu)) \vDash \cond{\bc{x}{E}{\pi}} P$ \newline \newline\newline\newline}
    & \multicolumn{1}{m{1cm}}{\iffdef \newline \newline\newline\newline}
    & 
      $(\gamma, D, (\mcal F, \mu)) \vDash \own{E}$ and $X_*\mu$ is absolutely continuous with respect to $\pi$ and
      for all measures $\mu^+ : \Sigma_\Omega \to \W$ satisfying $\mu^+|_{\mcal F} = \mu$ and all disintegrations $\{\mu^+_x\}_{x \in \mcal A}$ of $\mu^+$ along $(X, \pi)$
      and for $\pi$-almost-all $x \in A$,
      $((\gamma, x), D, (\mcal F, \mu^+_x|_{\mcal F})) \vDash P$
      where $X \deq E(\gamma) \circ D$
    \\
    \multicolumn{1}{m{3.7cm}}{$(\gamma, D, m) \vDash \set{P} M \set{X : \mcal A. Q}$\newline\ \newline\ \newline\  \newline\ \newline\ \newline\ \newline\ \newline\ \newline}
    & \multicolumn{1}{m{1cm}}{\iffdef \newline\ \newline\ \newline\  \newline\ \newline\ \newline\ \newline\ \newline\ \newline}
    & 
    \multicolumn{1}{m{8.3cm}}{
    for all $m_{\pre} \in \mcal M$ with $(\gamma, D, m_{\pre}) \vDash P$, $m_{\fr} \in \mcal M$ with $(\mcal F_0, \mu_0) \deq m_{\pre} \bullet m_{\fr}$ defined, measures $\mu^+_0 : \Sigma_\Omega \to [0, \infty]$ with $\mu^+_{0}|_{\mcal F_0} = \mu_0$,
    \newline if $\mu^+_0(\{\omega \in \Omega \alt \semo{M_\gamma}(D(\omega), \mcal A)\}) > 0$, then there exists
    \begin{enumerate}
      \item $X \in \msf{RV}(\mcal A)$, 
      \item $m_\post \in \mcal M$ with $(\mcal F_1, \mu_1) \deq m_\post \bullet m_\fr$ defined, and
      \item measure $\mu^+_1 : \Sigma_\Omega \to \W$ with $\mu^+_1|_{\mcal F_1} = \mu_1$
    \end{enumerate}
    s.t. $(\gamma, (D, X), m_\post) \vDash Q$,
    and for all $f: \Omega \to \mcal B$ and $U \in \Sigma_{\mcal B} \otimes \Sigma_{\mcal A}$, \newline
    \vspace{-10pt}
    \begin{center}
      $\int_\Omega \semo{M_\gamma}(D(\omega), \{x \in A \alt (f(\omega), x) \in U\})\,\mu^+_0(\dd{\omega}) =\qquad\quad \mu^+_1(\{\omega \in \Omega \alt (f(\omega), X(\omega)) \in U\})$
    \end{center}}
  \end{tabular}
  \vspace{-12pt}
  \caption{Semantics of $\logic$ assertions}
  \label{fig:basl:semantics}
\end{figure}

We define the semantics of \logic assertions through the satisfiability relation in \cref{fig:basl:semantics}. 
We write $(\gamma, D, m) \vDash P$ to denote that $P$ holds in state $m$ under the deterministic context $\gamma$ and the list of random variables $D$. 
The semantics of propositional, first-order and separation logic connectives and quantifiers ($\land$, $\lor$, $\Rightarrow$, $*$, $\wand$, $\forall$, $\exists$) is standard. 
In order for \logic to be sound, the `custom' probabilistic propositions, $\own{E}$, $\score{e}$, $\mbb E[E] = e$, $\cond{\bc{x}{E}{\pi}} P$ and $\{P\}\,M\,\{X.Q\}$, must satisfy \emph{Kripke monotonicity} as follows.

\begin{proposition}[Kripke monotonicity]
  Let $m \sqsubseteq m'$, $\Gamma; \Delta \vdash P$, $\gamma \in \semo{\Gamma}$ and $D \in \msf{RV}\semo{\Delta}$. Then $(\gamma, D, m) \vDash P$ implies $(\gamma, D, m') \vDash P$.
\end{proposition}

We next explain the intuitive meaning of the custom probabilistic assertions in \logic. 
The \emph{ownership} $\own{E}$ and \emph{distribution} $E \sim \pi$ assertions are expressed via measurability of random variables.
Specifically,  $(\gamma, D, m) \vDash \own{E}$ holds iff $E_{\gamma, D} \deq E(\gamma) \circ D$ (the expression $E$ applied to $(\gamma, D)$) is an $\mcal F$-measurable function. 
Similarly,  $(\gamma, D, m) \vDash E \sim \pi$ holds iff $E_{\gamma, D}$ is an $\mcal F$-measurable function and the pushforward of $E$ with respect to $\mu$ is $\pi$ ($\pi(\gamma) = ((E_{\gamma, D})_*\mu$)). 
The semantics of the expected value assertion $\mbb E[E] = e$ follows its usual interpretation in statistics: a random expression has expected value $e_\gamma$ if $E_{\gamma, D}$ integrates (with respect to the probability measure $\mu$) to $e_\gamma$.

The remaining three assertions, namely the \emph{likelihood proposition} $\score{e}$, the \emph{conditioning modality} $\cond{\bc{x}{E}{\pi}} P$ and the \emph{Hoare triple} $\{P\}\,M\,\{X. Q\}$ have non-trivial semantics, as we describe below. 

\paragraph{Likelihood $\score{e}$}
Recall from \cref{sec:overview} that $\score{e}$ asserts that the current state has \emph{likelihood} $e$. Indeed, a state $(\mcal F, \mu) \in \mcal M$ is more \emph{likely} if the total measure $\mu(\Omega)$ is higher. Hence,  $(\mcal F, \mu)$ has likelihood $e(\gamma)$ when $\mu(\Omega) = e(\gamma)$. In fact, the proposition $\msf{NormConst}$ (mentioned in \cref{sec:overview}) is simply defined as having a non-zero likelihood $k$: 
\[
  \msf{NormConst} \deq \exists k : (0, \infty).\score{k}
\]
Note that $\score{1}$ constitutes the multiplicative unit (unit of $*$) in \logic.
Specifically, let $(\gamma, D, (\mcal F, \mu)) \allowbreak \vDash X \sim \mbb P$ for a probability measure $\mbb P$; then $(\mcal F, \mu)$ satisfies $\score{1}$ as the probability measure is by definition normalised. 
Recall that in a Kripke resource model, the multiplicative unit is a proposition $I$ satisfying the following for any KRM $(\mcal M, \bullet, \mb 1, \sqsubseteq)$ \cite[Definition 2.5]{galmiche05}:
\[
  \tx{for all $m \in \mcal M$, } m \vDash I \iff e \sqsubseteq m
\]
Unfolding $\sqsubseteq$ in our KRM, we know that if $I$ is the multiplicative unit and $(\gamma, D, (\mcal F, \mu)) \vDash I$, then $\mu(\Omega) = 1$. Hence, $\score{1}$ is the unit and we define $\top_1 \deq \score{1}$. Recall that a proposition $P$ \emph{entails} $Q$ (written $P \vdash Q$) if $(\gamma, D, m) \vDash P$ implies $(\gamma, D, m) \vDash Q$, for all $(\gamma, D, m)$. With $\top_1$ being the multiplicative unit, we then know the bi-entailment $P * \top_1 \vdashv P$ holds.

\paragraph{\logic is partially affine}
The fact that $\top_1$ is the multiplicative unit has significant implications on $\logic$'s reasoning rules. 
In particular, this means we cannot \emph{forget} about `unnormalised propositions'. 
Recall that a separation logic is \emph{affine} if the entailment $P * Q \vdash P \land Q$ holds \cite{galmiche05}. 
For example, the \tsc{Iris} separation logic is affine \cite{jung18}, and we can \eg forget about pointers via the entailments $(x \hookrightarrow v * y \hookrightarrow v')\ \vdash\ (x \hookrightarrow v \land y \hookrightarrow v') \ \vdash\ x \hookrightarrow v$, 
i.e. given two addresses $x$ and $y$, weakening $*$ to $\land$ allows us to forget information about $y$. On the other hand, a separation logic is \emph{linear} (or \emph{boolean}) when the multiplicative unit is only satisfied by the empty element $\mb 1 \in \mcal M$, i.e. when $m \vDash I$ implies $m = \mb 1$ \cite{galmiche05}. 
Consequently, the entailment $P * Q \vdash P \land Q$ does not always hold, and we cannot forget information. In fact, this is the original approach to separation logic taken by \citet{ohearn99}.

\logic is neither affine nor linear -- the entailment $P * Q \vdash P \land Q$ holds in certain restricted cases. 
Indeed,  \logic is \emph{partially affine}, a category of logics first described by \citet{chargueraud20}, whereby only a class of (not necessarily all) assertions are designated as \ti{affine},  namely those that can be `dropped'. 
In \logic, we define an assertion to be affine, written $\affine P$, as follows:\\
\centerline{$
  \affine P \overset{
  \raisebox{-1pt}{$\scriptstyle\textnormal{\tiny def}$}
}{\iff} \tx{for all } \gamma, D,  \mcal F, \mu, \;  (\gamma, D, (\mcal F, \mu)) \vDash P \; \tx{implies } \mu(\Omega) = 1
$}\\
Intuitively, this means only normalised assertions (\eg $X \sim \mbb P$ for a probability measure $\mbb P$) are affine and assertions with unnormalised components (\eg $\score{e}$ when $e \neq 1$, or $\msf{NormConst}$) are not affine. Affine assertions enjoy the property that they can be dropped, as stated in \cref{prop:basl:affine} below.

\begin{proposition}[affine assertions]\label[proposition]{prop:basl:affine}
  The following entailments hold:
  \begin{mathpar}
    \inferrule[E-$*$-Weak$_1$]
      {\affine Q}
      {P * Q \vdash P} \and 
    \inferrule[E-$*$-Weak$_2$]
      {\affine P}
      {P * Q \vdash Q} \and
    \inferrule[E-$*$-Weak]
      {\affine P \\ \affine Q}
      {P * Q \vdash P \land Q}
  \end{mathpar}
\end{proposition}

\paragraph{Conditioning $\cond{\bc{x}{E}{\pi}} P$}
As explained in \cref{sec:overview}, the conditioning assertion $\bc{x}{E}{\pi}\;|\;P$ lets us use the assertion $P$ to describe the behaviour of a state $\mu$ assuming $E = x$ and $(\gamma, D, \mcal F, \mu) \vDash E \sim \pi$. 
Its semantics is as follows. 
To show that $P$ holds conditionally for almost every $x$, we require that $(\gamma, D, \mcal F, \mu^+_x) \vDash P$ holds for almost all $x$, where $\mu^+$ is a Borel measure extending $\mu$ and $\mu^+_x$ is the measure $\mu^+$ conditioned on $E = x$. The Borel measure extension condition is a technical condition to ensure the existence of the conditioned space $\{\mu^+_x\}_{x \in A}$, which are a family of $\sigma$-finite measures that exist by \cref{prop:basl:cond}. We can now see how the conditioning modality interacts with the likelihood assertion $\score{e}$ by revisiting \rrule{BayesCoin}. 
Suppose $X \sim \msf{Unif}(0, 1)$ and we assert the likelihood of $X = x$ to be $x$ as in the example of \cref{fig:overview:bayescoin-semantics}; 
then the state satisfies the proposition $\cond{\bc{x}{X}{\msf{Unif}(0, 1)}} \score{x}$. The resulting distribution is proportional to the $\msf{Beta}(2, 1)$ distribution because Bayes' theorem states that $\Pr[X \in U \alt \mi{Flip}_1 = 1] 
  = \int_0^1 \!\ch{U}(x) \cdot 2x\,\dd{x} = \msf{Beta}(U \alt 2, 1)$.
Now, in order to derive the desired postcondition ($X \sim \msf{Beta}(2, 1)$ with a normalising constant), we need to internalise a notion of Bayes' theorem within $\logic$.

\paragraph{The Internal Bayes' Theorem}
In disintegration theory, the following theorem states that the Radon-Nikodym derivative of the two absolutely-continuous measures $X_*\mu \ll \pi$ is almost surely equal to the total measure of a $\pi$-disintegration (\cref{lem:basl:density}). Using this result, we can internalise Bayes' theorem as a $\logic$ bi-entailment (\cref{thm:basl:bayes}). 
\begin{lemma}[\text{\cite[Theorem 2]{chang97}}]\label[lemma]{lem:basl:density}
  Let $\{\mu_x\}_{x \in A}$ be an $(X, \pi)$-disintegration of $\mu$. Then $\pi$ dominates the pushforward $X_*\mu$ and for $\pi$-almost-all $x \in A$, 
  $\frac{\dd{X_*\mu}}{\dd{\pi}}(x) = \mu_x(\Omega)$.
\end{lemma}
\begin{theorem}[internal Bayes' theorem]\label{thm:basl:bayes}
  For any Borel measurable function $f : \mcal A \to \Rnn$ and random expression $\Gamma; \Delta \vdashs{re} E : \mcal A$, the following bi-entailment holds:
  \[
    {\cond{\overbrace{\bc{x}{E}{\pi}}^{\textnormal{prior}}\,}\,\overbrace{\score{f(x)}\vphantom{\bc{x}{E}{\pi}}}^{\textnormal{likelihood}}\ \vdashv\ \overbrace{E \sim f \cdot \pi\vphantom{\bc{x}{E}{\pi}}}^{\textnormal{posterior}}}
  \]
\end{theorem}
\begin{proof}[Proof]
  Let $m = (\mcal F, \mu) \in \mcal M$, $(\gamma, D, m)$ be a configuration and $X(\omega) \deq E(\gamma)(D(\omega))$. 
To prove the $\vdash$ direction, we assume $(\gamma, D, (\mcal F, \mu)) \vDash \cond{\bc{x}{E}{\pi}} \score{f(x)}$. 
Then for any Borel extension $\mu^+$ and $(X, \pi)$-disintegration $\{\mu^+_x\}_{x \in A}$ we know $(\gamma, D, (\mcal F, \mu^+_x|_{\mcal F})) \!\vDash\! \score{f(x)}$ holds for $\pi$-almost-all $x \in A$. This implies $\mu^+_x(\Omega) = f(x)$ holds for $\pi$-almost-all $x \in A$, which implies for all $F \in \Sigma_A$:
  \begin{align*}
    X_*\mu(F)
    = X_*\mu^+(F)
    = \int_A \,\mu^+_x(X^{-1}(F))\,\pi(\dd{x})
    = \int_F \mu^+_x(\Omega)\,\pi(\dd{x})
    = \int_F f\,\dd{\pi}
    = (f \cdot \pi)(F).
  \end{align*}
  The first equality follows from our assumption that $\mu^+$ is a Borel extension of $\mu$. 
  The second equality is a disintegration axiom; 
  the third holds because $\ch{F}(x)\mu^+_x(\Omega)= X_*\mu^+_x(F)$; 
  and the last holds because for almost all $x$, we have $(\gamma, D, (\mcal F, \mu^+_x|_{\mcal F})) \vDash \score{f(x)}$. Hence, we know $(\gamma, D, (\mcal F, \mu)) \vDash E \sim f \cdot \pi$. 
  
  To prove the $\dashv$ direction, we assume $(\gamma, D, (\mcal F, \mu)) \vDash E \sim f \cdot \pi$ holds; 
  then $(\gamma, D, (\mcal F, \mu)) \vDash \msf{own}\ E$ by definition of $\vDash$. 
  Next, for any Borel extension $\mu^+$ and $(X, \pi)$-disintegration $\{\mu^+_x\}_{x \in A}$, we know $(\gamma, D, (\mcal F, \mu^+_x|_{\mcal F})) \vDash \score{f(x)}$ holds for $\pi$-almost-all $x \in A$ because for any $F \in \Sigma_A$:
  \begin{align*}
    \int_F \mu^+_x(\Omega)\,\pi(\dd{x})
    = \int_F \frac{\dd{X_*\mu}}{\dd{\pi}}\,\dd{\pi}
    = \int_F \frac{\dd{(f \cdot \pi)}}{\dd{\pi}}\,\dd{\pi} 
    = \int_F f\,\dd{\pi}.
  \end{align*}
  The first equality follows from \cref{lem:basl:density}; 
  the second follows from our assumption. %
  Note that $X_*\mu = f \cdot \pi$ is absolutely continuous with respect to $\pi$ as $(f \cdot \pi)(F) = 0$ for any $\pi$-null-set $F$.
\end{proof}

\paragraph{Hoare Triple $\{P\}\,M\,\{X. Q\}$}
The interpretation of a Hoare triple in \logic is similar to that of $\lilac$, but with one key difference: a \logic Hoare triple denotes \emph{partial correctness}. 
A \emph{partial correctness triple} $\{P\}\ M\ \{Q\}$ states that if a state satisfies $P$ and executing $M$ from that state \emph{terminates}, then the resulting state satisfies $Q$. 
One reason of assuming termination is due to its undecidability. 
Similarly,  determining whether a probabilistic program has a zero normalising constant is undecidable \citet[\S2.2.4]{staton20} -- 
this is not a relevant concern for existing probabilistic logics (including \lilac) as they do not support Bayesian updating.
As such, $\logic$ triples denote partial correctness. 
Specifically,  $\{P\}\ M\ \{X. Q\}$ holds iff starting from a state $m_{\pre} \deq (\mcal F_{\pre}, \mu_{\pre})$ satisfying the precondition $P$ and an arbitrary frame $m_\fr$ (such that $m_\pre \bullet m_\fr$ is defined),  
if `$\semo{M}$ normalises to a non-zero constant from $m_\pre \bullet m_\fr$', 
then there must exist a random variable $X$ and a postcondition state $m_\post \deq (\mcal F_\post, \mu_\post)$ that satisfies $Q$ and is compatible with the frame $m_\fr$ such that $m_\post \bullet m_\fr$ is also defined and `executing $M$ from $m_\pre$ results in $m_\post$'.

To express that `$\semo{M}$ normalises to a non-zero constant from $m_\pre \bullet m_\fr$', we require that $\mu^+_0(\{\omega \in \Omega \alt \semo{M_\gamma}(D(\omega), \mcal A)\}) > 0$, where $\mu^+_0$ is a Borel extension of $m_\pre \bullet m_\fr$. 
This means there exists a measurable set of $\Omega$ with non-zero-$\mu^+_0$ measure such that executing $\semo{M_\gamma}$ with random variables $D$ generated from the seed $\omega$ leads to a non-zero normalising constant. 
Similar to $\lilac$, we quantify over arbitrary extensions of random variables $D_\ext$ to prove a substitution lemma (see \cref{lem:append-logic:sub}).

To express `executing $M$ from the state $m_\pre$ results in $m_\post$', we require that the following hold for any measurable set $U$:
\begin{align*}
  \int_\Omega \semo{M_\gamma}(D(\omega), \{x \alt (D_\ext(\omega), D(\omega), x) \in U\})\,\mu^+_0(\dd{\omega}) = \mu_1^+\{\omega \alt (D_\ext(\omega), D(\omega), X(\omega)) \in U\}.
\end{align*}
where $\mu^+_0$ is a Borel measure that extends $m_\pre \bullet m_\fr$, and $\mu^+_1$ is a Borel measure that extends $\mu_\post \bullet m_\fr$. Comparing the $\logic$ semantics to the standard denotational semantics of $\lang$ in the category of $s$-finite kernels, the behaviour of the Hoare triple can be characterised by \cref{prop:basl:sfk}. Intuitively, suppose $\mu^+_0$ is a state that satisfies precondition $P$ and $\mu^+_1$ is a state that satisfies the postcondition $Q$. Then the resulting measure of $\semo{M_\gamma}$ with random variables $D$ along the random source $\mu^+_0$ is equal to the `output' random variable $X$ with the random source $\mu^+_1$, as stated below.
\begin{proposition}\label[proposition]{prop:basl:sfk}
  Let $(\gamma, D, \mu) \vDash P$, $\mcal M\mcal A$ be the space of all measures of $\mcal A$, and $\semo{-} : \mb{Syn} \to \mb{sfKrn}$ be the semantic functor  defined in \appendixref{append:ppl}. Suppose $\Delta \vdash M : \tau$ and $\{P\}\,M\{X. Q\}$ holds with respect to $\mu^+_0$ and $\mu^+_1$. Then the following diagram commutes:
  \[
  \begin{tikzcd}[column sep=large, row sep=small]
    \mcal M \Omega \ar[r, "D_*"] & \mcal M\semo{\Delta} \ar[r, "\semo{M_{\gamma}}_*"] & \mcal M \semo{\tau} \\
    \mb{1} \ar[u, "\mu^+_0\,"] \ar[r, "\mu^+_1"'] & \mcal M\Omega \ar[r, "X_*"'] & \mcal M\semo{\tau} \ar[u, equal]
  \end{tikzcd}
  \]
\end{proposition}
We finally show that the \logic proof system in \cref{fig:basl:proofrule} is sound, with the full proof given in \appendixref{append:proofrules}.
We write $\vDash \{P\}\,M\,\{X. Q\}$ to denote that $(\gamma, D, m) \vDash \{P\}\,M\,\{X. Q\}$ holds for all $(\gamma, D, m)$. 
\begin{theorem}[soundness]\label{thm:soundness}
  The \logic proof system is \emph{sound}: 
  for all $P, M, Q$, if $\vdash \{P\}\,M\{X. Q\}$ is derivable using the rules in \cref{fig:basl:proofrule}, then $\vDash \{P\}\,M\,\{X. Q\}$ holds.
\end{theorem}

\section{Conclusions, Related and Future Work}
\label{sec:rel}
\label{sec:con}

We developed \logic by extending probabilistic separation logic to verify Bayesian programs/statistical models. To this end, we devised a semantic model rich enough to encode probabilistic programming concepts such as conditional distributions, unnormalised distributions, Bayesian updating and improper priors. We then demonstrated the utility of $\logic$ by proving properties such as correlation, expected values and posterior distributions in various statistical models.

\paragraph{Related Work: Semantics of BPPLs}
The semantics of randomised languages is well-established: starting from the seminal work of \citet{kozen81} on linear operator semantics, there are numerous works on semantic domains for probability and non-determinism, \eg by \citet{jones89}. 
These led to the study of Bayesian inference from a programming language theory perspective \cite{gordon14,vandemeent21} with research on their operational \cite{borgstrom16} and denotational semantics \cite{dahlqvist19,huot23}. 
Based on the theory of concrete sheaves \cite{matache22}, \citet{heunen17,vakar19} developed a category known as ($\omega$-)quasi-Borel space (\textbf{Qbs}) for reasoning about higher-order, recursive Bayesian probabilistic programs and proved the existence of a strong monad of $s$-finite measures for a monadic semantics of higher-order BPPLs.
The $\logic$ proof system is underpinned by the category of $s$-finite kernels developed by \citet{staton17}.

\paragraph{Related Work: Probabilistic Separation Logics}
Separation logic (SL) developed a \emph{modular} theory for reasoning about computational resources \cite{ohearn99} for pointer-manipulating programs. 
This led to the development of abstract models for SL \cite{calcagno07,galmiche05,jung18}. 
\citeauthor{barthe19} gave a probabilistic interpretation of SL 
in $\tsc{Psl}$ (probabilistic SL) and proved the correctness of algorithms such as the one-time pad cipher. 

One of the overarching themes of probabilistic separation logics is the investigation of \emph{conditional probability}. 
\citet{bao21} developed \tsc{DIBI} based on bunched implications (BI) by extending $\tsc{Psl}$ with the `$\mathbf{;}$' connective for describing conditional independence of random variables, while \citet{bao21neg} developed \tsc{Lina} for reasoning about negative dependence of random variables.
\citet{li23} proposed $\lilac$,  a measure-theoretic interpretation of probabilistic SL that supports desirable features such as continuous distributions and `mathematical' random variables. 
$\lilac$ handles conditional independence by introducing the \emph{conditioning modality}, which was later adopted by \tsc{BlueBell} \cite{bao25} (a relational probabilistic SL) and \tsc{psOL} \cite{zilberstein24} (a concurrent probabilistic program logic). 
Recently, \citet{li24} developed a categorical model of $\lilac$ by drawing an analogy between random sampling and fresh name generations in the theory of nominal sets. 
$\logic$ generalises the model of $\lilac$ and brings a novel perspective of conditional reasoning by supporting Bayesian updating, hence giving an axiomatic semantics of probabilistic programming via SL. We summarise the features of probabilistic separation/BI logics in \cref{tbl:next:features}.

\begin{figure}[t]
\let\oldcheckmark\checkmark
\colorlet{darkgreen}{green!60!black}
\renewcommand{\checkmark}{{\color{darkgreen}\oldcheckmark}}
  \centering
  \begin{tabular}{cccccccc}
    \toprule
    &\tsc{Psl}
    & \tsc{Dibi}
    & \tsc{Lina}
    & \tsc{Lilac}
    & \tsc{psOL}
    & \tsc{BlueBell}
    & \logic
    \\ \midrule
    Discrete distribution
      & \checkmark & \checkmark & \checkmark & \checkmark & \checkmark & \checkmark & \checkmark
    \\
    Probabilistic independence
      & \checkmark & \checkmark & \checkmark & \checkmark & \checkmark & \checkmark & \checkmark
    \\
    Continuous distribution 
     &  &  & & \checkmark & & & \checkmark
    \\
    Negative dependence
      & & & \checkmark & & 
    \\
    Conditional independence 
      & & \checkmark & & \checkmark & \checkmark & \checkmark & \checkmark
    \\
    Conditioning modality 
      & & & & \checkmark & \checkmark & \checkmark & \checkmark
    \\
    Concurrency 
      & & & &  & \checkmark & &
    \\
    Relational reasoning
      & & & & & & \checkmark &
    \\ 
    \rowcolor{gray!10} \tb{Bayesian reasoning}
      & & & & & & & \checkmark
    \\ \bottomrule
  \end{tabular}
   \vspace{-10pt}
  \caption{Features of probabilistic separation/bunched logics}
  \label{tbl:next:features}
  \vspace{-10pt}
\end{figure}

\paragraph{Related Work: Program Logic for Bayesian Conditioning.}

Based on quasi-Borel spaces, \citet{sato19} derived a family of program logics known as \tsc{PPV}. Their semantic insight is that the logical assertions in the category $\mb{Qbs}$ corresponds to fibrations, and used a technique known as \emph{categorical $\top\top$-lifting} \cite{shinya18} to give semantics to predicates and derive logics for a condtioning construct known as $\kw{query}$. By contrast, our model is more closely inspired by measure theory, and we use a resource monoid of a well-behaved subset of $\sigma$-finite measure spaces over $[0, 1]^\N$ to give semantics to assertions. An interesting direction of future work is to determine whether our definition of Hoare triples is an instance of categorical $\top\top$-lifting. Practically, while \tsc{PPV} can handle higher-order functions, our approach, apart from a first-class handling of probabilistic independence, has the following advantages:
\begin{enumerate*}
  \item \logic has a general modality for expressing conditional distributions: \citet{sato19} develop a program logic on top of quasi-Borel spaces (\textbf{Qbs}), while the probability theory of \textbf{Qbs} is very promising, it is still under active development. For example, consider the (quasi-Borel) space of measurable real functions $\R^\R$, it is currently an open question whether conditioning makes sense for $\R^\R$. Our logic uses standard measure spaces, for which the probability theory is well-established. This allows us to derive a general conditioning modality compatible with Bayesian update. As a consequence, our logic can, via a first-class modality, express conditional probability distributions, which is useful for expressing properties such as conditional expectation/independence.
  \item Also, random variables (in the sense of measurable functions from the sample space) are first-class obejcts in \logic, as opposed to variable names in \tsc{PPV}. An immediate implication is that programs in \tsc{PPV} need to be written in a specific form to retain information about its distribution via the $\kw{query}$ construct.
\end{enumerate*}

\paragraph{Future Work}
We will consider three avenues of future research. 
First, we will mechanise \logic and its soundness proof in a theorem prover such as Rocq. 
Second, we will extend $\logic$ to support more sophisticated language features including 
\begin{enumerate*}
  \item mutable states;
	\item recursion; and
	\item higher-order functions.
\end{enumerate*}
These extensions will allow us to express programs in a class of  statistical models known as \emph{Bayesian nonparametric models} \cite{orbanz17,mak21}, namely statistical models with an unbounded number of random variables \cite{vandemeent21}.
Finally, we will apply \logic to develop program simplification/symbolic execution tools for probabilistic programs.

\begin{acks}
We thank the POPL 2026 reviewers and the members of the Veritas Lab and the Functional Programming Group at Imperial for their constructive feedback. We thank John Li for helpful discussions on the technical aspects of the Lilac separation logic. 
We thank Alberto Croquevielle for his insights on the measure-theoretic components of the proofs.
Azalea Raad is supported by a UKRI fellowship
MR/V024299/1, by the EPSRC grant EP/X037029/1, and by VeTSS.
\end{acks}

\bibliographystyle{ACM-Reference-Format}
\bibliography{main}

\clearpage

\clearpage
\appendix

\section{Disintegration}

Let $\Omega$ be a set and $\Sigma_\Omega$ be a set of subsets of $\Omega$ such that $\Sigma_\Omega$ is closed under countable intersection and complement. Then $\Sigma_\Omega$ is a \emph{$\sigma$-algebra} and the pair $(\Omega, \Sigma_\Omega)$ is a \emph{measurable space}. A \emph{measure} is a function $\mu : \Sigma_\Omega \to \W$ satisfying $\mu(\varnothing) = 0$ and $\mu(\biguplus_{i \in \N} U_i) = \sum_{i \in \N} \mu(U_i)$. We additionally call $\mu$ a \emph{probability measure} if $\mu(\Omega) = 1$, and we call $\mu$ a \emph{$\sigma$-finite measure} if there is a measurable cover $\{U_i \in \mcal F\}_{i \in \N}$ such that $\mu(U_i) < \infty$ for all $i \in \N$. A function $f : A \to B$ is $(\Sigma_A, \Sigma_B)$-measurable if for all $U \in \Sigma_B$, $f^{-1}(U) \in \Sigma_A$. We denote $\mb{Meas}$ the category of measurable spaces with the class of measurable spaces as objects and measurable functions as morphisms.

\begin{definition}
  Let $(A, \tau)$ be a topological space. The \emph{Borel $\sigma$-algebra} of $(A, \tau)$ is the set $A$ equipped with the smallest $\sigma$-algebra $\mcal B(A, \tau)$ that contains $\tau$.
\end{definition}

Since $\sigma$-algebras are closed under complements, notice that all closed sets are by definition part of $\mcal{B}(A, \tau)$. By convention, we write $\R$, $\Rnn$, $\W$ to mean the sets endowed with their usual topology. When we say a function $f : \Omega \to \R$ (or with $\Rnn$/$\W$) is \emph{Borel measurable} when $f$ is $(\Sigma_A, \mcal B(\R))$-measurable.

\begin{definition}
  Let $\mu : \Sigma_\Omega \to \W$ be a measure and $f : \Omega \to \W$ a Borel measurable function. Then the \emph{(Lebesgue) integral} of $f$ with respect to $\mu$ is defined by
  \[
    \int_\Omega f\,\dd{\mu} \deq \sup \Bigg\{\sum_{i = 1}^n \inf_{\omega \in U_i} f(\omega) \cdot \mu(U_i)\ \Bigg|\ \{U_i \in \mcal F\}_{i = 1}^n\ \textnormal{disjoint cover of $\Omega$} \Bigg\}.
  \]
  By convention, we write $\int_\Omega f(x)\,\mu(\dd{x})$ as a shorthand for $\int_\Omega (x \mapsto f(x))\,\dd{\mu}$.
\end{definition}

\begin{definition}
  Let $\mu : \Sigma_\Omega \to \W$ be a measure, $X : \Omega \to A$ a $(\Sigma_\Omega, \Sigma_A)$-measurable function and $\pi : \Sigma_A \to \W$ a measure. An \emph{$(X, \pi)$-disintegration of $\mu$} is a set of measures $\{\mu_x\}_{x \in A}$ satisfying the following axioms:
  \begin{itemize}
    \item (concentration) -- for $\pi$-almost-all $x \in A$, $\mu_x\{\omega \in \Omega \alt X(\omega) \neq x\} = 0$.
    \item (measurability) -- for all $(\Sigma_\Omega, \BW)$-measurable function $f : \Omega \to \W$, the function
    \[
      x \longmapsto \int_\Omega f\,\dd{\mu_x}
    \]
    is $(\Sigma_A, \W)$-measurable.
    \item (marginalisability) -- for all $(\Sigma_\Omega, \BW)$-measurable function $f : \Omega \to \W$, the following identity holds:
    \[
      \int_\Omega f\,\dd{\mu} = \int_A \int_\Omega f(\omega)\,\dd{\mu_x}(\omega)\,\dd{\pi}(x).
    \]
  \end{itemize}
  When $\mu_x$ is a probability measure $\pi$-almost-surely, we call $\{\mu_x\}$ the \emph{regular conditional distribution} of $\mu$ with respect to $X$ and $\pi$.
\end{definition}

A measurable space is a \emph{standard Borel space} if it is the Borel space of a Polish space, i.e. a topological space that is separable and completely metrisable. The \emph{Rokhlin-Simmons disintegration theorem} states that when the domain and codomain of a random variable $X$ is standard Borel, a disintegration with respect to a $\sigma$-finite measure must exist and is almost surely unique:

\begin{proposition}[\text{\cite{simmons12}}]\label[proposition]{lem:append-disint:exists}
  Let $\Omega$ and $\mcal A$ be standard Borel spaces, $\mu : \Sigma_{\Omega} \to \W$ be a $\sigma$-finite Borel measure, $X : \Omega \to \mcal A$ a $(\Sigma_\Omega, \Sigma_A)$-measurable function and $\pi : \Sigma_A \to \W$ a $\sigma$-finite measure that dominates $X_*\mu$. Then there exists a $\pi$-almost-surely-unique $(X, \pi)$-disintegration $\{\mu_x\}_{x \in A}$.
\end{proposition}
\begin{proof}
  The standard disintegration theorem requires the underlying measurable space to be
  \begin{enumerate*}
    \item countably generated, and
    \item contains all singletons, i.e. $\{x\} \in \Sigma_{\mcal A}$ for all $x \in \mcal{A}$.
  \end{enumerate*}
  This version of the disintegration theorem is more general than the standard one (in, for example, \citet[Theorem 1]{chang97}) when we require $\mcal A$ to be a standard Borel space. Apart from the original proof, an alternative proof can be found in \citet{vakar19}, and the result is an instantiation of Theorem 13, which establishes a disintegration theorem for $s$-finite kernels. In the proof, there is an extra condition that the \emph{top $0$-$\infty$ sets} $\infty[\mu]$, $\infty[\pi]$ of $\mu$ and $\pi$ satisfy the condition
  \[
    \mu(X^{-1}(\infty[\pi]) \setminus \infty[\mu]) = 0.
  \]
  Since we assumed $\mu$ and $\pi$ to be $\sigma$-finite, $\infty[\mu]$ and $\infty[\pi]$ are the largest null sets of $\mu$ and $\pi$ \citet[Theorem 8]{vakar19}, then the condition is satisfied via the assumption that $X_*\mu$ is absolutely continuous with respect to $\pi$, and we have
  \begin{align*}
    \mu(X^{-1}(\infty[\pi]) \setminus \infty[\mu])
    &\le \mu(X^{-1}(\infty[\pi])) \\
    &= X_*\mu(\infty[\pi]) \\
    &= 0.
  \end{align*}
  Then implies $\mu(X^{-1}(\infty[\pi]) \setminus \infty[\mu]) = 0$.
\end{proof}
\newpage
\section{Syntax and semantics of $\lang$}\label{append:ppl}

As defined in \citet{staton17}, $\lang$ is a typed first-order language equipped with two effects: $\kp{sample}(\mbb P)$, which samples from a probability distribution $\mbb P$, and the soft conditioning effect $\kp{score}(\ell)$, which `scales' the current distribution according to a non-negative number $\ell$. Let $\Delta$ be a context $[X_1 : \mcal A_1, ..., X_n : \mcal A_n]$ such that $X_i$ is a name and $\mcal A_i = (A_i, \Sigma_i)$ is a measurable space.
\begin{definition}
  The \emph{syntax} of $\lang$ terms and types are defined by the following grammar:
  \begin{align*}
    \mi{Term} \ni M &\ddeq () \alt \pv{x} \alt n \alt r \alt f(M) \alt (M, M) \alt M.\kw{1} \alt M.\kw{2} \\
    &\quad|\,\ \kw{true} \alt \kw{false} \alt \kr{if} M \kb{then} M \kb{else} M \\
    &\quad|\,\ \kp{sample}(M) \alt \kp{score}(M) \alt \kp{return}(M) \\
    &\quad|\,\ \kr{let} \pv{x} = M \kb{in} M \\
    \mi{Type} \ni \tau &\ddeq \tunit \alt \tnat \alt \treal \alt \tbool \alt \tau \times \tau \alt \tprob(\tau) \tagi{types}
  \end{align*}
  where $X$ ranges over a countably infinite set of names, $n \in \N$, $r \in \R$ and $f$ ranges over all measurable functions.
  Additionally, we define a few shorthands, all of which are standard:
  \begin{align*}
    M; N &\deq \kr{let} \_ = M \kb{in} N \tag{sequencing} \\
    \kp{observe}(M) &\deq \kp{score}(\kr{if} M \kb{then} 1 \kb{else} 0). \tag{hard constraint}
  \end{align*}
  A term $M$ is \emph{well-typed} under context $\Delta$ if there is a type $\tau$ such that $\Delta \vdashp M : \tau$, where $\vdashp$ is defined in \cref{fig:append-ppl:typing}.
\end{definition}

\begin{figure}[t]
  \begin{mdframed}
    \small
    \tb{Deterministic typing judgement:} $\Delta \vdashd M : \tau$ \\
    \tb{Probabilistic typing judgement:} $\Delta \vdashp M : \tau$ \\
    \begin{mathpar}
      \inferrule[PT-Unit]
        {\ }
        {\Delta \vdashd () : \tunit} \and
      \inferrule[PT-Var]
        {\ }
        {\Delta \vdashd \pv{x} : \tau}\ (\pv x : \tau \in \Delta) \and
      \inferrule[PT-Nat]
        {\ }
        {\Delta \vdashd n : \tnat} \and
      \inferrule[PT-Real]
        {\ }
        {\Delta \vdashd r : \treal} \and
      \inferrule[PT-Func]
        {\Delta \vdashd M : \sigma}
        {\Delta \vdashd f(M) : \tau}\ (f : \semo{\sigma} \to \semo{\tau})\and
      \inferrule[PT-Pair]
        {(\Delta \vdashd M_i : \tau_i)_{i = 1, 2}}
        {\Delta \vdashd (M_1, M_2) : \tau_1 \times \tau_2} \and
      \inferrule[PT-Fst]
        {\Delta \vdashd M : \sigma \times \tau}
        {\Delta \vdashd M.\kw{1} : \sigma} \and
      \inferrule[PT-Snd]
        {\Delta \vdashd M : \sigma \times \tau}
        {\Delta \vdashd M.\kw{2} : \tau} \and
      \inferrule[PT-True]
        {\ }
        {\Delta \vdashd \kw{true} : \tbool} \and
      \inferrule[PT-False]
        {\ }
        {\Delta \vdashd \kw{false} : \tbool} \and
      \inferrule[PT-If]
        {\Delta \vdashd M : \tbool \\
         \Delta \vdashz N_1 : \tau \\
         \Delta \vdashz N_2 : \tau}
        {\Delta \vdashz \kr{if} M \kb{then} N_1 \kb{else} N_2 : \tau}\ (\msf{z} \in \{\msf{p}, \msf{d}\}) \and
      \inferrule[PT-Sample]
        {\Delta \vdashd M : \tprob(\tau)}
        {\Delta \vdashp \kp{sample}(M) : \tau} \and
      \inferrule[PT-Score]
        {\Delta \vdashd : M : \treal}
        {\Delta \vdashp \kp{score}(M) : \tunit} \and
      \inferrule[PT-Dirac]
        {\Delta \vdashd M : \tau}
        {\Delta \vdashp \kp{return}(M) : \tau} \and
      \inferrule[PT-Let]
        {\Delta \vdashp M : \sigma \\ \Delta, X : \sigma \vdashp N : \tau}
        {\Delta \vdashp \kr{let} X = M \kb{in} N : \tau}
    \end{mathpar}
  \end{mdframed}
  \caption{Typing judgements for $\lang$}\label{fig:append-ppl:typing}
\end{figure}

Fixing two measurable spaces $\mcal A = (A, \Sigma_A)$ and $\mcal B = (B, \Sigma_B)$. Recall that a measure is \emph{$s$-finite} if it is the countable sum of finite measures, i.e. $\mu : \Sigma_A \to \W$ is $s$-finite if there exists a sequence of finite measures $\{\mu_i : \Sigma_A \to \W\}_{i \in \N}$ such that $\mu(U) = \sum_{i \in \N} \mu_i(U)$. A \emph{finite kernel} from $\mcal A$ to $\mcal B$ is a map $\kappa : A \times \Sigma_B \to \W$ such that $\kappa({-}, U)$ is Borel measurable for all $U \in \Sigma_B$, $\kappa(x, {-})$ is a finite measure for all $x \in A$, and the family of measures is uniformly finite, i.e. $\sup_{x \in A} \kappa(x, B) < \infty$. An \emph{$s$-finite kernel} between $\mcal A$ and $\mcal B$ is a map $\kappa : A \times \Sigma_B \to \W$ such that $\kappa = \sum_{i \in \N} \kappa_i$ for a sequence of finite kernels $\{\kappa_i\}_{i \in \N}$. The category of $s$-finite kernels $\mb{sfKrn}$ has measurable spaces as objects and $s$-finite kernels as morphisms. We write $\mcal A \rightsquigarrow \mcal B$ for the set $\mb{sfKrn}(\mcal A, \mcal B)$. Given $\kappa : \mcal A \rightsquigarrow \mcal B$ and $\ell : \mcal B \rightsquigarrow \mcal C$, the composition operator is defined by
\[
  (\ell \circ_{\mb{sfKrn}} \kappa)(x, U) \deq \int_B \ell(y, U)\,\kappa(x, \dd{y}).
\]

\begin{definition}\label[definition]{defn:append-ppl:sem}
  Let $\mb{Syn}$ be the syntactic category of $\lang$ terms and types.
  The denotational semantics of $\lang$ is defined below by the semantic functor $\semo{-} : \mb{Syn} \to \mb{sfKrn}$, i.e. a typing judgement $\Delta \vdashp M : \tau$ denotes an $s$-finite kernel $\semo{M} : \semo{\Delta} \rightsquigarrow \semo{\tau}$. The denotation of types is defined inductively as follows:
  \[
  \begin{array}{ll}
    \begin{aligned}
      \semo{\tunit} &\deq (\set{\star}, \mcal P(\set{\star})) \\
      \semo{\tnat} &\deq (\N, \mcal P(\N)) \\
      \semo{\tbool} &\deq (\mbb B, \mcal P(\mbb B))
    \end{aligned} &
    \begin{aligned}
      \semo{\treal} &\deq (\mbb R, \mcal B(\R)) \\
      \semo{\tprob(\tau)} &\deq \mcal G\semo{\tau}  \\
      \semo{\sigma \times \tau} &\deq \semo{\sigma} \times \semo{\tau}
    \end{aligned}
  \end{array}
  \]
  where $\mcal G : \mb{Meas} \to \mb{Meas}$ is the \emph{Giry functor} that maps $\mcal A$ to a measurable space of probability measures. Note that every $\lang$ type denotes a \emph{standard Borel space}, which is a measurable space generated by the open sets of a Polish topology. The denotation of context is standard: $\Delta = [\pv x_1 : \tau_1, ..., \pv x_n : \tau_n]$ is interpreted as the product measurable space $\semo{\Delta} \deq \prod_{i = 1}^n \semo{\tau_i}$. The denotation of deterministic and probabilistic $\lang$ terms are listed in \cref{fig:append-ppl:densem-det} and \cref{fig:append-ppl:densem-prob} respectively.
\end{definition}

\begin{figure}
  \begin{mdframed}
  \small
  \tb{Deterministic terms:} $\Delta \vdashd M : \tau \implies \semo{M} : \semo{\Delta} \to \semo{\tau}$
  \begin{mathpar}
    \inferrule
      { }
      {\semo{()}(\delta) \deq ()} \and
    \inferrule
      { }
      {\semo{\pv x}(\delta) \deq \delta_{\pv x}} \and
    \inferrule
      { }
      {\semo{n}(\delta) \deq n} \and
    \inferrule
      { }
      {\semo{r}(\delta) \deq r} \and
    \inferrule
      {\semo{M} : \semo{\Delta} \to \semo{\sigma} \\
       f : \semo{\sigma} \to \semo{\tau}}
      {\semo{f(M)}(\delta) \deq f(\semo{M}(\delta))} \and
    \inferrule
      {(\semo{M_i} : \semo{\Delta} \to \semo{\tau_i})_{i = 1, 2}}
      {\semo{(M_1, M_2)}(\delta) \deq
       (\semo{M_1}(\delta), \semo{M_2}(\delta))} \and
    \inferrule
      {\semo{M} : \semo{\Delta} \to \semo{\sigma \times \tau}}
      {\semo{M.\kw{1}}(\delta) \deq \msf{proj}_1(\semo{M}(\delta))} \and
    \inferrule
      {\semo{M} : \semo{\Delta} \to \semo{\sigma \times \tau}}
      {\semo{M.\kw{2}}(\delta) \deq \msf{proj}_2(\semo{M}(\delta))} \and
    \inferrule
      {\ }
      {\semo{\kw{true}}(\delta) \deq 1} \and
      \inferrule
      {\ }
      {\semo{\kw{false}}(\delta) \deq 0} \and \and
    \inferrule
      {\semo{M} : \semo{\Delta} \to \semo{\sigma} \\
       \semo{N_1} : \semo{\Delta} \times \semo{\sigma} \to \semo{\tau} \\
       \semo{N_2} : \semo{\Delta} \to \semo{\tau}}
      {\semo{\kr{if} M \kb{then} N_1 \kb{else} N_2}(\delta) \deq
      \msf{if}\ \semo{M}  = \msf{inj}_2(x)\ \msf{then}\ \semo{N_1}(\delta, x)\ \msf{else}\ \semo{N_2}(\delta)}
  \end{mathpar}
  \end{mdframed}
  \caption{Denotation of deterministic $\lang$ terms}
  \label{fig:append-ppl:densem-det}
\end{figure}

\begin{figure}
  \begin{mdframed}
    \small
    \noindent \tb{Probabilistic terms:} $\Delta \vdashp M : \tau \implies \semo{M} \in \semo{\Delta} \rightsquigarrow \semo{\tau}$
    $\color{gray}\ \subseteq \{\mu : \semo{\Delta} \times \Sigma_{\semo{\tau}} \to \W\}$

    \begin{mathpar}
      \inferrule
        {\semo{M} : |\semo{\Delta}| \to |\mcal G(\semo{\tau})|}
        {\semo{\kp{sample}(M)}(\delta, U) \deq \semo{M}(\delta)(U)} \and
      \inferrule
        {\semo{M} : |\semo{\Delta}| \to \R}
        {\semo{\kp{score}(M)}(\delta, U) \deq
         \max(0, \semo{M}(\delta)) \cdot \ch{U}(\star)} \and
      \inferrule
        {\semo{M} : |\semo{\Delta}| \to |\semo{\tau}|}
        {\semo{\kp{return}(M)}(\delta, U) \deq \ch{U}(\semo{M}(\delta))} \and
      \inferrule
        {\semo{M} : \semo{\Delta} \rightsquigarrow \semo{\sigma} \\
         \semo{N} : \semo{\Delta} \times \semo{\sigma} \rightsquigarrow \semo{\tau}}
        {\semo{\kr{let} \pv x = M \kb{in} N}(\delta, U) \deq \textstyle\int_{\semo{\sigma}} \semo{N}((\delta, x), U) \semo{M}(\delta, dx)} \and
      \inferrule
        {\semo{M} : |\semo{\Delta}| \to |\semo{\sigma}| \\
         \semo{N_1} : \semo{\Delta} \times \semo{\sigma} \rightsquigarrow \semo{\tau} \\
         \semo{N_2} : \semo{\Delta} \rightsquigarrow \semo{\tau}}
        {\semo{\kr{if} M \kb{then} N_1 \kb{else} N_2}(\delta, U) \deq 
        \msf{if}\ \semo{M}(\delta)\ \msf{then}\ \semo{N_1}(\delta, U)\ \msf{else}\ \semo{N_2}(\delta, U)}
    \end{mathpar}
  \end{mdframed}
  \caption{Denotation of probabilistic $\lang$ terms}
  \label{fig:append-ppl:densem-prob}
\end{figure}

In the formulation of program logics, it is often the case that a program contains some metavariables. We distinguish these names by using lowercase letters, as opposed to uppercase letters for random names. For example, $\kr{let} X = \kp{sample}(\msf{Unif}(0, a)) \kb{in} X + Y$ contains three kinds of names: the bound random name $X : \treal$, the free random name $Y : \treal$, and a metavariable $a : \treal$. Technically speaking, the above (pre-)term takes an ambient metavariable context $\Gamma$ as argument and outputs another term open under a context $\Delta \deq [Y : \treal]$. To formally model programs that may contain metavariables in the separation logic, we consider a new judgement $\vdash_{\msf{prog}}$, introduced by the following rule:
\begin{mathpar}
  \inferrule[T-Prog]
    {\forall \gamma_i \in \mi{Term}.\ \Delta \vdash_{\msf{z}_i} \gamma_i : \tau_i \implies
     \Delta \vdashp M[\gamma_1/x_1, ..., \gamma_n/x_n] : \tau}
    {\Gamma, x_1 : \tau_1, ..., x_n : \tau_n ; \Delta \vdash_{\msf{prog}} M[x_1, ..., x_n] : \tau}\ (\msf{z}_i \in \set{\msf{d}, \msf{p}})
\end{mathpar}
where $M[x_1, ..., x_n] \in \mi{Term}[x_1, ..., x_n]$. For example, here is a program with a metavariable $x : \R$:
\[
  \inferrule
    {\forall \gamma.\ \Delta \vdashd \gamma : \treal \implies \Delta \vdashp \overbrace{(\kr{let} X = \kp{sample}(\msf{Unif}(0, x)) \kb{in} X + Y)}^{M[x]}[\gamma/x] : \treal}
    {[x : \treal] ; \Delta \vdash_{\msf{prog}} \kr{let} X = \kp{sample}(\msf{Unif}(0, x)) \kb{in} X + Y : \treal}
\]
Intuitively, the above expression is a term of $\treal$ with a free variable $Y : \treal$ and a metavariable $x \in \R$. If $\Gamma ; \Delta \vdash_{\msf{prog}} M : A$, we say that $M$ is a program of type $A$ under the program variable context $\Delta$ and contains metavariables specified in $\Gamma$. In cases where it is clear, we simply refer to $M$ as a program. Notice that we write $\R$ instead of the measurable space $(\R, \BR)$ as a shorthand. We denote $\tsf{fv}(M)$ the set of free random variables for a well-formed program (that may include metavariables) $\Gamma ; \Delta \vdash_{\msf{prog}} M : \mcal A$. For instance, $\msf{fv}(\kr{let} X = \kp{sample}(\msf{Bern}(a)) \kb{in} X + Y) = \set{Y}$, $X$ is not included because it is bound to $\kw{let}$, $a$ is not included because it is a metavariable.

\newpage
\section{Resource Monoid for Randomness}\label{append:krm}

To develop the Kripke resource monoid for $\logic$, we fix the Hilbert cube as our underlying measurable space, i.e. we define $(\Omega, \BO) \deq ([0, 1]^\N, \mcal B[0, 1]^\N)$, where $\mcal B[0, 1]^\N$ is the Borel $\sigma$-algebra generated by iterating the product measure on Borel sets of $[0, 1]$, i.e. $\mcal B[0, 1]^\N \deq \bigotimes_{i \in \N} \mcal B[0, 1]$. We also adopt  standard conventions in measure theory -- multiplication $(\cdot) : [0, \infty] \times [0, \infty] \to [0, \infty]$ is defined to satisfy $0 \cdot \infty = \infty \cdot 0 = 0$. Recall that a measure $\mu : \mcal F \to \W$ is \emph{$\sigma$-finite} if there exists a sequence of disjoint measurable sets $\{E_i \in \mcal F\}_{i \in \N}$ such that $\biguplus_{i \in \N} E_i = \Omega$ and $\mu(E_i) < \infty$ for every $i \in \N$. Every finite measure is trivially $\sigma$-finite. The Lebesgue measure on $(\R, \BR)$ that returns the length of a subset (e.g. $\lambda_\R([a, b]) = b - a$) and the counting measure on $(\N, \mcal P(\N))$ that returns the cardinality of the subset (e.g. $\#_\N(\{2025, 2026, 2027\}) = 3$) are examples of $\sigma$-finite measures with total measure infinity.

\begin{definition}
  A \emph{random generator} is a pair $(\mcal F, \mu)$ such that $(\Omega, \mcal F, \mu)$ is a $\sigma$-finite measure space, $\mcal F \subseteq \Sigma_\Omega$ and $\mu(\Omega) > 0$. We write $\mcal M_1$ for the set of such pairs. That is, we define 
  \[
    \mcal M_1 \deq \set{(\mcal F, \mu) \alt (\Omega, \mcal F, \mu)\ \tx{$\sigma$-finite measure space}, \mcal F \subseteq \Sigma_\Omega, \mu(\Omega) > 0}.
  \]
\end{definition}

Notice that e used the notation $\mcal M_1$ instead of $\mcal M$ because $\mcal M_1$ is not the final definition -- it needs to be further restricted in order to have desirable properties, namely the existence of disintegration and Borel measure extensions.

\begin{definition}[\text{\cite[Definition 2.2]{li23}}]
  Let $(\Omega, \mcal H, \rho)$ be a $\sigma$-finite measure space. We say that $(\mcal H, \rho)$ is the \emph{independent combination} of $(\mcal F, \mu)$ and $(\mcal G, \nu)$ if 
  \begin{enumerate*}
    \item $\mcal H$ is the smallest $\sigma$-algebra containing $\mcal F$ and $\mcal G$, and
    \item for all $F \in \mcal F$ and $G \in \mcal G$, $\rho(F \cap G) = \mu(F)\nu(G)$.
  \end{enumerate*}
\end{definition}

Recall that a \emph{$\pi$-system} over a set $A$ is a collection of subsets $\Pi \subseteq \mcal P(A)$ such that $\Pi$ is closed under intersection. Similarly, a \emph{$\lambda$-system} over $A$ is a collection of subsets $\Lambda \subseteq \mcal P(A)$ such that $\varnothing \in \Lambda$, $\Lambda$ is closed under complements and countable disjoint union. Let $\mcal C_i \subseteq \mcal P(\Omega)$ be a set of subsets of $\Omega$ for $i \in \{1, ..., n\}$. We write $\gen{ \mcal C_1, ..., \mcal C_n }$ for the smallest $\sigma$-algebra generated by $\mcal C_1 \cup ... \cup \mcal C_n$.

\begin{proposition}[\text{\cite[Lemma B.2]{li23}}]\label[proposition]{prop:append-pcm:pisys}
  Let $\mcal F, \mcal G$ be $\sigma$-algebras of $\Omega$. Then $\Pi \subseteq \mcal P(A)$  defined by $\Pi \deq \set{F \cap G \alt F \in \mcal F, G \in \mcal G}$ is a $\pi$-system that generates the smallest $\sigma$-algebra that contains $\mcal F$ and $\mcal G$.
\end{proposition}

\begin{proposition}[\text{\cite[Theorem 10.3]{billingsley95}}]\label[proposition]{prop:append-pcm:unique-meas}
  Let $\Pi \subseteq \mcal P(\Omega)$ be a $\pi$-system, $\gen{\Pi}$ be the smallest $\sigma$-algebra containing $\Pi$, and $\mu, \nu : \gen{\Pi} \to \W$ be $\sigma$-finite measures. If $\mu|_{\Pi} = \nu|_{\Pi}$ and there exists a sequence $\{U_i \in \Pi\}_{i\in \N}$ such that $\mu(U_i) < \infty$ for all $i \in \N$ and $\bigcup_{i \in \N} U_i = \Omega$, then $\mu = \nu$.
\end{proposition}

\begin{lemma}[uniqueness]\label[lemma]{lem:append-pcm:uniqueness}
  Let $(\mcal H, \rho)$ and $(\mcal H', \rho')$ be independent combinations of $(\mcal F, \mu)$ and $(\mcal G, \nu)$, then $\mcal H = \mcal H'$ and $\rho = \rho'$.
\end{lemma}
\begin{proof}
  The $\sigma$-algebras $\mcal H$ and $\mcal H'$ are equal as both are defined to be the smallest $\sigma$-algebra containing $\mcal F$ and $\mcal G$. By \cref{prop:append-pcm:unique-meas}, $\rho$ and $\rho'$ are equal when
  \begin{enumerate*}
    \item there is a $\pi$-system $\Pi$ such that $\Pi$ generates $\mcal H$ and $\rho|_{\Pi} = \rho'|_{\Pi}$, and
    \item there is a sequence $\{W_i \in \Pi\}_{i \in \N}$ such that $\bigcup_{i \in \N} W_i = \Omega$ and $\rho(W_i) < \infty$ for all $i \in \N$.
  \end{enumerate*}
  By \cref{prop:append-pcm:pisys}, the $\pi$-system 
  \[ 
    \Pi \deq \{F \cap G \alt F \in \mcal F, G \in \mcal G\} 
  \]
  generates $\gen{ \mcal F, \mcal G}$. To show (1), notice that or any $F \in \mcal F$ and $G \in \mcal G$, $\rho(F \cap G) = \mu(F)\nu(G) = \rho'(F \cap G)$. For (2), notice that there are two sequences of disjoint covers $\{U_i \in \mcal F\}_{i \in \N}$ and $\{V_i \in \mcal G\}_{i \in \N}$ such that $\mu(U_i), \nu(V_i) < \infty$. Consider the countable set $\{U_i \cap V_j \in \mcal C\}_{i, j \in \N}$, clearly $\bigcup_{i, j} U_i \cap V_j = \Omega$ and for all $i, j \in \N$, $\rho(U_i \cap V_i) = \mu(U_i)\nu(V_i) < \infty$ by assumption.
\end{proof}

\begin{lemma}[closure]\label[lemma]{lem:append-pcm:closure}
  If $(\mcal H, \rho)$ is the independent combination of $(\mcal F, \mu) \in \mcal M_1$ and $(\mcal G, \nu) \in \mcal M_1$, then $(\mcal H, \rho) \in \mcal M_1$.
\end{lemma}
\begin{proof}
  We need to show that (1) $\mcal H \subseteq \Sigma_\Omega$, (2) $\rho(\Omega) > 0$, and (3) $\rho : \mcal H \to [0, \infty]$ is $\sigma$-finite. For (1), $\mcal F \subseteq \Sigma_\Omega$ and $\mcal G \subseteq \Sigma_\Omega$ implies $\mcal F \cup \mcal G \subseteq \Sigma_\Omega$, hence the generated $\sigma$-algebra $\mcal H$ is a subset of $\Sigma_\Omega$. For (2), $\rho(\Omega) > 0$ because $\rho(\Omega \cap \Omega) = \mu(\Omega) \cdot \nu(\Omega) > 0$ by definition of $\mcal M_1$; For (3), the sequences of disjoint covers $\{U_i \in \mcal F\}_{i \in \N}$ and $\{V_i \in \mcal G\}_{i \in \N}$ allow us to define the countable set $\{W_{i, j} \in \mcal H\}_{i, j \in \N}$ with $W_{i, j} \deq U_i \cap V_j$. Notice that $\bigcup_{i, j \in \N} W_{i, j} = \Omega$ and $\rho(W_{i, j}) = \mu(U_i) \nu(V_j) < \infty$, which means $\rho$ is $\sigma$-finite.
\end{proof}

Due to the uniqueness and closure lemmas (\cref{lem:append-pcm:uniqueness,lem:append-pcm:closure}), there exists a unique partial function $(\oplus) : \mcal M_1 \times \mcal M_1 \pto \mcal M_1$ that maps two elements to their independent combination when it exists. Moreover, let $\mu_{\mb 1} : \set{\varnothing, \Omega} \to [0, \infty]$ be the trivial probability measure with $\mu_{\mb 1}(\Omega) \deq 1$ and $\mb 1 \deq (\set{\varnothing, \Omega}, \mu_{\mb 1})$. Then $(\mcal M_1, \oplus, \mb 1)$ is a partial commutative monoid:

\begin{theorem}[PCM]\label{thm:append-pcm:pcm}
  $(\mcal M_1, \oplus, \mb 1)$ is a partial commutative monoid.
\end{theorem}
\begin{proof}
  To prove identity, we need to show that when $m \deq (\mcal F, \mu)$, then $\mb 1 \oplus m$ is defined and $\mb 1 \oplus m = m$. For the $\sigma$-algebra, compute that $\gen{ \set{\varnothing, \Omega}, \mcal F } = \mcal F$. For the measure, notice that for all $E \in \set{\varnothing, \Omega}$ and $F \in \mcal F$, $\mu(E \cap F) = \mu_{\mb 1}(E)\mu(F)$ -- when $E = \varnothing$, we have $\mu(E \cap F) = 0 = \mu_{\mb 1}(E)\mu(F)$ (recall that $0 \cdot \infty$ is defined to be $0$ in measure theory). When $E = \Omega$, we have $\mu(E \cap F) = \mu(F) = \mu_{\mb 1}(E)\mu(F)$. Hence, the combination $\mb 1 \oplus m$ is defined and is equal to $m$.

  To prove commutativity, let $m_1 \deq (\mcal F_1, \mu_1)$ and $m_2 \deq (\mcal F_2, \mu_2)$, we need to show that if $m_1 \oplus m_2$ is defined then $m_2 \oplus m_1$ is defined and $m_1 \oplus m_2 = m_2 \oplus m_1$. For the $\sigma$-algebra, compute that $\gen{ \mcal F_1, \mcal F_2} = \gen{ \mcal F_2, \mcal F_1 }$. For the measures, let $m_{12} \deq (\mcal F_{12}, \mu_{12}) \deq m_1 \oplus m_2$, notice that for all $F_1 \in \mcal F_1$ and $F_2 \in \mcal F_2$, we have 
  \begin{align*}
    \mu_{12}(F_2 \cap F_1) 
    &= \mu_{12}(F_1 \cap F_2) \\
    &= \mu_1(F_1)\mu_2(F_2) \\
    &= \mu_2(F_2)\mu_1(F_1).
  \end{align*}
  Hence, $m_{12}$ witnesses the independent combination of $m_2$ and $m_1$.

  To prove associativity, let $m_1 \deq (\mcal F_1, \mu_1)$, $m_2 \deq (\mcal F_2, \mu_2)$, $m_3 \deq (\mcal F_3, \mu_3)$, $m_{12} \deq (\mcal F_{12}, \mu_{12}) \deq m_1 \oplus m_2$, $m_{(12)3} \deq (\mcal F_{(12)3}, \mu_{(12)3}) \deq m_{12} \oplus m_3$. We need to show that if $m_1 \oplus m_2$ and $(m_1 \oplus m_2) \oplus m_3$ are defined, then $m_2 \oplus m_3$ and $m_1 \oplus (m_2 \oplus m_3)$ are defined, and $(m_1 \oplus m_2) \oplus m_3 = m_1 \oplus (m_2 \oplus m_3)$. 
  We first show $m_2 \oplus m_3$ is defined: any element $\mcal M_1$ is $\sigma$-finite and has non-zero total measure by definition. Hence, there exists a set $V \in \mcal F_1$ such that $0 < \mu_1(V) < \infty$. Consider a function $\rho : \gen{ \mcal F_2, \mcal F_3 } \to [0, \infty]$ defined by 
  \[ 
    \rho(F_{23}) \deq \frac{\mu_{(12)3}(V \cap F_{23})}{\mu_{1}(V)}.
  \]
  Notice that $\rho$ is a measure: when $F_{23} = \varnothing$, we have $\rho(F_{23}) = 0$, when $F_{23} = \biguplus_{i \in \N} F_i$, we have 
  \begin{align*} 
    \rho(\biguplus_{i \in \N} F_i) 
    &= \sum_{i \in \N} \frac{\mu_{(12)3}(V \cap F_i)}{\mu_1(V)} \\
    &= \sum_{i \in \N} \rho(F_i).
  \end{align*} 
  Also, $\rho$ is the independent combination measure of $F_2 \in \mcal F_2$ and $F_3 \in \mcal F_3$ because 
  \begin{align*}
  \rho(F_2 \cap F_3) 
  &= \frac{\mu_{(12)3}(V \cap F_2 \cap F_3)}{\mu_1(V)} \\
  &= \frac{\mu_1(V)\mu_2(F_2)\mu_3(F_3)}{\mu_1(V)} \\
  &= \mu_2(F_2)\mu_3(F_3).
  \end{align*}
  Hence, $m_2 \oplus m_3$ is defined and is equal to $(\gen{ \mcal F_2, \mcal F_3 }, \rho)$. Note that the choice of $V$ does not matter, they all yield the same measure by the uniqueness lemma (\cref{lem:append-pcm:uniqueness}). We write $m_{23} \deq (\mcal F_{23}, \mu_{23}) \deq m_2 \oplus m_3$.

  We now show that $m_1 \oplus (m_2 \oplus m_3)$ is defined by proving $\mcal F_{1(23)} = \gen{ \mcal F_{12}, \mcal F_3 }$ and $\mu_{1(23)} = \mu_{(12)3}$. This means we need to show that $\mu_{(12)3}$ is the independent combination of $\mu_{1}$ and $\mu_{23}$, i.e. proving for all $F_1 \in \mcal F_1$ and $F_{23} \in \mcal F_{23}$, $\mu_{(12)3}(F_1 \cap F_{23}) = \mu_1(F_1)\mu_{23}(F_{23})$. We first define $\Lambda, \Lambda' \subseteq \mcal P(\Omega)$ by 
  \begin{align*}
    \Lambda &\deq \left\{F_{23} \in \mcal F_{23}\ \middle|\ \textnormal{for all}\ F_1 \in \mcal F_1, \mu_{(12)3}(F_1 \cap F_{23}) = \mu_1(F_1)\mu_{23}(F_{23}) \right\}, \\
    \Lambda' &\deq \left\{F_{23} \in \mcal F_{23}\ \middle|\ 
      \begin{aligned}
        &\tx{$\textnormal{for all}\ F_1 \in \mcal F_1$, ($\mu_1(F_1) < \infty$ and $\mu_{23}(F_{23}) < \infty$)}\\
        &\tx{$\tx{implies } \mu_{(12)3}(F_1 \cap F_{23}) = \mu_1(F_1)\mu_{23}(F_{23})$}
      \end{aligned}
    \right\}.
  \end{align*}
  We now show that $\Lambda = \Lambda'$ and $\Lambda$ forms a $\lambda$-system -- $\Lambda \subseteq \Lambda'$ is trivially true. To show that $\Lambda' \subseteq \Lambda$, let $F_{23} \in \Lambda'$, we consider two cases:
  \begin{enumerate*}
    \item when $\mu_{23}(F_{23}) < \infty$, and
    \item when $\mu_{23}(F_{23}) = \infty$.
  \end{enumerate*}
  For (1), let $F_1 \in \mcal F_1$ and $\set{U_i \in \mcal F_1}_{i \in \N}$ be a disjoint cover of $\Omega$ with $\mu_1(F_1) < \infty$ (which exists because $\mu_1$ is $\sigma$-finite). Then
  \begin{align*}
    \mu_{(12)3}(F_1 \cap F_{23})
    &= \sum_{i \in \N} \mu_{(12)3}(F_1 \cap U_i \cap F_{23}) \tag{additivity} \\
    &= \sum_{i \in \N} \mu_1(F_1 \cap U_i)\mu_{23}(F_{23}) \tag{$\mu_1(F_1 \cap U_i) < \infty$} \\
    &= \mu_1(F_1)\mu_{23}(F_{23}). \tag{additivity}
  \end{align*}
  For (2), consider a disjoint cover $\{V_i \in \mcal F_{23}\}_{i \in \N}$ with $\mu_{23}(V_i) < \infty$. Then
  \begin{align*}
    \mu_{(12)3}(F_1 \cap F_{23})
    &= \sum_{i \in \N}\sum_{j \in \N} \mu_{(12)3}(F_1 \cap U_i \cap F_{23} \cap V_j) \tag{$\sigma$-finite measure} \\
    &= \sum_{i \in \N}\sum_{j \in \N} \mu_1(F_1 \cap U_i)\mu_{23}(F_{23} \cap V_j) \tag{$\mu_{23}(F_{23} \cap V_j) < \infty$} \\
    &= \sum_{i \in \N} \mu_1(F_1 \cap U_i)\sum_{j \in \N}\mu_{23}(F_{23} \cap V_j) \tag{distributivity in $\W$} \\
    &= \mu_1(F_1)\mu_{23}(F_{23}). \tag{additivity}
  \end{align*}
  Hence, $\Lambda = \Lambda'$. To show that $\Lambda$ forms a $\lambda$-system, we need to prove that 
  \begin{enumerate*}[label=(\roman*)]
    \item $\varnothing \in \Lambda'$, 
    \item $\Lambda'$ is closed under complements, and
    \item $\Lambda'$ is closed under countable disjoint unions.
  \end{enumerate*}
  For (i), it is true because $\mu_{(12)3}(\varnothing \cap F_{23}) = 0 = \mu_1(\varnothing)\mu_{23}(F_{23})$; For (ii), suppose $F_{23} \in \Lambda'$, then for all $F_1 \in \mcal F_1$, if $0 < \mu_1(F_1) < \infty$, then
  \begin{align*}
    \mu_{(12)3}(F_1 \cap (\Omega \setminus F_{23}))
    &= \mu_1(F_1) \cdot \frac{\mu_{(12)3}(F_1 \cap (\Omega \setminus F_{23}))}{\mu_1(F_1)} \\
    &= \mu_1(F_1)\mu_{23}(\Omega \setminus F_{23}).
  \end{align*}
  The second equality holds because we assumed $\mu_1(F_1) > 0$, which means the measure 
  \[
    F_{23} \mapsto \frac{\mu_{(12)3}(F_1 \cap F_{23})}{\mu_1(F_1)}
  \]
  witnesses the independent combination of $\mcal F_2$ and $\mcal F_3$. By \cref{lem:append-pcm:uniqueness}, we know 
  $$\mu_{23} = \frac{\mu_{(12)3}(F_1 \cap {-})}{\mu_1(F_1)}.$$
  Next, assuming $\mu_{1}(F_1) = 0$, then
  \begin{align*}
    \mu_{(12)3}(F_1 \cap (\Omega \setminus F_{23}))
    &\le \mu_{(12)3}(F_1) \\
    &= \mu_1(F_1)\mu_{23}(\Omega) \\
    &= 0.
  \end{align*}
  Hence, $\mu_{(12)3}(F_1 \cap (\Omega \setminus F_{23})) = 0 = \mu_1(F_1)\mu_{23}(\Omega \setminus F_{23})$ and (ii) holds.
  For (iii), assuming $\{F_i \in \Lambda'\}_{i \in \N}$ is a sequence of disjoint sets, then for any $G \in \mcal F_1$ with $\mu_1(G) < \infty$ and $\mu_{23}(\biguplus_{i \in \N} F_i) < \infty$, we have
  \begin{align*}
    \mu_{(12)3}\biggl(G \cap \biguplus_{i \in \N} F_i\biggr)
    &= \sum_{i \in \N} \mu_{(12)3}(G \cap F_i) \\
    &= \sum_{i \in \N} \mu_1(G)\mu_{23}(F_i) \\
    &= \mu_1(G)\mu_{23}\biggl(\biguplus_{i \in \N} F_i\biggr).
  \end{align*}
  Hence, $\Lambda = \Lambda'$ forms a $\lambda$-system. By the $\pi$-$\lambda$ theorem, to show that $\mcal F_{23} \subseteq \Lambda'$, it suffices to show that $\Pi \subseteq \Lambda'$ for some $\pi$-system $\Pi$ that generates $\mcal F_{23}$. By \cref{prop:append-pcm:pisys}, we know that $\Pi \deq \set{F_2 \cap F_3 \alt F_2 \in \mcal F_2, F_3 \in \mcal F_3}$ generates $\mcal F_{23}$. As a consequence, if $\Pi \subseteq \Lambda'$, then $\mcal F_{23} \subseteq \Lambda$. Consider any element $F_2 \cap F_3 \in \Pi$, we need to show that for all $F_1 \in \mcal F_1$, if $\mu_1(F_1) < \infty$ and $\mu_{23}(F_2 \cap F_3) < \infty$, then $\mu_{(12)3}(F_1 \cap (F_2 \cap F_3)) = \mu_1(F_1)\mu_{23}(F_2 \cap F_3)$, which is trivially true by definition of independent combination. This means $\mcal F_{23} = \Lambda' = \Lambda$, which implies associativity. Hence, $(\mcal M_1, \oplus, \mb 1)$ is a partial commutative monoid.
\end{proof}

\begin{definition}[\text{\cite[Definition 2.6]{li23}}]
  A $\sigma$-algebra $\mcal F$ of $\Omega$ is said to have \emph{finite footprint} if every element is of the form $F' \times [0, 1]^\N$ for some $F' \subseteq [0, 1]^n$ and $n \in \N$. A random variable $X : (\Omega, \Sigma_\Omega) \to (A, \Sigma_A)$ is said to have finite footprint if its pullback $\sigma$-algebra $\{X^{-1}(F) \alt F \in \Sigma_A\}$ has finite footprint. We denote $\msf{RV}(A, \Sigma_A)$ the set of random variables from $(\Omega, \Sigma_\Omega)$ with finite footprint.
\end{definition}

\begin{definition}
  Recall that a $\sigma$-algebra $\mcal F$ is countably generated if there exists a set of subsets $\mcal C \subseteq \mcal P(\Omega)$ such $\gen{ \mcal C } = \mcal F$. We say that \emph{random generator} $\mcal M$ is the restriction of $\mcal M_1$ where the $\sigma$-algebras are countably generated and have finite footprint, i.e. the set is defined by
  \[
    \mcal M \deq \setc{(\mcal F, \mu)}{
      \begin{aligned}
        &(\Omega, \mcal F, \mu)\ \tx{$\sigma$-finite measure space} \\
        &\mcal F \subseteq \Sigma_\Omega, \mu(\Omega) > 0 \\
        &\mcal F\ \tx{finite footprint and countably generated}
      \end{aligned}}\!.
  \]
\end{definition}

\begin{proposition}[Borel measure extension]\label[proposition]{prop:append-pcm:borel-ext}
  Let $(\mcal F, \mu) \in \mcal M$. Then there exists a $\sigma$-finite Borel measure $\mu^+ : \Sigma_\Omega \to [0, \infty]$ satisfying $\mu^+|_{\mcal F} = \mu$.
\end{proposition}
\begin{proof}
  Since $(\Omega, \Sigma_\Omega)$ is a standard Borel space and $\mcal F$ is a countably generated sub-$\sigma$-algebra of $\Sigma_\Omega$. By \citet[Proposition 433K]{fremlin11}, the extension $\mu^+ : \Sigma_\Omega \to [0, \infty]$ exists. By $\sigma$-finiteness of $\mu$, there is a sequence of sets $\set{F_i \in \mcal F \subseteq \Sigma_\Omega}_{i \in \N}$ that satisfies $\bigcup_{i \in \N} F_i = \Omega$ and $\mu^+(F_i) = \mu^+|_{\mcal F}(F_i) = \mu(F_i) < \infty$.
\end{proof}

\begin{lemma}[$\logic$ PCM]\label[lemma]{lem:append-pcm:basl-pcm}
  Let $(\bullet) : \mcal M \times \mcal M \pto \mcal M$ be a partial function defined by
  \[
    m_1 \bullet m_2 \deq \begin{cases}
      m_1 \oplus m_2 & \tx{if $m_1 \oplus m_2 \in \mcal M$} \\
      \tx{undefined} & \tx{otherwise.}
    \end{cases}
  \]
  Then $(\mcal M, \bullet, \mb 1)$ is a partial commutative monoid.
\end{lemma}
\begin{proof}
  Notice that $\mb 1 \in \mcal M$ because it has finite footprint $0$ and the $\sigma$-algebra is finite, which means countably generated.
  Next, we show that $(\oplus)$ is closed under $\mcal M$: Let $m_1 \deq (\mcal F_1, \mu_1)$ and $m_2 \deq (\mcal F_1, \mu_2)\in \mcal M$, assuming $(\mcal F_{12}, \mu_{12}) \deq m_1 \bullet m_2$ exists and $\mcal F, \mcal G$ are generated by $\{F_i\}_{i \in \N}$ and $\{G_i\}_{i \in \N}$ respectively, then $\mcal F_{12}$ is generated by $\{F_i\}_{i \in \N} \cup \{G_i\}_{i \in \N}$, which is a countable set. We also need to check that $\mcal F_{12}$ has finite footprint when $\mcal F_1$ and $\mcal F_2$ have finite footprints, this property is analogous to \citet[Lemma B.7]{li23}, where $\max(m, n)$ witnesses the finite footprint of $\mcal F_{12}$ when $\mcal F_1$ and $\mcal F_2$ have finite footprint $m$ and $n$ respectively.
\end{proof}

\begin{definition}[KRM]
  A \emph{(partial commutative) Kripke resource monoid} is a quadruple $(A, \cdot, 1, \sqsubseteq)$ such that $(A, \cdot, 1)$ is a partial commutative monoid, $(\sqsubseteq) \subseteq A \times A$ is a partial order such that $(\cdot)$ is bifunctorial with respect to $(\sqsubseteq)$, i.e. if $m_1 \sqsubseteq n_1$, $m_2 \sqsubseteq n_2$, and both $m_1 \cdot m_2$ and $n_1 \cdot n_2$ are defined, then $m_1 \cdot m_2 \sqsubseteq n_1 \cdot n_2$.
\end{definition}

\begin{theorem}[$\logic$ KRM]
  Let $(\sqsubseteq) \subseteq \mcal M \times \mcal M$ be a partial order defined by
  \[
    (\mcal F_1, \mu_1) \sqsubseteq (\mcal F_2, \mu_2)\ \textnormal{if and only if}\ \mcal F_1 \subseteq \mcal F_2\ \textnormal{and}\ \mu_2|_{\mcal F_1} = \mu_1.
  \]
  Then $(\mcal M, \bullet, \mb 1, \sqsubseteq)$ is a Kripke resource monoid.
\end{theorem}
\begin{proof}
  $(\mcal M, \bullet, \mb 1)$ is a PCM by \cref{lem:append-pcm:basl-pcm}. For $(\sqsubseteq)$, it suffices to check that $(\bullet)$ is bifunctorial with respect to $(\sqsubseteq)$. Let $m_i \deq (\mcal F_i, \mu_i)$ and $n_i = (\mcal G_i, \nu_i)$ for $i = 1, 2$. Suppose $m_1 \sqsubseteq n_1$,  $m_2 \sqsubseteq n_2$  and $(\mcal G_{12}, \mu_{12}) \deq n_1 \bullet n_2$ is defined, then by definition of independent combination, we know for all $G_1 \in \mcal G_1$ and $G_2 \in \mcal G_2$, $\mu_{12}(G_1 \cap G_2) \deq \mu_1(G_1) \cdot \mu_2(G_2)$. Since $\mcal F_1 \subseteq \mcal G_1$ and $\mcal F_2 \subseteq \mcal G_2$ by assumption, we know $\mu_{12}|_{\gen{\mcal F_1, \mcal F_2}}(F_1 \cap F_2) = \mu_1(F_1) \cdot \mu_2(F_2)$ for all $F_1 \in \mcal F_1$ and $F_2 \in \mcal F_2$. This means $\mu_{12}|_{\gen{\mcal F_1, \mcal F_2}}$ witnesses the independent combination of $m_1$ and $m_2$, which implies $m_1 \bullet m_2 = (\gen{ \mcal F_1, \mcal F_2 }, \mu_{12}|_{\gen{ \mcal F_1, \mcal F_2}}) \sqsubseteq n_1 \bullet n_2$.
\end{proof}

\begin{lemma}\label[lemma]{lem:append-pcm:int-indep-one}
  Let $m_i \deq (\mcal F_i, \mu_i) \in \mcal M$ for $i = 1, 2, 3$. Suppose $m_1 \bullet m_2$ is defined and $m_3 \sqsupseteq m_1 \bullet m_2$. Let $f : \Omega \to \W$ be an $(\mcal F_1, \BW)$-measurable function and $F_2 \in \mcal F_2$. Then
  \[
    \int_{F_2} f \,\dd{\mu_3} = \int_\Omega f\,\dd{\mu_1} \cdot \mu_2(F_2).
  \]
\end{lemma}
\begin{proof}
  Suppose $f = \ch{E_1}$ for some $E_1 \in \mcal F_1$, then $\int_{F_2} \ch{E_1}\,\dd{\mu_3} = \mu_3(F_1 \cap F_2) = \mu_1(F_1)\mu_2(F_2) = \int_\Omega \ch{E_1}\,\dd{\mu_1} \cdot \mu_2(F_2)$. Suppose $f$ is a $\Rnn$-valued simple function $\sum_{i = 1}^n c_i \ch{E_i}$, then 
  \begin{align*}
    \int_{F_2} \sum_{i = 1}^n c_i\ch{E_i}\,\dd{\mu_3} 
    &= \sum_{i = 1}^n c_i \int_{F_2} \ch{E_i}\,\dd{\mu_3} \\
    &= \sum_{i = 1}^n c_i \mu_1(E_i) \mu_2(F_2)  \\
    &= \int_\Omega \sum_{i = 1}^n c_i\ch{E_i}\,\dd{\mu_1} \cdot \mu_2(E_2).
  \end{align*}
  By the simple function approximation theorem for $\W$, every measurable function $f : \Omega \to \W$ has a sequence of $\Rnn$-simple measurable functions $\{f_i : \Omega \to \Rnn\}_{i \in \N}$ that converges pointwise to $f$, i.e. $f = \lim_{n \to \infty} f_n$. By the monotone convergence theorem, we have
  \begin{align*}
    \int_{F_2} \lim_{n \to \infty} f_n\,\dd{\mu_3}
    &= \lim_{n \to \infty} \int_{F_2} f_n\,\dd{\mu_3} \\
    &= \lim_{n \to \infty} \int_{\Omega} f_n\,\dd{\mu_1} \cdot \mu_2(F_2) \\
    &= \int_\Omega \lim_{n \to \infty} f_n\,\dd{\mu_1} \cdot \mu_2(F_2) \\
    &= \int_\Omega f\,\dd{\mu_1} \cdot \mu_2(F_2).
  \end{align*}
  Hence, the property holds for all measurable functions.
\end{proof}

\begin{lemma}\label[lemma]{lem:append-pcm:int-indep-two}
  Let $m_i \deq (\mcal F_i, \mu_i) \in \mcal M$ for $i = 1, 2$. Suppose $m_1 \bullet m_2$ is defined and $m_3 \sqsupseteq m_1 \bullet m_2$. Let $f_i : \Omega \to \W$ be an $(\mcal F_i, \BW)$-measurable function. Then
  \[
    \int_\Omega f_1 f_2\,\dd{\mu_3} = \int_\Omega f_1\,\dd{\mu_1} \cdot \int_\Omega f_2\,\dd{\mu_2}.
  \]
\end{lemma}
\begin{proof} 
  Suppose $f_2 = \ch{F}$ for some $E \in \mcal F_2$, then we apply \cref{lem:append-pcm:int-indep-one} and yield $\int_\Omega f_1 \ch{E}\,\dd{\mu_3} = \int_\Omega f_1\,\dd{\mu_1} \cdot \int_\Omega \ch{E}\,\dd{\mu_2}$. The next steps are similar to the proof of \cref{lem:append-pcm:int-indep-one}, suppose $f_2 = \sum_{i = 1}^n c_i \ch{E_i}$, then linearity of integral in $\W$ ensures the property holds. For the limiting case $f_2 = \lim_{n \to \infty} f'_n$, we apply the monotone convergence theorem.
\end{proof}

\begin{lemma}\label[lemma]{lem:append-pcm:int-indep-n}
  Let $m_i \deq (\mcal F_i, \mu_i) \in \mcal M$ for $i = 1, ..., n$. Suppose $m_1 \bullet ... \bullet m_n$ is defined and $m \sqsupseteq m_1 \bullet ... \bullet m_n$. Let $f_i : \Omega \to \W$ be an $(\mcal F_i, \BW)$-measurable function. Then
  \begin{align*}
    \int_\Omega \prod_{i = 1}^n f_i\,\dd{\mu} 
    = \prod_{i = 1}^n \int_\Omega f_i\,\dd{\mu_i}.
  \end{align*}
\end{lemma}
\begin{proof}
  By induction on $n \in \N_1$. The base case is equal by definition. For the inductive case, suppose $m_1 \bullet ... \bullet m_n$ is defined, then $m_1 \bullet ... \bullet m_{n - 1}$ is defined with the $\sigma$-algebra $\mcal F_{1..n - 1} \deq \gen{ \mcal F_1, ..., \mcal F_{n - 1} }$. It follows from \cref{lem:append-pcm:int-indep-two} that
  \begin{align*}
    \int_\Omega f_n\prod_{i = 1}^{n - 1} f_i\,\dd{\mu}
    &= \int_\Omega f_n\,\dd{\mu_n} \cdot \int_{\Omega} \prod_{i = 1}^{n - 1} f_i\,\dd{\mu} \\
    &= \prod_{i = 1}^{n} \int_\Omega f_i\,\dd{\mu_i}. \qedhere
  \end{align*}
\end{proof}

\begin{proposition}[mutual independence]
  Let $m_i \deq (\mcal F_i, \mu_i) \in \mcal M$ and $X_i : \Omega \to A_i$ for $i = 1, ..., n$ and suppose $(\mcal F, \mu) \deq m_1 \bullet ... \bullet m_n$ is defined. If for $i = 1, ..., n$, $X_i$ is $(\mcal F_i, \Sigma_{A_i})$-measurable and $E_i \in \Sigma_{A_i}$, then
  \[
    \mu\{X_1 \in E_1, ..., X_n \in E_n\} = \prod_{i = 1}^n \mu_i\{X_i \in E_i\}.
  \]
\end{proposition}
\begin{proof}
  By direct computation:
  \begin{align*}
    \mu\{X_1 \in E_1, ..., X_n \in E_n\}
    &= \mu\{\omega \in \Omega \alt \forall i \in \{1, ..., n\}, X_i(\omega) \in E_i\} \\
    &= \int_{\Omega} \ch{\bigcap_{i = 1}^n X_i^{-1}(E_i)}\,\dd{\mu} \\
    &= \int_\Omega \prod_{i = 1}^n \ch{X_i^{-1}(E_i)} \,\dd{\mu} \\
    &= \prod_{i = 1}^n \int_\Omega \ch{X_i^{-1}(E_i)}\,\dd{\mu_i} \tag{\cref{lem:append-pcm:int-indep-n}} \\
    &= \prod_{i = 1}^n \mu_i\{X_i \in E_i\}. \qedhere
  \end{align*}
\end{proof}

\begin{lemma}\label[lemma]{lem:append-pcm:inverse-radon-nikodym}
  Let $\varphi : \Omega \to \Rnn$ be an $(\mcal F, \mcal B\Rnn)$-measurable function and $\mu : \mcal F \to [0, \infty]$ a $\sigma$-finite measure. Then the induced measure $(\varphi \cdot \mu) : \mcal F \to \W$ defined by
  \[
    (\varphi \cdot \mu)(U) \deq \int_U \varphi\,\dd{\mu}
  \]
  is $\sigma$-finite. Moreover, if $(\mcal F, \mu) \in \mcal M$ and $(\varphi \cdot \mu)(\Omega) > 0$, then $(\mcal F, \varphi \cdot \mu) \in \mcal M$.
\end{lemma}
\begin{proof}
  Since $\mu$ is $\sigma$-finite, there is a cover $\set{U_i \in \mcal F}_{i \in \N}$ with $\mu(U_i) < \infty$. For each $i \in \N$, we define the sequence $\{U_{i, j} \subseteq \Omega\}_{j \in \N}$ with $U_{i, j} \deq \set{\omega \in U_i \alt \varphi(\omega) \le j}$. Notice that
  \begin{enumerate*}
    \item $\bigcup_{j \in \N} U_{i, j} = U_i$, and
    \item $U_{i, j} \in \mcal F$:
  \end{enumerate*}
  (1) is true as $\varphi$ has finite range, meaning that for any $\omega \in \Omega$, there exists an $N \in \N$ such that $\varphi(\omega) < N$. Hence, $\bigcup_{j \in \N} U_{i, j}$ `covers' all of $U_i$. For (2), notice that $U_{i, j} = U_i \cap \varphi^{-1}([0, j])$. By measurability of $\varphi$, we know $\varphi^{-1}([0, j]) \in \mcal F$, which implies $U_i \cap \varphi^{-1}([0, j]) = U_i \in \mcal F$. Next, notice that the restriction $\varphi|_{U_{i, j}} : U_{i, j} \to \Rnn$ is bounded above by $j$. Since $\mu(U_i) < \infty$, $U_{i, j} \in \mcal F$ and $U_{i, j} \subseteq U_i$, we know $\mu(U_{i, j}) \le \mu(U_i) < \infty$. Then
  \begin{align*}
    (\varphi \cdot \mu)(U_{i, j}) 
    &= \int_{U_{i, j}} \varphi\,\dd{\mu} \\
    &\le \mu(U_{i, j}) \cdot \sup \varphi|_{U_{i, j}} \\
    &\le \mu(U_i) \cdot j \\
    &< \infty.
  \end{align*}
  This implies $\{U_{i, j} \in \mcal F\}_{i, j \in \N}$ is a countable cover of $\mcal F$-measurable sets with each $(\varphi \cdot \mu)(U_{i, j}) < \infty$, which means $\varphi \cdot \mu$ is $\sigma$-finite. Assuming $(\mcal F, \mu) \in \mcal M$ and $(\varphi \cdot \mu)(\Omega) > 0$, since we have proven that $\varphi \cdot \mu$ is $\sigma$-finite and $\mcal F$ stays the same, we know $(\mcal F, \varphi \cdot \mu) \in \mcal M$.
\end{proof}

\begin{theorem}[conditional]\label{thm:append-pcm:cond}
  Let $(\mcal F, \mu) \in \mcal M$, $X : \Omega \to \mcal A$ be a $(\BO, \Sigma_A)$-measurable function and $\pi : \Sigma_A \to \W$ a $\sigma$-finite measure that dominates $X_*\mu$. Then there exists a measure $\mu^+ : \BO \to \W$ satisfying $\mu^+|_{\mcal F} = \mu$. Further, the $(X, \pi)$-disintegration $\{\mu^+_{x}\}_{x \in A}$ of $\mu^+$ exists and $(\mcal F, \mu^+_x|_{\mcal F}) \in \mcal M$ for $\pi$-almost-all $x \in A$.
\end{theorem}

\begin{proof}
  Readers familiar with disintegration will note that this is a disintegration theorem in disguise. However, there are several non-trivialities. For existence of a Borel measure, since $\mu : \mcal F \to \W$ is assumed to be a countably-generated $\sigma$-finite measure on a sub-Borel $\sigma$-algebra $\mcal F$, the extension $\mu^+$ exists following from the result of \citet[Proposition 433K]{fremlin11}.
  We then apply the Rokhlin-Simmons disintegration theorem (\cref{lem:append-disint:exists}) and obtain the conditional measure $\{\mu^+_x\}_{x \in A}$ and restrict them to the $\sigma$-algebra $\mcal F$.
\end{proof}
\newpage
\section{Syntax and semantics of \logic}

In this section, we range $\Gamma$ and $\Delta$ over two kinds of context: the \emph{deterministic context} $\Gamma = [x_1 : A_1, ..., x_n : A_n]$, which is a list of names $x_i$ and their underlying \emph{sets} $A_i$, and the \emph{probabilistic context} $\Delta = [X_1 : \mcal A_1, ..., X_n : \mcal A_n]$, which is a list of names $X_i$ and their underlying \emph{measurable spaces} $\mcal A_i = (A_i, \Sigma_{A_i})$. The contexts admit a standard interpretation (in the sense of denotational semantics): $\sem{\Gamma}$ is interpreted as the ($\mb{Set}$-)cartesian product by $A_1 \times ... \times A_n$, while $\sem{\Delta}$ is interpreted as the product measurable space $\mcal A_1 \times ... \times \mcal A_n$.

\begin{definition}
  The \emph{syntax} of a $\logic$ proposition is defined by the following grammar:
  \begin{align*}
    P &\ddeq \top \alt \bot \alt P \land P \alt P \lor P \alt P \Rightarrow P \alt P * P \alt P \wand P \\
    &\quad\alt \forall x : A. P \alt \exists x : A. P \alt \forallrv X : \mcal A. P \alt \existsrv X : \mcal A. P \\
    &\quad\alt E \sim \pi \alt \mbb E[E] = e \alt \own{E} \alt (\cond{\bc{x}{E}{\pi}} P) \alt \score{e} \\
    &\quad\alt \set{P} M \set{X : \mcal A. P}
  \end{align*}
  where $\pi$, $E$, $e$, $M$ range over maps defined in \cref{fig:basl:typing}, along with the definition of a well-typed proposition, i.e. a formula $P$ is \emph{well-typed} under $\Gamma; \Delta$ if $\Gamma; \Delta \vdash P$.
\end{definition}
\begin{figure}[h]
\begin{mdframed}
  \begin{mathpar}
    \inferrule[T-RandExpr]
      {E \in \sem{\Gamma} \to \mfn{\sem{\Delta}} {\mcal A}}
      {\Gamma ; \Delta \vdash_{\msf{re}} E : \mcal A} \and
    \inferrule[T-DetExpr]
      {e \in \sem{\Gamma} \to A}
      {\Gamma \vdash_{\msf{de}} e : A} \and
    \inferrule[T-Meas]
      {\pi \in \sem{\Gamma} \to M_\sigma(\mcal A)}
      {\Gamma \vdash_{\msf{meas}} \pi : \mcal A} \and
    \inferrule[T-True]
      {\ }
      {\Gamma ; \Delta \vdash \top} \and
    \inferrule[T-False]
      {\ }
      {\Gamma ; \Delta \vdash \bot} \and
    \inferrule[T-BinModality]
      {\Gamma ; \Delta \vdash P \\
       \Gamma ; \Delta \vdash Q}
      {\Gamma ; \Delta \vdash P \odot Q} \and
    \inferrule[T-DetQuant]
      {\Gamma, x : A ; \Delta \vdash P}
      {\Gamma ; \Delta \vdash \mcal Q x : A. P} \and
    \inferrule[T-RandQuant]
      {\Gamma ; \Delta, X : \mcal A \vdash P}
      {\Gamma ; \Delta \vdash \mcal Q_{\msf{rv}} X : \mcal A. P} \and
    \inferrule[T-Expectation]
      {\Gamma ; \Delta \vdash_{\msf{re}} E : (\R, \BR) \\
       \Gamma ; \Delta \vdash_{\msf{de}} e : \R}
      {\Gamma ; \Delta \vdash \mbb E[E] = e} \and
    \inferrule[T-Dist]
      {\Gamma ; \Delta \vdash_{\msf{re}} E : \mcal A \\
       \Gamma \vdash_{\msf{meas}} \pi : \mcal A}
      {\Gamma ; \Delta \vdash E \sim \pi}
    \and
    \inferrule[T-Ownership]
      {\Gamma ; \Delta \vdash_{\msf{re}} E : \mcal A}
      {\Gamma ; \Delta \vdash \own E} \and
    \inferrule[T-Conditioning]
      {\Gamma ; \Delta \vdash_{\msf{re}} E : \mcal A \\\\
       \Gamma \vdash_{\msf{meas}} \pi : \mcal A \\
       \Gamma, x : A ; \Delta \vdash P}
      {\Gamma ; \Delta \vdash \cond{\bc{x}{E}{\pi}}\, P} \and
    \inferrule[T-Likelihood]
      {\Gamma \vdash_{\msf{de}} e : \Rnn}
      {\Gamma ; \Delta \vdash \score{e}} \and
    \inferrule[T-Hoare]
      {\Gamma ; \Delta \vdash P \\
       \Gamma ; \Delta \vdash_{\msf{prog}} M : \mcal A \\
       \Gamma ; \Delta, X : \mcal A \vdash Q}
      {\Gamma ; \Delta \vdash \set{P} M \set{X : \mcal A. Q}} \and
  \end{mathpar}
  where ${\odot} \in \set{\land, \lor, \Rightarrow, *, \wand}$, $\mcal Q \in \set{\forall, \exists}$ and $M_\sigma(\mcal A)$ is the set of $\sigma$-finite measures over $\mcal A$.
\end{mdframed}
  \caption{Typing judgements for $\logic$ propositions}\label{fig:append-logic:typing}
\end{figure}

\begin{lemma}[Kripke monotonicity]\label[lemma]{lem:append-logic:kripke}
  Let $\gamma \in \sem{\Gamma}$ and $D \in \sem{\Delta}$. If $(\gamma, D, m) \vDash P$ and $m \sqsubseteq m'$, then $(\gamma, D, m') \vDash P$.
\end{lemma}
\begin{proof}
  By structural induction on $P$. The cases $\top$, $\bot$, $\land$, $\Rightarrow$, $\lor$, $*$, $\wand$, $\forall$, $\exists$, $\forallrv$, $\existsrv$ are straightforward. We focus on the probabilistic propositions. Let $(\mcal F, \mu) \deq m$ and $(\mcal F', \mu') \deq m'$. When $P = E \sim \mbb P$, $P = \msf{rv}\ E$, or $P = \mbb E[E] = e$, notice that having a finer $\sigma$-algebra $\mcal F'$ preserves measurability and the integral, and consequently the distribution as well. When $P = \score{e}$, $\mu'(\Omega) = \mu(\Omega)$ by definition of $\sqsubseteq$. When $P = \{P'\} M \{X. Q\}$, as the semantics does not depend on $m$, it holds for $m'$ as well. When $P = \cond{\bc{x}{E}{\pi}} P'$, any Borel extension $\mu'^+$ of $\mu'$ is a Borel extension of $\mu$, and consequently $X_*\mu'^+$ is absolutely continuous with respect to $\pi$.
\end{proof}

\begin{definition}[§B.4.1, \cite{li23}]
  A \emph{substitution} from $(\Gamma, \Delta)$ to $(\Gamma', \Delta')$ is a pair $(s, S)$ where $s : \sem{\Gamma'} \to \sem{\Gamma}$ and $S \in \mfn{\sem{\Delta'}}{\sem{\Delta}}$. Given a well-formed $\logic$ formula $\Gamma ; \Delta \vdash P$, we define the \emph{syntactic substitution} $\Gamma' ; \Delta' \vdash P[s, S]$ inductively as follows:
  \[
  \small
  \arraycolsep=1.4pt\def\arraystretch{4}
  \begin{array}{ll}
    \inferrule[S-Form]
      {s \in \semo{\Gamma'} \to \semo{\Gamma} \\
       S \in \mfn{\semo{\Delta'}}{\semo{\Delta}}} 
      {\Gamma'; \Delta' \vdash (s, S) : \Gamma; \Delta} &
      \begin{tikzcd}
        (\semo{\Gamma'}, \semo{\Delta'}) \ar[r, "(s{,} S)"] & (\semo{\Gamma}, \semo{\Delta})
      \end{tikzcd}
    \\
    \inferrule[S-RandExpr]
      {\Gamma; \Delta \vdashs{re} E : \mcal A \\
       \Gamma'; \Delta' \vdash (s, S) : \Gamma; \Delta}
      {\Gamma'; \Delta' \vdashs{re} E[s, S] : \mcal A} &
      E[s, S](\gamma) \deq 
        \begin{tikzcd}
          \semo{\Delta'}\ar[r, "S"] & \semo{\Delta} \ar[r, "E(s(\gamma))"] & \mcal{A}
        \end{tikzcd}
      \\
    \inferrule[S-DetExpr]
      {\Gamma \vdashs{de} e : A \\
       s \in \semo{\Gamma'} \to \semo{\Gamma}}
      {\Gamma' \vdashs{de} e[s] : A} & 
    e[s] \deq
    \begin{tikzcd}
      \semo{\Gamma'} \ar[r, "s"] 
      & \semo{\Gamma} \ar[r, "e"] 
      & A 
    \end{tikzcd}
    \\
    \inferrule[S-Meas]
      {\Gamma \vdashs{meas} \pi : \mcal A \\
       s \in \semo{\Gamma'} \to \semo{\Gamma}}
      {\Gamma' \vdashs{meas} \pi[s] : \mcal A} & 
    \pi[s] \deq
      \begin{tikzcd}
        \semo{\Gamma'} \ar[r, "s"]  
        & \semo{\Gamma} \ar[r, "\pi"]
        & M_\sigma(\mcal A)
      \end{tikzcd}
    \\
    \inferrule[S-Prog]
      {\Gamma; \Delta \vdashs{prog} M : \mcal A \\
       \Gamma'; \Delta' \vdash (s, S) : \Gamma; \Delta}
      {\Gamma'; \Delta' \vdashs{prog} M[s, S] : \mcal A} 
    & 
    M[s, S](\gamma) \deq 
    \begin{tikzcd}
      (\semo{\Gamma'}, \semo{\Delta'}) 
        \ar[r, "1 \times S"]
      & (\semo{\Gamma'}, \semo{\Delta})
        \ar[r, "M(s(\gamma))"]
      & (\semo{\Gamma}, \semo{\Delta})
    \end{tikzcd}
    \\
    \inferrule[S-Prop]
      {\Gamma; \Delta \vdash P \\
       \Gamma'; \Delta' \vdash (s, S) : \Gamma; \Delta}
      {\Gamma; \Delta \vdash P[s, S]} & \text{$P[s, S]$ is defined inductively below}
  \end{array}
  \]
  \[
  \small
  \begin{array}{ll}
    \begin{aligned}
      \top[s, S] &\deq \top \\ 
      \bot[s, S] &\deq \bot \\
      (E \sim \pi)[s, S] &\deq E[s, S] \sim \pi[s] \\
      (\own{E})[s, S] &\deq \own {E[s, S]} \\
      (\score{e})[s, S] &\deq \score{e[s]}
    \end{aligned} &
    \begin{aligned}
      (P \mathbin{\mcal B} Q)[s, S] &\deq P[s, S] \mathbin{\mcal B} Q[s, S] \\
      (\mcal Q x : A. P)[s, S] &\deq \mcal Q x : A. P[s \times 1_A, S] \\
      (\mcal Q_{\msf{rv}} X : \mcal A. P)[s, S] &\deq \mcal Q_{\msf{rv}} X : A. P[s, S \times 1_{\mcal A}] \\
      (\mbb E[E] = e)[s, S] &\deq \mbb E[E[s, S]] = e[s]
    \end{aligned}
  \end{array}
  \]
  \vspace{-10pt}
  \begin{align*}
    \scriptsize
    (\set{P} M \set{X : \mcal A. Q})[s, S] &\deq \set{P[s, S]} M[s, S] \set{X : \mcal A. Q[s, S \times 1_{\mcal A}]} \\
      (\cond{\bc{x}{E}{\pi} : \mcal A} P)[s, S] &\deq \cond{\bc{x}{E[s, S]}{\pi[s]}} P[s \times 1_A, S]
  \end{align*}
  where $\mcal B \in \set{\land, \lor, \to, *, \wand}$ and $\mcal Q \in \set{\forall, \exists}$.
\end{definition}

\begin{lemma}[substitution]\label[lemma]{lem:append-logic:sub}
  Let $(\gamma, D, m)$ be a configuration and $P$ a well-formed proposition. Then 
  \[
    (\gamma, D, m) \vDash P[s, S] \iff (s(\gamma), S \circ D, m) \vDash P.
  \]
\end{lemma}
\begin{proof}
  Most cases follow from structural induction, we focus on the case for Hoare triple $\set{P} M \set{X : \mcal A. Q}$. The proof for $\set{P} M \set{X : \mcal A. Q}$ is similar to the proof in $\lilac$ (see \citet[Lemma B.10]{li23}), except for the `normalising' assumption, which is equivalent as both sides yield the following condition: 
  \[
    \int_\Omega \semo{M_{s(\gamma)}}(S(D(\omega)), A)\,\mu^+_0(\dd{\omega}) > 0.
  \]
  The rest of the proof follows from instantiating the extension $D_\ext'$ to $(D_\ext, D)$.   \end{proof}

\begin{proposition}[entailments]
  Define semantic entailment $P \vdash Q$ to mean $(\gamma, D, m) \vDash P$ implies $(\gamma, D, m) \vDash Q$ for all $(\gamma, D, m)$.
  The following entailments are sound:
  \begin{mathpar}
    \inferrule[E-$*$-Ident]
      {}
      {P * \top_1 \vdash P} \and
    \inferrule[E-$*$-Comm]
      {}
      {P * Q \vdashv Q * P} \and
    \inferrule[E-$*$-Assoc]
      {}
      {(P * Q) * R \vdashv P * (Q * R)} \and
    \inferrule[E-True]
      {\affine P}
      {P \vdash \top_1} \and
    \inferrule[E-False]
      {}
      {\score{0} \vdash \bot} \\
    \inferrule[E-$*$-Weak$_1$]
      {\affine Q}
      {P * Q \vdash P} \and
    \inferrule[E-$*$-Weak$_2$]
      {\affine P}
      {P * Q \vdash Q} \and
    \inferrule[E-$*$-Weak]
      {\affine P \\ \affine Q}
      {P * Q \vdash P \land Q} \and
    \inferrule[E-$*$-Cong]
      {P \vdash Q}
      {P * R \vdash Q * R} \and
    \inferrule[E-$*$-AdjointL]
      {P * Q \vdash R}
      {P \vdash Q \wand R} \and
    \inferrule[E-$*$-AdjointR]
      {P \vdash Q \wand R}
      {P * Q \vdash R} \and
    \inferrule[E-$\land$-Elim$_1$]
      {}
      {P \land Q \vdash P} \and
    \inferrule[E-$\land$-Elim$_2$]
      {}
      {P \land Q \vdash Q} \and
    \inferrule[E-$\lor$-Intro$_1$]
      {}
      {P \vdash P \lor Q} \and
    \inferrule[E-$\lor$-Intro$_2$]
      {}
      {Q \vdash P \lor Q} \and
    \inferrule[E-NormConst]
      {}
      {\msf{NormConst} * \msf{NormConst} \vdash \msf{NormConst}}
    \end{mathpar}
\end{proposition}
\begin{proof}
  We focus on rules that involve affine propositions. The rule \rrule{E-True} follows from definition of an affine proposition -- $P$ being affine and $(\gamma, D, (\mcal F, \mu)) \vDash P$ imply $\mu(\Omega) = 1$, which satisfies of $\top_1 = \score{1}$. For \rrule{E-$*$-Weak$_1$}, we assume $(\gamma, D, m) \vDash P * Q$ and $m_1 \bullet m_2 \sqsubseteq m$ such that $(\gamma, D, m_1) \vDash P$ and $(\gamma, D, m_2) \vDash Q$. Notice that $m_2$ being a probability space implies $m_1 \sqsubseteq m_2$ -- by definition of independent combination, we have $\mu(F_1) = \mu(F_1 \cap \Omega) = \mu_1(F_1)\mu_2(\Omega) = \mu_1(F_1)$ for any $F_1 \in \mcal F_1$. Then, notice that Kripke monotonicity holds (\cref{lem:append-logic:kripke}), which means $(\gamma, D, m_1) \vDash P$ implies $(\gamma, D, m) \vDash P$. The proof for \rrule{E-$*$-Weak$_2$} is similar. For \rrule{E-$*$-Weak}, it follows directly from \rrule{E-$*$-Weak$_1$} and \rrule{E-$*$-Weak$_2$}.

  For \rrule{E-NormConst}, suppose $(\gamma, D, (\mcal F, \mu)) \vDash \exists k : (0, \infty). \score{k} * \exists k' : (0, \infty). \score{k'}$, then there exists $m_1 = (\mcal F_1, \mu_1), m_2 = (\mcal F_2, \mu_2) \in \mcal M$ such that $m_1 \bullet m_2 \sqsubseteq m$ and $(\gamma, D, (\mcal F, m_1)) \vDash \score{k}$ for some $k \in (0, \infty)$ and $(\gamma, D, (\mcal F, m_2)) \vDash \score{k'}$ for some $k' \in (0, \infty)$. This implies $\mu_1(\Omega) = k$ and $\mu_2(\Omega) = k'$. By definition of independent combination, we know $\mu(\Omega) = \mu(\Omega \cap \Omega) = \mu_1(\Omega) \mu_2(\Omega) = k \cdot k'$, which means $(\gamma, D, (\mcal F, \mu)) \vDash \score{k \cdot k'}$, implying $\msf{NormConst}$.
\end{proof}

\begin{lemma}[\text{\cite[Theorem 2]{chang97}}]\label[lemma]{lem:append-logic:dendis}
  Let $\set{\mu_x}_{x \in A}$ be an $(X, \pi)$-disintegration of $\mu : \Sigma_\Omega \to \W$. Then $X_*\mu$ is absolutely continuous with respect to $\pi$ and for $\pi$-almost-all $x \in A$, 
  \[
    \frac{\dd{X_*\mu(x)}}{\dd{\pi}} = \mu_x(\Omega).
  \]
  Also, if $\pi = X_*\mu$, then $\mu_x$ is a probability measure for $\pi$-almost-all $x \in A$.
\end{lemma}

\begin{theorem}[Bayes' theorem]\label{lem:append-logic:repr}
  The following entailment is sound:
  \begin{mathpar}
    \mprset{fraction={---}}
    \inferrule[E-Bayes]
      {}
      {E \sim f \cdot \pi \vdashv \cond{\bc{x}{E}{\pi}} \score{f(x)}}
  \end{mathpar}
\end{theorem}
\begin{proof}
  Let $(\gamma, D, m)$ be a configuration, $m = (\mcal F, \mu)$ and 
  write $X(\omega) \deq E(\gamma)(D(\omega))$. We prove the entailment in two directions. From left to right, we assume $(\gamma, D, (\mcal F, \mu)) \vDash E \sim f \cdot \pi$, then $(\gamma, D, (\mcal F, \mu)) \vDash \msf{own}\ E$ by definition of $\vDash$. Next, for any Borel extension $\mu^+$ and any $(X, \pi)$-disintegration $\{\mu^+_x\}_{x \in A}$, we know $(\gamma, D, (\mcal F, \mu^+_x|_{\mcal F})) \vDash \score{f(x)}$ holds for $\pi$-almost-all $x \in A$ because for any $F \in \Sigma_A$, 
  \begin{align*}
    \int_F \mu^+_x(\Omega)\,\pi(\dd{x})
    &= \int_F \frac{\dd{X_*\mu(x)}}{\dd{\pi}}\,\pi(\dd{x}) \tag{\cref{lem:append-logic:dendis}} \\
    &= \int_F \frac{\dd{(f \cdot \pi)}(x)}{\dd{\pi}}\,\pi(\dd{x}) \tag{$(\gamma, D, m) \vDash E \sim f \cdot \pi$} \\
    &= \int_F f(x)\,\pi(\dd{x}). \tag{$\frac{\dd{(f \cdot \pi)}}{\dd{\pi}} = f$ almost surely}
  \end{align*}
  Also, $X_*\mu = f \cdot \pi$ is absolutely continuous with respect to $\pi$. Let $F$ be a $\pi$-null-set, then $(f \cdot \pi)(F) = 0$. From right to left, we assume $(\gamma, D, (\mcal F, \mu)) \vDash \cond{\bc{x}{E}{\pi}} \score{f(x)}$, then for any Borel extension $\mu^+$ and $(X, \pi)$-disintegration $\{\mu^+_x\}_{x \in A}$, we know the following statement holds for $\pi$-almost-all $x \in A$:
  \[
    (\gamma, D, (\mcal F, \mu^+_x|_{\mcal F})) \vDash \score{f(x)}.
  \]
  This implies $\mu^+_x(\Omega) = f(x)$ holds for $\pi$-almost-all $x \in A$, which then implies for all $F \in \Sigma_A$,
  \begin{align*}
    X_*\mu(F)
    &= X_*\mu^+(F) \tag{Borel extension} \\
    &= \int_{\Omega} \ch{X^{-1}(F)}\,\dd{\mu^+} \tag{by definition} \\
    &= \int_A \,\mu_x(X^{-1}(F))\,\pi(\dd{x}) \tag{$(X, \pi)$-disintegration} \\
    &= \int_F \mu_x(\Omega)\,\pi(\dd{x}) \tag{$\ch{F}(x)\mu_x(\Omega)= \mu_x(X^{-1}(E))$} \\
    &= \int_F f\,\dd{\pi} \tag{$(\gamma, D, (\mcal F, \mu^+_x|_{\mcal F})) \vDash \score{f(x)}$ a.e.} \\
    &= (f \cdot \pi)(F), \tag{by definition}
  \end{align*}
  which yields $(\gamma, D, (\mcal F, \mu)) \vDash E \sim f \cdot \pi$.
\end{proof}

\newpage
\section{Proof Rules}\label{append:proofrules}

\begin{theorem}[soundness]
The following Hoare triples are sound, \noindent i.e. for all proof rules defined below, $\vdash \{P\}\ M\ \{X : \mcal A.Q\}$ implies $\vDash \{P\}\ M\ \{X : \mcal A. Q\}$.

\begin{figure}[h]
  \begin{mdframed}
    \small
    \begin{mathpar}
      \inferrule[H-Sample]
        {}
        {\{\topp\}\,\kp{sample}(\mbb P)\,\{X : \R. X \sim \mbb P\}} \and
      \inferrule[H-CondSample]
        {}
        {\{X \sim \mbb P\}\,\kp{sample}(p(X))\,\{Y : \R. \cond{\bc{x}{X}{\mbb P}} Y \sim p(x)\}} \and
      \inferrule[H-Score]
        {}
        {\{X \sim \pi\}\,\kp{score}(f(X))\,\{X \sim f \cdot \pi\}} \and
      \inferrule[H-CondScore]
        {}
        {\{\cond{\bc{x}{X}{\pi}} Y \sim p(x)\}\,\kp{score}(f(Y)) \{\cond{\bc{x}{X}{\pi}} Y \sim f \cdot p(x)\}} \and
      \inferrule[H-Observe]
        {}
        {\{X \sim \pi\}\,\kp{observe}(P(X))\,\{X \sim \ell_P \cdot \pi\}} \and
      \inferrule[H-CondObserve]
        {}
        {\{\cond{\bc{x}{X}{\pi}} Y \sim p(x)\}\,\kp{observe}(P(Y)) \,\{\cond{\bc{x}{X}{\pi}} Y \sim \ell_P \cdot p(x)\}} \and
      \inferrule[H-Return]
        {}
        {\{Q[\semo{M}/X]\}\,\kpr{return} M\,\{X : \mcal A. Q\}} \and
      \inferrule[H-Frame]
        {\vdash \set{P} M \set{X : \mcal A. Q}}
        {\vdash \set{P * F} M \set{X : \mcal A. Q * F}}\ (X \notin \msf{fv}(F)) \and
      \inferrule[H-Let]
        {\vdash \{P\}\, M \,\{X : \mcal A. Q\} \\
        \vdash \forallrv X : \mcal A. \set{Q} N \set{Y : \mcal B. R}}
        {\vdash \set{P} \kr{let} X = M \kb{in} N \set{Y : \mcal B. R}} \and
      \inferrule[H-Cons]
        {P' \vdash P \\
          \vdash \set{P} M \set{X : \mcal A. Q} \\
          Q \vdash Q'}
        {\vdash \{P'\}\,M\,\{X : \mcal A. Q'\}}
    \end{mathpar}
  \end{mdframed}
  \vspace{-10pt}
\end{figure}.
\end{theorem}

\begin{lemma}\label[lemma]{lem:append-hoare:foot}
  Let $(\mcal F, \mu) \in \mcal M$ such that $m \in \N$ witnesses its finite footprint. For any $n \ge m$, define $\mcal F|_n \deq \pi_{1..n}[\mcal F] = \{\{\omega_{1..n} \alt \omega \in F\} \alt F \in \mcal F\}$ and $\mu|_n : \mcal F|_n \to \W$ by $\mu|_n(F) \deq \mu(F \times \Omega)$. Then: 
  \begin{enumerate}
    \item $\mcal F|_n$ is a sub-$\sigma$-algebra of $\mcal B[0, 1]^n$,
    \item the map $\pi_{1..n} : \Omega \to [0, 1]^n$ is $(\mcal F, \mcal F|_n)$-measurable,
    \item $\mu|_n = \pi_{1..n*}\mu$, and
    \item for any standard Borel $(A, \Sigma_A)$ and $(\mcal F, \Sigma_A)$-measurable $f : \Omega \to A$, the map $f|_n : [0, 1]^n \to A$ defined by $f|_n(\omega_{1..n}) \deq f(\omega_{1..n}, x)$ for some $x \in \Omega$
    is $(\mcal F|_n, \Sigma_A)$-measurable. Also, for all $\omega_{1..n} \in [0, 1]^n$, $x, y \in \Omega$, $f(\omega_{1..n}, x) = f(\omega_{1..n}, y)$.
  \end{enumerate}
\end{lemma}
\begin{proof}
  For (1), we must first show that $\mcal F|_n$ is a $\sigma$-algebra. Since $\varnothing = \varnothing \times \Omega \in \mcal F$, $\varnothing \in \mcal F|_n$. Next, for every $F' \in \mcal F|_n$, we know $F' \times \Omega \in \mcal F$, which implies
  \[
    \Omega \setminus (F' \times \Omega) = ([0, 1]^n \setminus F') \times \Omega
  \] 
  and $[0, 1]^n \setminus F' \in \mcal F|_n$. For $\{F'_{i} \in \mcal F|_n\}_{i \in \N}$, notice that 
  $\biguplus_{i \in \N} (F'_i \times \Omega) = (\biguplus_{i \in \N} F'_i) \times \Omega$,
  which implies $\biguplus_{i \in \N} F'_i \in \mcal F|_n$; For (2), suppose $F' \in \mcal F|_n$, then 
  \begin{align*}
    \pi^{-1}_{1..n}(F') 
    &= \{\omega \in \Omega \alt \omega_{1..n} \in F'\} \\
    &= \{\omega \in \Omega \alt \omega \in F' \times \Omega \} \\
    &\in \mcal F.
  \end{align*}
  For (3), since $\pi_{1..n} : \Omega \to [0, 1]^n$ is $(\mcal F, \mcal F|_n)$-measurable, the pushforward measure $\pi_{1..n*}\mu : \mcal F|_n \to \W$ is well-defined. Moreover, 
  \begin{align*}
    \mu|_n(F') 
    &= \mu(F' \times \Omega) \\
    &= \mu(\pi_{1..n}^{-1}(F')) \\
    &= \pi_{1..n*}\mu(F').
  \end{align*}
  For (4), notice that the domain of $f$ is measurably isomorphic to the product with $(\Omega, \mcal F) \cong ([0, 1]^n, \mcal F|_n) \times (\Omega, \{\varnothing, \Omega\})$. Recall that the slice of a product measurable function is measurable, which means the function
  \[ 
    \omega_{1..n} \longmapsto f(\omega_{1..n}, x)
  \] 
  is $(\mcal F|_n, \Sigma_A)$-measurable for any $x \in \Omega$.
    Next, assume by contradiction that there is some $y \in A$ and $\gamma \in [0, 1]^n$ such that $x \neq y$ and $f(\gamma, x) \neq f(\gamma, y)$. Observe that $f(\gamma, {-}) : \Omega \to A$ is $(\{\varnothing, \Omega\}, \Sigma_A)$-measurable because it is a product $\sigma$-algebra. Since $(A, \Sigma_A)$ is standard Borel, it contains all singleton sets and we have $E \deq f^{-1}(\gamma, \{f(\gamma, x)\}) \in \{\varnothing, \Omega\}$. Notice that $x \in E$, which means $E = \Omega$ by definition of $(\{\varnothing, \Omega\}, \Sigma_A)$-measurable functions. This implies $y \in E$ and $f(\gamma, x) = f(\gamma, y)$, which is a contradiction.
\end{proof}

\begin{lemma}\label[lemma]{lem:append-hoare:sample}
  The following Hoare triple is sound:
  \[
    \inferrule[H-Sample]
    {}
    {\vdash \{\topp\}\ \kp{sample}(\mbb P) \set{X : \R. X \sim \mbb P}}
  \]
\end{lemma}
\begin{proof}
  Let $m_\pre \deq (\mcal F_\pre, \mu_\pre)$, $(\gamma, D, m_\pre)$ be a configuration, $(\gamma, D, m_\pre) \vDash \topp$ and $m_\fr \in \mcal M$ such that $(\mcal F_0, \mu_0) \deq m_\pre \bullet m_\fr$ is defined. Consider a Borel measure $\mu_0^+$ that extends $m_\pre \bullet m_\fr$ (i.e. $\mu^+_0|_{\mcal F_0} = \mu_0$), a probabilistic context $\Delta_\ext$ and its random variables $D_\ext \in \msf{RV}\semo{\Delta_\ext}$. Notice that the condition regarding the program integrating to a positive number is always true because $\mu_0(\Omega)$ is non-zero by definition of $\mcal M$:
  \[
    \int_\Omega \semo{\kp{sample}\ \mbb P}(D(\omega), A)\,\mu_0^+(\dd{\omega}) = \int_\Omega 1\,\mu_0^+(\dd{\omega}) = \mu_0^+(\dd{\omega}) > 0.
  \]
  
  Denote $n$ the maximum dimension of the Hilbert cube used in $m_\pre$, $m_\fr$, $D$ and $D_\ext$, i.e. $n$ is the maximum of the witness of their finite footprint property. We define the $\sigma$-algebra $\mcal F_\post$ and its measure $\mu_\post$ by
  \begin{align*}
    \mcal F_\post &\deq \mcal F_\pre|_{n} \otimes \mcal B[0, 1] \otimes \{\varnothing, \Omega\} \\
    \mu_\post &\deq \mu_\pre|_n \otimes \lambda_{[0, 1]} \otimes p
  \end{align*}
  where $p : \{\varnothing, \Omega\} \to \W$ is the trivial probability measure with $p(\Omega) = 1$. We first check that the above definition is well-formed: if $\mcal F$ is a $\sigma$-algebra with finite footprint $m$, then for any $n \ge m$, its restriction to $n$ dimensions 
  \[
    \mcal F|_n \deq \pi_{1..n}(\mcal F) \deq \{\{\pi_{1..n}(\omega) \alt \omega \in F \} \alt F \in \mcal F\}
  \]
  is a $\sigma$-algebra as every dimension greater $n$ can be written as $[0, 1]^\N$. Similarly, the restriction $\mu|_n : \mcal F|_n \to \W$ is defined by
  \[
    \mu|_n(F) \deq \mu(F \times [0, 1]^\N).
  \]
  By the randomisation lemma (\cref{lem:append-hoare:rand}), there exists a $(\mcal B[0, 1], \BR)$-measurable function $f : [0, 1] \to \R$ such that $f_*\lambda_{[0, 1]} = \mbb P(\gamma)$. Define $X \in \msf{RV}(\mcal A)$ by $X(\omega) \deq f(\omega_{n + 1}) = (f \circ \pi_{n + 1})(\omega)$, which is the composition of two measurable functions, and hence, is measurable. Notice that $X$ has finite footprint $n + 1$ as the pullback $\sigma$-algebra is of the form
  \begin{align*}
    X^{-1}(\BR)
    &= \{\pi_{n + 1}^{-1}(f^{-1}(U)) \alt U \in \BR\} \\
    &= \{\{(\omega_1, \omega_2, ...) \in \Omega \alt \omega_{n + 1} \in f^{-1}(U)\} \alt U \in \BR \} \\
    &= \{[0, 1]^n \times f^{-1}(U) \times \Omega \alt U \in \BR\}.
  \end{align*}
  Also, define $\mcal F_1 \deq \langle \mcal F_\post, \mcal F_\fr \rangle$ and $\mu_1 : \mcal F_1 \to \W$ by $\mu_1 \deq \mu_0|_n \otimes \lambda_{[0, 1]} \otimes p$. Notice that $(\mcal F_1, \mu_1)$ is well-defined: since $n$ witnesses the finite footprint of $\mcal F_\fr$, $\mu_0|_n$ is a well-defined measure on $\mcal F_0|_n$, which is a $\sigma$-algebra of $[0, 1]^n$. Additionally, $(\mcal F_1, \mu_1) \deq m_\post \bullet m_\fr$ because for all $F_\post \in \mcal F_\post$ and $F_\fr \in \mcal F_\fr$, we have $F_\post = F_\post|_{n + 1} \times [0, 1]^\N$, $F_\fr = F_\fr|_{n} \times [0, 1]^\N$ and 
  \begin{align*}
    \mu_1(F_\post \cap F_\fr)
    &= (\mu_0|_n \otimes \lambda_{[0, 1]} \otimes p)(F_\post \cap F_\fr) \\
    &= (\mu_\pre|_n \otimes \lambda_{[0, 1]} \otimes p)(F_\post) \cdot \mu_\fr({F_\fr}|_{n + 1}) \\
    &= \mu_\post(F_\post) \cdot \mu_\fr(F_\fr).
  \end{align*}
  Also, $(\gamma, (D, X), m_\post) \vDash X \sim \mbb P$: notice that $X = f \circ \pi_{n + 1}$ is $(\mcal F_\post, \Sigma_A)$-measurable because $\pi_{n + 1}$ is $(\mcal F_\post, \mcal B[0, 1])$ by construction of the product measure and $f$ is $(\mcal B[0, 1], \Sigma_A)$-measurable by assumption, and for every $E \in \Sigma_A$, we have
  \begin{align*}
    X_*\mu_\post(E) 
    &= \mu_\post(\{\omega \in \Omega \alt \omega_{n + 1} \in f^{-1}(E)\}) \\
    &= \mu_\post([0, 1]^n \times f^{-1}(E) \times [0, 1]^\N) \\
    &= \mu_\pre([0, 1]^\N) \cdot f_*\lambda_{[0, 1]}(E) \\
    &= \mbb P(\gamma)(E). \tag{$\mu_\pre([0, 1]^\N) = 1$ by assumption}
  \end{align*}
  Finally, consider a Borel measure $\mu_1^+ : \Sigma_\Omega \to \W$ 
  that satisfies $\mu^+_1|_{\mcal F_1} = \mu_1$ (which is guaranteed to exist by \cref{prop:append-pcm:borel-ext}), then the following identity holds for all $U \in \Sigma_{\semo{\Delta_\ext}} \otimes \Sigma_{\semo{\Delta}} \otimes \Sigma_A$:
  \begin{align*}
    \int_\Omega &\int_\R \ch{U}(D_\ext(\omega), D(\omega), x)\,\semo{\kp{sample}(\mbb P_\gamma)}(D(\omega), \dd{x})\,\mu_0^+(\dd{\omega}) \\
    &= \int_\Omega \int_\R \ch{U}(D_\ext(\omega), D(\omega), x)\,\mbb P_\gamma(\dd{x})\,\mu_0^+(\dd{\omega}) \tag{semantics of $\kp{sample}$} \\
    &= \int_{\Omega} \int_\R \ch{U}(D_\ext|_n(\pi_{1..n}(\omega)), D|_n(\pi_{1..n}(\omega)), x)\,\mbb P_\gamma(\dd{x})\,\mu_0^+(\dd{\omega}) \tag{\cref{lem:append-hoare:foot}} \\
    &= \int_{[0, 1]^n} \int_{[0, 1]} \ch{U}(D_\ext|_n, D|_n, f(x))\, \lambda_{[0, 1]}(\dd{x}) \,\dd{\mu_0|_n} \tag{\cref{lem:append-hoare:foot}} \\
    &= \int_{[0, 1]^{n + 1}} \ch{U}(D_\ext|_n, D|_n, f \circ \pi_{n + 1}) \circ \pi_{1..{n + 1}}\, \dd{(\mu_0|_n \otimes \lambda_{[0, 1]})} \tag{Fubini} \\
    &= \int_{\Omega} \ch{U}(D_\ext, D, X)\,\dd{\mu_1^+}, \tag{by definition of $\mu_1$}
  \end{align*}
  which yields the postcondition $X \sim \mbb P$.
\end{proof}

\begin{lemma}
  The following Hoare triple is sound:
  \[
    \inferrule[H-Score]
    {}
    {\vdash \set{X \sim \pi} \kp{score}(f(X)) \set{X \sim f \cdot \pi}}
  \]
\end{lemma}
\begin{proof}
  Let $(\gamma, (D, X), \_)$ be a configuration, $m_\pre \in \mcal M$ such that $(\gamma, (D, X), m_\pre) \vDash X \sim \pi$, and all $m_\fr$ with $(\mcal F_0, \mu_0) \deq m_\pre \bullet m_\fr$ defined, and all measures $\mu_0^+$ satisfying $\mu_0^+|_{\mcal F_0} = \mu_0$, and all probabilistic contexts $\Delta_\ext$ and all $D_{\ext} \in \msf{RV}\semo{\Delta_\ext}$. We assume
  \[
    \int_{\Omega}\semo{f(X)}(D(\omega), 1)\,\mu_0^+(\dd{\omega}) = \int_{\Omega} f(X(\omega))\,\mu_0^+(\dd{\omega}) > 0,
  \]
  Define $Y \in \msf{RV}(\mb 1)$ by $Y(\omega) \deq \star$, $\mcal F_\post = \mcal F_\pre$ and $\mu_\post : \mcal F_{\post} \to \W$ by
  \[
    \mu_\post(F) \deq \int_F f(X(\omega))\,\mu_\pre(\dd{\omega}).
  \]
  Notice that $m_\post$ is well-defined: $f \circ X : \Omega \to \W$ is $(\mcal F_\pre, \BW)$-measurable since $X$ is assumed to be $(\mcal F_\pre, \Sigma_A)$-measurable via $(\gamma, D, m_\pre) \vDash X \sim \pi$ and $f$ is a $(\Sigma_A, \BW)$-measurable function. Moreover, there is a measure $\mu_1 : \langle\mcal F_\post, \mcal F_{\fr}\rangle \to \W$ defined by
  \[
    \mu_1(F) \deq \int_F f(X(\omega))\,\mu_0(\dd{\omega})
  \]
  such that $\mu_1$ witnesses the independent combination of $m_\post$ and $m_\fr$: for all $F_1 \in \mcal F_\post = \mcal F_\pre$ and $F_2 \in \mcal F_\fr$, we have
  \begin{align*}
    \mu_1(F_1 \cap F_2) 
    &= \int_{F_1 \cap F_2} f \circ X\,\dd{\mu_0} \tag{definition of $\mu_1$} \\
    &= \int_\Omega \ch{F_1} \cdot (f \circ X) \cdot \ch{F_2}\,\dd{\mu_0} \tag{$\ch{F_1 \cap F_2} = \ch{F_1} \cdot \ch{F_2}$} \\
    &= \int_{\Omega} \ch{F_1} \cdot (f \circ X)\,\dd{\mu_\pre} \cdot \int_{\Omega} \ch{F_2} \,\dd{\mu_\fr} \tag{\cref{lem:append-pcm:int-indep-two}} \\
    &= \mu_\post(F_1) \cdot \mu_\fr(F_2). \tag{definition of $\mu_\post$}
  \end{align*}
  We now show that $(\gamma, (D, X, Y), m_\post) \vDash \cond{\bc{x}{X}{\pi}} \left(\score{f(x)}\right)$: write $\pi_\gamma \deq \pi(\gamma)$, $\pi_\gamma$ is a $\sigma$-finite measure by assumption, and $X_*\mu_\post$ is absolutely continuous with respect to $\pi_\gamma$ because for every $E \in \Sigma_A$ such that $\pi_\gamma(E) = 0$, we know
  \begin{align*}
    X_*\mu_\post(E)
    &= \mu_\post(X^{-1}(E)) \\
    &= \int_{X^{-1}(E)} f \circ X \,\dd{\mu_\pre} \tag{definition of $m_\post$} \\
    &= \int_{E} f\,\dd{X_*\mu_{\pre}} \tag{change of variable} \\
    &= \int_{E} f\,\dd{\pi_\gamma} \tag{$(\gamma, D, m_\pre) \vDash X \sim \pi$} \\
    &= 0. \tag{$\int_E\,\dd{\pi_\gamma} = 0$}
  \end{align*}
  Consider an arbitrary Borel measure extension $\mu_\post^+ : \BW \to \W$ and a disintegration $\{\mu^+_{\post, x}\}_{x \in A}$ of $\mu^+_\post$ along $X$ and $\pi_\gamma$. Notice that, via the above derivation, that $X_*\mu_\post$ has a density $f$ with respect to $\pi_\gamma$. By \cref{lem:append-logic:dendis}, we know for $\pi$-almost-all $x \in A$, 
  \begin{align*}
    \mu^+_{\post, x}|_{\mcal F_\post}(\Omega) 
    &= \mu^+_{\post, x}(\Omega) \\
    &= \frac{\dd{X_*\mu_\post^+}}{\dd{\pi_\gamma}}(x) \\
    &= \frac{\dd{X_*\mu_\post}}{\dd{\pi_\gamma}}(x)  \\
    &= f(x).
  \end{align*}
  Hence, $((\gamma, x), (D, Y), m_{\post, x}) \vDash \score{f(x)}$ and $(\gamma, (D, Y), m_\post) \vDash \cond{\bc{x}{X}{\pi}} \score{f(x)}$. Finally, consider a Borel measure $\mu_1^+ : \BW \to \W$ defined by 
  \[
    \mu_1^+(U) \deq \int_U f \circ X \,\dd{\mu_0^+}.
  \]
  Notice that $\mu_1^+|_{\mcal F_1}(F) = \int_F f \circ X\,\dd{\mu_0^+} = \int_F f \circ X\,\dd{\mu_0} = \mu_1(F)$ and $f \circ X$ is the density of $\mu_1^+$ with respect to $\mu_0^+$. Hence, the following identity holds:
  \begin{align*}
    \int_\Omega &\int_{\mcal A} \ch{U}(D_\ext(\omega), D(\omega), y)\,\semo{\kp{score}(f(X))}(D(\omega), \dd{y})\,\mu_0^+(\dd{\omega}) \\
    &= \int_\Omega \ch{U}(D_\ext(\omega), D(\omega), \star) f(X(\omega))\,\mu_0^+(\dd{\omega}) \tag{semantics of $\kp{score}$} \\
    &= \int_\Omega \ch{U}(D_\ext(\omega), D(\omega), \star) \frac{\dd{\mu^+_1}(\omega)}{\dd{\mu^+_0}}\,\mu_0^+(\dd{\omega}) \tag{density} \\
    &= \int_\Omega \ch{U}(D_\ext(\omega), D(\omega), \star)\,\mu_1^+(\dd{\omega}) \\
    &= \int_\Omega \ch{U}(D_\ext(\omega), D(\omega), Y(\omega))\,\mu_1^+(\dd{\omega}).
  \end{align*}
  Notice that $\msf{own}\ X$ also holds under $(\gamma, (D, X, Y), m_\post)$. By \cref{lem:append-logic:repr}, we know $\cond{\bc{x}{X}{\pi}} \score{f(x)} \vdash X \sim f \cdot \pi$. Now, we apply the consequence rule $\tsc{H{-}Cons}$ (\cref{lem:append-hoare:cons}), which yields the desired triple.
\end{proof}

\begin{lemma}[randomisation; \text{\cite[Proposition 24]{heunen17}}]\label[lemma]{lem:append-hoare:rand}
  Let $(A, \Sigma_A)$ be a measurable space and $\kappa : A \times \BR \to \I$ a probability kernel. Then there is a measurable function $f : A \times \R \to \R$ such that
  \[
    \kappa(x, U) = \lambda_{\I}\{r \in \R \alt f(x, r) \in U\}.
  \]
\end{lemma}

\begin{lemma}\label[lemma]{lem:append-hoare:disindep}
  Let $(A, \Sigma_A)$ and $(B, \Sigma_B)$ be standard Borel spaces, $(\mcal F, \mu) \in \mcal M$ such that $\mu$ is a probability measure, $X : \Omega \to A$ an $(\mcal F, \Sigma_A)$-measurable function and $Y : \Omega \to B$ an $(\mcal F, \Sigma_B)$-measurable function. For any Borel extension $\mu^+ : \BO \to \I$ and its $(X, X_*\mu^+)$-disintegration $\{\mu^+_x\}_{x \in A}$, $X$ and $Y$ are $\mu^+_x$-independent for $X_*\mu^+$-almost-all $x \in A$. Moreover, if $X$ and $Y$ are $\mu$-independent, then $Y_*\mu^+_x = Y_*\mu$ holds for $X_*\mu^+$-almost-all $x \in A$.
\end{lemma}
\begin{proof}
  Define $\pi \deq X_*\mu$. We need to show that for $\pi$-almost-all $x \in A$, $U \in \Sigma_A$ and $V \in \Sigma_B$, the following identity holds:
  \[
    \mu^+_x\{X \in U, Y \in V\} = \mu^+_x\{X \in U\} \cdot \mu^+_x\{Y \in V\}.
  \]
  Suppose $x \notin U$, then both L.H.S. and R.H.S. are zero because $\{X \in U\} = \{\omega \alt X(\omega) \in U\} \subseteq \{\omega \alt X(\omega) \neq x\} = \{X \neq x\}$, which means $\mu^+_x\{X \in U\} = 0$ by the concentration axiom. Suppose $x \in U$, then $\{X \in U\} = \{X = x\} \uplus \{X \in U \setminus \{x\}\}$. By additivity, we know
  \[
    \mu^+_x\{X \in U\} = \mu^+_x\{X = x\} + \mu^+_x\{X \in U \setminus \{x\}\} = 1 + 0 = 1.
  \]
  Next, compute that
  \begin{align*}
    \mu^+_x\{X \in U, Y \in V\}
    &= \mu^+_x(X^{-1}(U) \cap Y^{-1}(V)) \tag{by definition} \\
    &= \mu^+_x((X^{-1}(\{x\}) \uplus X^{-1}(U \setminus \{x\})) \cap Y^{-1}(V)) \\
    &= \mu^+_x((X^{-1}(\{x\}) \cap Y^{-1}(V)) \uplus (X^{-1}(U \setminus \{x\}) \cap Y^{-1}(V))) \\
    &= \mu^+_x\{X = x, Y \in V\} + \mu^+_x\{X \in U \setminus \{x\}, Y \in V\} \tag{additivity} \\
    &= \mu^+_x\{X = x, Y \in V\} \tag{$\mu^+_x\{X \neq x\} = 0$} \\
    &= \mu^+_x\{X = x, Y \in V\} + \mu^+_x\{X \neq x, Y \in V\} \tag{$\mu^+_x\{X \neq x\} = 0$} \\
    &= \mu^+_x\{Y \in V\} \tag{additivity} \\
    &= \mu^+_x\{X \in U\} \cdot \mu^+_x\{Y \in V\}. \tag{$\mu^+_x\{X \in U\} = 1$}
  \end{align*}
  Next, assuming $X$ and $Y$ are $\mu$-independent, by the above derivation they must be $\mu^+_x$-independent as well. Suppose $U \in \Sigma_A$, then $\mu^+_x\{X \in U\} = \ch{U}(x)$ and $\mu^+_{-}\{X \in U\} = \ch{U}$. This implies
  \begin{align*}
    \int_U Y_*\mu^+_x(V)\,\pi(\dd{x}) 
    &= \int_A \ch{U}(x) \cdot \mu^+_x\{Y \in V\}\,\pi(\dd{x}) \tag{by definition} \\
    &= \int_A \mu^+_x\{X \in U\} \cdot \mu^+_x\{Y \in V\}\,\pi(\dd{x}) \tag{$\mu^+_x\{X \in U\} = 1$} \\
    &= \int_A \mu^+_x\{X \in U, Y \in V\}\,\pi(\dd{x}) \tag{see above} \\
    &= \mu^+\{X \in U, Y \in V\} \tag{disintegration} \\
    &= \mu\{X \in U, Y \in V\} \tag{$\mu^+$ extends $\mu$} \\
    &= \mu\{X \in U\} \cdot \mu\{Y \in V\} \tag{$\mu$-independence} \\
    &= \int_U Y_*\mu(V)\,\pi(\dd{x}). \tag{$\pi = X_*\mu$}
  \end{align*}
  Hence, $Y_*\mu^+_x$ and $Y_*\mu$ are $\pi$-almost-surely equal.
\end{proof}

\begin{lemma}\label[lemma]{lem:append-hoare:disintsub}
  Let $(A, \Sigma_A), (B, \Sigma_B)$ be measurable spaces, $(\mcal F, \mu) \in \mcal M$, $X : \Omega \to A$ an $(\mcal F, \Sigma_A)$-measurable function and $Y : \Omega \to B$ an $(\mcal F, \Sigma_B)$-measurable function. Then for any Borel extension $\mu^+$ and $(X, X_*\mu^+)$-disintegration $\{\mu^+_x\}_{x \in A}$, the following identity holds for all $(\Sigma_A \otimes \Sigma_B, \BW)$-measurable $f : A \times B \to \W$ and almost all $x \in A$:
  \begin{align*}
    \int_\Omega f(X(\omega), Y(\omega))\,\mu^+_x(\dd{\omega}) = \int_B f(x, y)\,Y_*\mu^+_x(\dd{y}).
  \end{align*}
\end{lemma}
\begin{proof}
  For $X_*\mu$-almost-all $x \in A$, the following identity holds:
  \begin{align*}
    \int_\Omega f(X(\omega), Y(\omega))\,\mu^+_x(\dd{\omega})
    &= \int_{A \times B} f\,\dd{(X, Y)_*\mu^+_x} \tag{change of variable} \\
    &= \int_{A \times B} f\,\dd{(X_*\mu_x^+ \otimes Y_*\mu^+_x)} \tag{\cref{lem:append-hoare:disindep}} \\
    &= \int_{A \times B} f\,\dd{(\delta_x \otimes Y_*\mu^+_x)} \tag{disintegration} \\
    &= \int_{A \times B} f\,\dd{(\delta_x \otimes p)} \tag{\cref{lem:append-hoare:disindep}} \\
    &= \int_{B} f(x, y)\,p(\dd{y}). \tag{$\int_A f(-, y)\,\dd{\delta_x} = f(x, y)$}
  \end{align*}
  This completes the proof.
\end{proof}

\begin{lemma}
  Let $(A, \Sigma_A)$ be a standard Borel space, $p : (A, \Sigma_A) \to \mcal G(\R)$ be a probability kernel. Then the following triple is sound:
  \[
    \inferrule[H-CondSample]
      {}
      {\{X \sim \pi\}\ \kp{sample}(p(X))\ \{Y : \R.\ \cond{\bc{x}{X}{\pi}} Y \sim p(x)\}}
  \]
\end{lemma}
\begin{proof}
  Let $m_\pre \deq (\mcal F_{\pre}, \mu_\pre) \in \mcal M$ such that $(\gamma, D, m_\pre) \vDash X \sim \pi$, $m_\fr \in \mcal M$ such that $(\mcal F_0, \mu_0) \deq m_\pre \bullet m_\fr$ is defined, $\mu^+_0 : \BO \to \W$ a measure satisfying $\mu^+_0|_{\mcal F_0} = \mu_0$, and all probabilistic contexts $\Delta_\ext$ and all $D_\ext \in \msf{RV}\semo{\Delta_\ext}$. Notice the assumption that the program integrates to non-zero is always true:
  \[
    \int_\Omega \semo{\kp{sample}(p(X))}((D, X)(\omega), \R)\,\mu^+_0(\dd{\omega}) = \int_\Omega 1\,\mu^+_0(\dd{\omega}) = \mu^+_0(\Omega)
  \]
  Denote $n$ the maximum dimension of $m_\pre \bullet m_\fr$, $D$ and $D_\ext$'s finite footprint. Define a measurable function $Y : \Omega \to \R$ by
  \[
    Y(\omega) \deq f(X(\omega), \pi_{n + 1}(\omega))
  \]
  where $f : A \times [0, 1] \to \R$ is the $(\Sigma_A \otimes \BI, \BR)$-measurable function obtained via \cref{lem:append-hoare:rand} that generates the distribution via a number uniformly distributed in $[0, 1]$. Next, we define $\mcal F_{\post} \deq \mcal F_\pre|_n \otimes \BI \otimes \{\varnothing, \Omega\}$ and $m_\post \deq (\mcal F_\post, \mu_\post)$. Notice that
  \begin{align*}
    \mcal F_\post 
    &= \mcal \langle\mcal F_\pre|_n \times \{\varnothing, \Omega\}, \{\varnothing, [0, 1]^n\} \times \BI \times \{\varnothing, \Omega\}\rangle \\
    &= \sigma\{F' \times U \times \Omega \alt F' \in \mcal F_\pre|_n, U \in \BI\}.
  \end{align*}
  Define $\mu_\post$ to be the product measure $\mu_\post \deq \mu_\pre|_n \otimes \lambda_{\I} \otimes p$ where $p$ is the trivial probability measure on $\{\varnothing, \Omega\}$. By definition of a product measure, we know when $F \in \mcal F_\pre$ and $U \in \BI$, $\mu_\post(F|_n \times U \times \Omega) = \mu_\pre(F) \cdot \lambda_{\I}(U)$.

  We now show that $(\gamma, (D, X, Y), m_\post) \vDash \cond{\bc{x}{X}{\pi}} Y \sim p(x)$. Let $\mu^+_\post : \BO \to \W$ be a Borel extension of $\mu_\post$ and consider any $(X, \pi)$-disintegration $\{\mu^+_{\post, x}\}_{x \in A}$, we need to show that the following holds for $\pi$-almost-all $x \in A$:
  \[
    ((\gamma, x), (D, X, Y), (\mcal F_\post, \mu^+_{\post, x}|_{\mcal F_\post})) \vDash Y \sim p(x).
  \]
    Also, $X_*\mu^+_{\post} \ll \pi$ since $X$ is $(\mcal F_\pre, \Sigma_A)$-measurable and $(\gamma, (D, X, Y), m_\pre) \vDash X \sim \pi$, which implies for any $\pi$-null-set $E \in \Sigma_A$, we have $X_*\mu^+_\post(E) = X_*\mu_\pre(E) = \pi(E) = 0$. Next, notice that $X : \Omega \to A$ and $\pi_{n + 1} : \Omega \to \I$ are $\mu_\post$-independent: for any $E \in \Sigma_A$ and $F \in \BI$, we know
  \begin{align*}
    \mu_\post\{X \in E, \pi_{n + 1} \in F\}
    &= \mu_\post(X^{-1}(E) \cap \pi_{n + 1}^{-1}(F)) \\
    &= \mu_\post((E'|_n \times \Omega) \cap ([0, 1]^n \times F' \times \Omega)) \tag{measurability} \\
    &= \mu_\post(E'|_n \times F' \times \Omega) \\
    &= \mu_\pre(E') \cdot \lambda_{\I}(F') \tag{see above} \\
    &= \mu_\post(E') \cdot \mu_\post([0, 1]^n \times F' \times \Omega) \tag{$\mu_\post|_{\mcal F_\pre} = \mu_\pre$} \\
    &= \mu_\post\{X \in E\} \cdot \mu_\post\{\pi_{n + 1} \in F\}.
  \end{align*}
  Hence, $p(x) = Y_*\mu^+_{\post, x}|_{\mcal F_\post}$ for almost all $x \in A$: let $E \in \BR$ and $F \in \Sigma_A$, we can then calculate
  \begin{align*}
    \int_F Y_*\mu^+_{\post, x}(E)\,\pi(\dd{x})
    &= \int_F \int_\Omega \ch{E}(f(X(\omega), \omega_{n + 1})) \,\mu^+_{\post, x}(\dd{\omega}) \,\pi(\dd{x}) \\
    &= \int_F \int_\Omega \ch{E}(f(x, r))\,\pi_{{n + 1}*}\mu^+_{\post, x}(\dd{r})\,\pi(\dd{x})
    \tag{\cref{lem:append-hoare:disintsub}} \\
    &= \int_F \int_\Omega \ch{E}(f(x, r))\,\lambda_{\I}(\dd{r}) \,\pi(\dd{x}) \tag{\cref{lem:append-hoare:disintsub}} \\
    &= \int_F \lambda_{\I}\{r \in \I \alt f(x, r) \in E\} \,\pi(\dd{x}) \\
    &= \int_F p(x, E)\,\pi(\dd{x}). \tag{\cref{lem:append-hoare:rand}}
  \end{align*}
  Hence, $(\gamma, x), (D, X, Y), (\mcal F_\post, \mu^+_{\post, x}|_{\mcal F_\post}) \vDash Y \sim p(x)$ holds for $\pi$-almost-all $x \in A$. Notice that $\mu_1$ is always defined since $n$ witnesses the finite footprint of the frame $(\mcal F_0, \mu_0)$. Choose an extension $\mu_1^+ : \BO \to \W$ that satisfies $\mu^+_1|_{\mcal B[0, 1]^{n + 1} \otimes \{\varnothing, \Omega\}} = \mu^+_0|_n \otimes \lambda_{\I} \otimes p$. Then the following identity holds:
  \begin{align*}
    \int_\Omega &\int_{\R} \ch{U}(D_\ext(\omega), D(\omega), y) \semo{p(X)}((D, X)(\omega), \dd{y}) \,\mu^+_0(\dd{\omega}) \\
    &= \int_\Omega \int_\R \ch{U}(D_\ext(\omega), D(\omega), y)\,p(X(\omega), \dd{y})\,\mu^+_0(\dd{\omega}) \\
    &= \int_\Omega p(X(\omega), E_{\omega})\,\mu^+_0(\dd{\omega}) \\
    &= \int_\Omega \lambda_{\I}\{r \in \R \alt f(X(\omega), r) \in E_\omega\}\,\mu^+_0(\dd{\omega}) \tag{\cref{lem:append-hoare:rand}} \\
    &= \int_{[0, 1]^n} \lambda_{\I}\{r \in \R \alt f(X|_n(\omega_{1..n}), r) \in E_{\omega_{1..n}}\}\,\mu^+_0|_n(\dd{\omega_{1..n}}) \tag{\cref{lem:append-hoare:foot}} \\
    &= \int_{[0, 1]^{n + 1}} \ch{E_{\omega_{1..n}}}(f(X(\omega_{1..n}), r))\,(\mu^+_0|_n \otimes \lambda_{\I})(\dd{\omega_{1..n}}, \dd{r}) \tag{Fubini} \\
    &= \int_\Omega \ch{E_\omega}(Y(\omega))\,\mu^+_1(\dd{\omega}) \\
    &= \int_\Omega \ch{U}(D_\ext(\omega), (D, X)(\omega), Y(\omega))\,\mu^+_1(\dd{\omega}),
  \end{align*}
  where $E_{\omega} \deq \{y \in \R \alt (D_\ext(\omega), D(\omega), y) \in U\}$.
\end{proof}

\begin{lemma}\label[lemma]{lem:append-hoare:indmeaext}
  Let $(\mcal F, \mu) \in \mcal M$, $f : \Omega \to \Rnn$ an $(\mcal F, \BRnn)$-measurable function and $(f \cdot \mu)^+ : \BO \to \W$ a Borel measure that extends a $\sigma$-finite measure $f \cdot \mu : \mcal F \to \W$. Then there is a Borel measure $\nu$ such that $\nu$ extends $\mu$ and
  \[
    (f \cdot \mu)^+ = f \cdot \nu.
  \]
  Further, let $X : \Omega \to A$ be a $(\Sigma_\Omega, \Sigma_A)$-measurable function and $\{\nu_x\}_{x \in A}$ an $(X, \pi)$-disintegration of $\nu$. Then $\{f \cdot \nu_x\}_{x \in A}$ is an $(X, \pi)$-disintegration of $(f \cdot \mu)^+$.
\end{lemma}
\begin{proof}
  First, we define a Borel measure $\pi : \BO \to \W$ by
  \[
    \pi(U) \deq \int_U f'\,\dd{(f \cdot \mu)^+}\qquad f'(\omega) \deq 
    \ch{(0, \infty)}(f(\omega)) \cdot \frac{1}{f(\omega)}.
  \]
  By \cref{lem:append-pcm:inverse-radon-nikodym}, the measure $\pi$ is $\sigma$-finite.
  Notice that for any $F \in \mcal F$, $\pi$ satisfies
  \begin{align*}
    \pi(F) 
    &= \int_F f'\,\dd{(f \cdot \mu)^+} \\
    &= \int_F f'\,\dd{(f \cdot \mu)} \\
    &= \int_F f' \cdot f\,\dd{\mu} \\
    &= \int_F \ch{f^{-1}(0, \infty)}\,\dd{\mu} \\
    &= \mu(F \cap f^{-1}(0, \infty)).
  \end{align*}
  By \cref{prop:append-pcm:borel-ext}, $\mu$ has a Borel measure extension $\mu^+ : \Sigma_\Omega \to \W$. We can therefore define a $\sigma$-finite Borel measure $\pi' : \Sigma_\Omega \to \W$ by $\pi'(U) \deq \mu^+(U \cap f^{-1}\{0\})$. Now, the sum of two $\sigma$-finite measures remains a $\sigma$-finite measure and we define $\nu : \Sigma_\Omega \to \W$ by $\nu \deq \pi + \pi'$. Notice that for any $F \in \mcal F$, we have
  \begin{align*}
    \nu(F) 
    &= \pi(F) + \pi'(F) \\
    &= \mu(F \cap f^{-1}(0, \infty)) + \mu(F \cap f^{-1}\{0\}) \\
    &= \pi(F).
  \end{align*}
  Hence, $\nu$ is a Borel extension of $\mu$. Next, notice that for any $U \in \Sigma_\Omega$, we have
  \begin{align*}
    (f \cdot \nu)(U) 
    &= \int_U f\,\dd{(\pi + \pi')} \\
    &= \int_U f\,\dd{\pi} + \int_U f\,\dd{\pi'} \\
    &= \int_{U}f\,\dd{\pi} + \int_{U} f\,{\mu^+(\dd{\omega}\cap f^{-1}\{0\})} \\
    &= \int_U f\,\dd{\pi} \\
    &= (f \cdot \pi)(U).
  \end{align*}
  Also, since $f^{-1}\{0\}$ is a $(f \cdot \mu)^+$-null-set, the following identity holds:
  \begin{align*}
    (f \cdot \mu)^+(U) 
    &= (f \cdot \mu)^+(U \cap f^{-1}(0, \infty)) 
    + (f \cdot \mu)^+(U \cap f^{-1}\{0\}) \\
    &= (f \cdot \mu)^+(U \cap f^{-1}(0, \infty)).
  \end{align*}
  Hence, to show $(f \cdot \mu)^+ = f\cdot \nu$, it suffices to prove that $(f \cdot \mu)^+(U \cap f^{-1}(0, \infty)) = (f \cdot \pi)(U)$. Notice that
  \begin{align*}
    (f \cdot \mu)^+(U \cap f^{-1}(0, \infty))
    &= \int_{U \cap f^{-1}(0, \infty)} \dd{(f \cdot \mu)^+} \\
    &= \int_U f \cdot f'\,\dd{(f \cdot \mu)^+} \\
    &= \int_U f\,\dd{(f' \cdot (f \cdot \mu)^+)} \\
    &= (f \cdot \pi)(U).
  \end{align*}

  Next, we show that any $(X, \pi)$-disintegration of $\nu$ induces an $(X, \pi)$-disintegration of $(f \cdot \mu)^+$. Let $\{\nu_x : \Sigma_\Omega \to \W\}_{x \in A}$ be an $(X, \pi)$-disintegration of $\nu$. Then $\{f \cdot \nu_x\}_{x \in A}$ is an $(X, \pi)$-disintegration of $f \cdot \nu$. Notice that the concentration property of disintegration holds:
  \begin{align*}
    (f \cdot \nu_x)\{X \neq x\} &= \int_{\{X \neq x\}} f\,\,\dd{\nu_x} \\
    &\le \sup f \cdot \nu_x\{X \neq x\} \\
    &= 0.
  \end{align*}
  Also, for every $x \in A$ and $(\Sigma_\Omega, \BW)$-measurable $f : \Omega \to \W$, the map $x \longmapsto \int_{\Omega} f'\,\dd{(f \cdot \nu_x)}$ is measurable -- since $x \longmapsto \int_{\Omega} f''\,\dd{\nu_x}$ is measurable for any $(\Sigma_\Omega, \BW)$-measurable $f'' : \Omega \to \W$, instantiating $f'' = f' \cdot f$ makes $f'$ measurable. Finally, notice that the `nesting' property of disintegration holds:
  \begin{align*}
    \int_{\Omega} f'\,\dd{(f \cdot \nu)}
    &= \int_\Omega f' \cdot f\,\dd{\nu} \\
    &= \int_A \int_\Omega f' \cdot f\,\dd{\nu_x}\,\mu(\dd{x}) \\
    &= \int_A \int_\Omega f'\,\dd{(f \cdot \nu_x)}\,\mu(\dd{x})
  \end{align*}
  for all $(\Sigma_\Omega, \BW)$-measurable $f' : \Omega \to \W$, which concludes the proof.
\end{proof}

\begin{lemma}
  Let $(A, \Sigma_A)$ be a standard Borel space, $p : (A, \Sigma_A) \to \mcal G(\R)$ be a probability kernel. Then the following triple is sound:
  \[
    \inferrule[H-CondScore]{}
      {\{\cond{\bc{x}{X}{\pi}} Y \sim p(x)\}\ \kp{score}(f(Y))\ \{\cond{\bc{x}{X}{\pi}} Y \sim f \cdot p(x)\}}
  \]
\end{lemma}
\begin{proof}
  Let $m_\pre \deq (\mcal F_{\pre}, \mu_\pre) \in \mcal M$ such that $(\gamma, D, m_\pre) \vDash \cond{\bc{x}{X}{\pi}} Y \sim p(x)$, $m_\fr \in \mcal M$ such that $(\mcal F_0, \mu_0) \deq m_\pre \bullet m_\fr$ is defined, $\mu^+_0 : \BO \to \W$ a measure satisfying $\mu^+_0|_{\mcal F_0} = \mu_0$, and all probabilistic contexts $\Delta_\ext$ and all $D_\ext \in \msf{RV}\semo{\Delta_\ext}$. We assume that the following is true:
  \[
    \int_\Omega \semo{\kp{score}(f(Y))}(D, X, Y, \mb 1)\,\dd{\mu^+_0} = \int_\Omega f(Y)\,\dd{\mu^+_0} > 0.
  \]
  Define $Z : \Omega \to \mb 1$ by $Z(\omega) \deq \star$, $\mcal F_\post \deq \mcal F_\pre$ and $\mu_\post : \mcal F_\post \to \W$ by $\mu_\post \deq (f \circ Y) \cdot \mu_\pre$. We now show that $(\gamma, (D, X, Y, Z), m_\post) \vDash \cond{\bc{x}{X}{\pi}} Y \sim (f \cdot p)(x)$. Let $\mu^+_\post : \BO \to \W$ be a Borel extension of $\mu_\post$ and consider any $(X, \pi)$-disintegration $\{\mu^+_{\post, x}\}_{x \in A}$, we need to show that the following holds for $\pi$-almost-all $x \in A$:
  \[
    ((\gamma, x), (D, X, Y, Z), (\mcal F_\pre, \mu^+_{\post, x}|_{\mcal F_\pre})) \vDash Y \sim f \cdot p(x).
  \]
  The function $Y$ is $(\mcal F_\pre, \BR)$-measurable by assumption. Also, for any $U \in \BR$, the following identity holds:
  \begin{align*}
    Y_*\mu^+_{\post, x}|_{\mcal F_\pre}(U)
    &= \mu^+_{\post, x}(Y^{-1}(U)) \\
    &= ((f \circ Y) \cdot \mu_{\pre, x})^+(Y^{-1}(U)) \tag{by definition} \\
    &= \int_{Y^{-1}(U)} f(Y(\omega))\,\mu^+_{\pre, x}(\dd{\omega}) \tag{see below} \\
    &= \int_U f\,\dd{Y_*\mu^+_{\pre, x}} \tag{change of variable} \\
    &= \int_U f(y)\,p(x, \dd{y}) \tag{\cref{lem:append-hoare:disintsub}} \\
    &= (f \cdot p(x))(U). \tag{by definition}
  \end{align*}
  where $\mu^+_{\pre, x}$ is the disintegration of a Borel measure $\mu^+_\pre$ satisfying $((f \circ Y) \cdot \mu_\pre)^+ = (f \circ Y) \cdot \mu^+_\pre$ (which exists by \cref{lem:append-hoare:indmeaext}). Notice that$(\mcal F_1, \mu_1) \deq m_\post \bullet m_\fr$ is well-defined: $\mcal F_1 = \mcal F_0$ because $\langle\mcal F_\post, \mcal F_\fr\rangle = \langle\mcal F_\pre, \mcal F_\fr\rangle = \mcal F_0$, and $\mu_1 = (f \circ Y) \cdot \mu_0$. For any $F \in \mcal F_\post$ and $G \in \mcal F_\fr$, we have
  \begin{align*}
    \mu_1(F \cap G)
    &= \int_{F \cap G} f \circ Y\,\dd{\mu_0} \\
    &= \int_F f \circ Y \dd{\mu_\pre} \cdot \mu_\fr(G) \tag{\cref{lem:append-pcm:int-indep-two}} \\
    &= \mu_\post(F) \cdot \mu_\fr(G).
  \end{align*}
  Define $\mu^+_1 \deq (f \circ Y) \cdot \mu_0$. Then for any $U \in \Sigma_{\semo{\Delta_\ext}} \otimes \Sigma_{\semo{\Delta, X, Y}}$, we have
  \[
    \int_\Omega \ch{U}(D_\ext(\omega), (D, X, Y)(\omega), {*})\,f(Y(\omega))\,\mu^+_0(\dd{\omega}) = \int_\Omega \ch{U}(D_\ext(\omega), D(\omega), *)\,\mu^+_1(\dd{\omega}),
  \]
  which yields the postcondition $\cond{\bc{x}{X}{\pi}} Y \sim f \cdot p(x)$.
\end{proof}

\begin{lemma}
  The following Hoare triple is sound:
  \[
    \inferrule[H-Ret]
      {}
      {\{P[\semo{M}/X]\}\ \kpr{return} M\ \{X. P\}}
  \]
\end{lemma}
\begin{proof}
  Let $m_\pre \deq (\mcal F_\pre, \mu_\pre) \in \mcal M$, $(\gamma, D, m_\pre) \vDash P[\semo{M}/X]$, $m_\fr \deq (\mcal F_\fr, \mu_\fr)$ such that $m_0 \deq m_\pre \bullet m_\fr$ is defined, and all $m_\fr$ with $(\mcal F_0, \mu_0) \deq m_\pre \bullet m_\fr$ defined, and all measures $\mu_0^+$ satisfying $\mu_0^+|_{\mcal F_0} = \mu_0$, and all probabilistic contexts $\Delta_\ext$ and all $D_{\ext} \in \msf{RV}\semo{\Delta_\ext}$. By the substitution lemma (\cref{lem:append-logic:sub}), we know
  \begin{align*}
    (\gamma, D, m_\pre) \vDash P[\semo{M}/X] \iff 
    (\gamma, (D, \semo{M}(\gamma)), m_\pre) \vDash P.
  \end{align*}
  Define $X : \Omega \to A$ by $X(\omega) \deq \semo{M(\gamma)}(D(\omega))$, $m_\post \deq m_\pre$, and $\mu^+_1 \deq \mu^+_0$. Then $(\gamma, (D, X), m_\pre) \vDash P$ and for all $U \in \semo{\Delta_\ext} \otimes \semo{\Delta} \otimes \semo{A}$, the following identity holds:
  \begin{align*}
    \int_\Omega &\int_A \ch{U}(D_\ext(\omega), D(\omega), x)\,\semo{\kpr{return}M(\gamma)}(D(\omega), \dd{x})\,\mu^+_0(\dd{\omega}) \\
    &= \int_\Omega \ch{U}(D_\ext(\omega), D(\omega), \semo{M(\gamma)}(D(\omega)))\,\mu^+_0(\dd{\omega}) \tag{semantics of $\kp{return}$} \\
    &= \int_\Omega \ch{U}(D_\ext(\omega), D(\omega), X(\omega))\,\mu^+_1(\dd{\omega}). \tag{by definition}
  \end{align*}
  which yields the postcondition $P$.
\end{proof}

\begin{lemma}\label[lemma]{lem:append-hoare:posmeasposint}
  Let $\mu : \Sigma_A \to \W$ be a measure and $f : A \to \W$ be a $(\Sigma_A, \mcal B\W)$-measurable function. Then $\int_A f\,\dd{\mu} > 0$ if and only if there exists $U \in \Sigma_A$ such that $\mu(U) > 0$ and $f(x) > 0$ for all $x \in U$.
\end{lemma}
\begin{proof}
  Since the codomain of $f$ is $\W$, $f(x) = 0$ holds for almost all $x \in A$ iff $\int_A f\,\dd{\mu} = 0$. This means $\mu\!\set{x \in A \alt f(x) \neq 0} = 0$ iff $\int_A f\,\dd{\mu} = 0$. Negating L.H.S. yields $\mu\!\set{x \in A \alt f(x) \neq 0} \neq 0$, which is equivalent to $\mu\!\set{x \in A \alt f(x) > 0} > 0$. By contrapositivity, we know 
  \[
  \int_A f\,\dd{\mu} > 0\quad\textnormal{ iff }\quad
  \mu\!\set{x \in A \alt f(x) > 0} > 0.
  \]  
  This proves the direction from left to right by defining $U \deq \set{x \in A \alt f(x) > 0}$. From right to left, assuming such a set $U$ exists, then $0 < \mu(U) = \mu\!\set{x \in U \alt f(x) > 0} \le \mu\!\set{x \in A \alt f(x) > 0}\!$. By the above derivation, we have $\int_A f\,\dd{\mu} > 0$.
\end{proof}

\begin{corollary}\label[corollary]{cor:append-hoare:posmeasposkern}
  Let $\mu : \Sigma_\Omega \to \W$ be a measure, $\kappa : \Omega \times \Sigma_A \to \W$ a measure kernel, and $\varphi : \Omega \times A \to \W$ a measurable function. Then
  \[
    \int_\Omega\int_A \varphi(\omega, x)\,\kappa(\omega, \dd{x})\,\mu(\dd{\omega}) > 0 \implies \int_\Omega \kappa(\omega, A)\,\mu(\dd{\omega}) > 0.
  \]
\end{corollary}
\begin{proof}
  By \cref{lem:append-hoare:posmeasposint}, we know there exists $B \in \Sigma_\Omega$ such that $\mu(B) > 0$ and $\int_A \varphi(\omega, x)\,\kappa(\omega, \dd{x}) > 0$ for all $\omega \in B$. Also, we know there exists, for every $\omega \in B$, a set $U_\omega \in \Sigma_A$ such that $\kappa(\omega, U_\omega) > 0$ and $\varphi(\omega, x) > 0$ for all $x \in U_\omega$. To show that $\int_\Omega \kappa(\omega, A)\,\mu(\dd{\omega}) > 0$, notice that $\kappa(\omega, A) \ge \kappa(\omega, U_\omega) > 0$ for all $\omega \in B$. By \cref{lem:append-hoare:posmeasposint}, the desired inequality holds.
\end{proof}

\begin{lemma}\label[lemma]{lem:append-hoare:continuation}
  Let $\mu, \nu : \Sigma_\Omega \to \W$ be measures and $\kappa : \Omega \times \Sigma_A \to \W$ a measure kernel. If for every $U \in \Sigma_\Omega \otimes \Sigma_A$ and $(\Sigma_\Omega, \Sigma_A)$-measurable $X : \Omega \to A$,
  \[
    \int_\Omega \int_A \ch{U}(\omega, x)\,\kappa(\omega, \dd{x})\,\mu(\dd{\omega}) 
    = \int_\Omega \ch{U}(\omega, X(\omega))\,\nu(\dd{\omega}),
  \]
  then for any $(\Sigma_\Omega \otimes \Sigma_A, \mcal B\W)$-measurable $\varphi : \Omega \times A \to \W$, the following identity holds:
  \[
    \int_\Omega \int_A \varphi(\omega, x)\,\kappa(\omega, \dd{x})\,\mu(\dd{\omega})
    = \int_\Omega \varphi(\omega, X(\omega))\,\nu(\dd{\omega}).
  \]
\end{lemma}
\begin{proof}
  For every Borel set $V \in \mcal B\W$, the following identity holds by assumption:
  \begin{align*}
    \int_\Omega &\int_A \ch V(\varphi(\omega, x))\,\kappa(\omega, \dd{x})\,\mu(\dd{\omega}) \\
    &= \int_\Omega \int_A \ch{\varphi^{-1}(V)}(\omega, x)\,\kappa(\omega, \dd{x})\,\mu(\dd{\omega}) \tag{inverse image} \\
    &= \int_\Omega \ch{\varphi^{-1}(V)}(\omega, X(\omega))\,\nu(\dd{\omega}) \tag{assumption} \\
    &= \int_\Omega \ch V(\varphi(\omega, X(\omega)))\,\nu(\dd{\omega}). \tag{inverse image}
  \end{align*}
  Assume that $\varphi = \ch{W}$ for some measurable $W \in \Sigma_\Omega \otimes \Sigma_A$, then by instantiating $V \deq \set{1}$, we know $\ch{V} \circ \ch{W} = \ch{W}$, which implies
  \[
    \int_\Omega \int_A \ch{W}(\omega, x)\,\kappa(\omega, \dd{x})\,\mu(\dd{\omega}) = \int_\Omega \ch{W}(\omega, X(\omega))\,\nu(\dd{\omega}).
  \]
  Hence, all characteristic functions satisfy the property. Next, assume that $\varphi = \sum_{i = 1}^n c_i\ch{W_i}$ with $c_i \in \Rnn$ and $\set{W_i \in \Sigma_\Omega \otimes \Sigma_A}_{i = 1}^n$ is a disjoint measurable partition. We can assume that $c_i \neq c_j$ when $i \neq j$ (if $i \neq j$ and $c_i = c_j$, $\varphi$ can alternatively be represented by $\sum_{i = 1}^{n - 1} c_i\ch{W'_i}$ for some partition $\{W'_k\}_{k = 1}^{n - 1}$ with $W'_i \deq W_i \cup W_j$ and `removing' $W_j$). Our assumption is now
  \begin{align*}
    \int_\Omega \int_A \ch{V}\biggl(
      \sum_{i = 1}^n c_i \ch{W_i}(\omega, x)
    \biggr)\,\kappa(\omega, \dd{x})\,\mu(\dd{\omega}) = \int_\Omega \ch{V}\biggl(
      \sum_{i = 1}^n c_i \ch{W_i}(\omega, X(\omega))
    \biggr)\,\nu(\dd{\omega}).
  \end{align*}
  for all $V \in \mcal B\W$. For each $i = 1, ..., n$, instantiating $V \deq \set{c_i}$ implies $\ch{V}(\sum_{i = 1}^n c_i \ch{W_i}(\omega, x)) = \ch{W_i}(\omega, x)$ for all $\omega \in \Omega$ and $x \in A$, which then implies
  \begin{align*}
    \int_\Omega \int_A \ch{W_i}(\omega, x)\,\kappa(\omega, \dd{x}) = \int_\Omega \ch{W_i}(\omega, X(\omega))\,\nu(\dd{\omega}).
  \end{align*}    
  Since the integrands are both non-negative, multiplying both sides by $c_i$ yields
  \begin{align*}
    \int_\Omega \int_A c_i\ch{W_i}(\omega, x)\,\kappa(\omega, \dd{x})\,\mu(\dd{\omega}) = \int_\Omega c_i\ch{W_i}(\omega, X(\omega))\,\nu(\dd{\omega}).  \end{align*}
  An integral is always defined, and hence, linear, when the codomain is $\W$ \cite[Theorem 3.16]{axler19}, which implies
  \begin{align*}
    \int_\Omega &\int_A 
      \sum_{i = 1}^{n} c_i \ch{W_i}(\omega, x)\,\kappa(\omega, \dd{x})\,\mu(\dd{\omega}) \\
    &= 
    \sum_{i = 1}^n \int_\Omega \int_A c_i \ch{W_i}(\omega, x)\,\kappa(\omega, \dd{x})\,\mu(\dd{\omega})
    \\
    &= \sum_{i = 1}^n \int_\Omega c_i \ch{W_i}(\omega, X(\omega))\,\nu(\dd{\omega}) \tag{see above} \\
    &= \int_\Omega \sum_{i = 1}^n c_i \ch{W_i}(\omega, X(\omega))\,\nu(\dd{\omega}).
  \end{align*}
  Hence, every non-negative measurable simple function $\varphi : \Omega \to \Rnn$ satisfies the desired property. Next, we assume $\varphi : \Omega \to \W$ is an arbitrary $(\Sigma_\Omega \otimes \Sigma_A, \mcal B\W)$-measurable function. By the simple function approximation theorem for $\W$ \cite[Theorem 2.89]{axler19}, there exists a sequence of measurable simple functions $\set{\varphi_i : \Omega \to \Rnn}_{i \in \N}$ such that $\varphi_i \le \varphi_{i + 1}$ for all $i \in \N$ and $\varphi_i$ converges pointwise to $\varphi$, i.e. $\lim_{i \to \infty}\varphi_i(\omega, x) = \varphi(\omega, x)$ for all $\omega \in \Omega$ and $x \in A$. Then
  \begin{align*}
    \int_{\Omega} &\int_A \lim_{i \to \infty} \varphi_i(\omega, x)\,\kappa(\omega, \dd{x})\,\mu(\dd{\omega}) \\
    &= \lim_{i \to \infty} \int_\Omega \int_A \varphi_i(\omega, x)\,\kappa(\omega, \dd{x})\,\mu(\dd{\omega}) \tag{monotone convergence} \\
    &= \lim_{i \to \infty} \int_\Omega \varphi_i(\omega, X(\omega))\,\nu(\dd{\omega}) \tag{see above} \\
    &= \int_{\Omega} \lim_{i \to \infty} \varphi_i(\omega, X(\omega))\,\nu(\dd{\omega}). \tag{monotone convergence}
  \end{align*}
  Hence, every $(\Sigma_\Omega \otimes \Sigma_A, \mcal B\W)$-measurable function satisfies the property.
\end{proof}

\begin{lemma}\label[lemma]{lem:append-hoare:letintegral}
  Let $\mu : \Sigma_A \to \W$ be a measure, $\kappa : A \times \Sigma_B \to \W$ a measure kernel. Consider a measure $\nu : \Sigma_B \to \W$ defined by $\nu(U) \deq \int_A \kappa(x, U)\,\mu(\dd{x})$, then for any $(\Sigma_B, \mcal B\W)$-measurable $\varphi : B \to \W$, 
  \[
    \int_B \varphi\,\dd{\nu} = \int_A \int_B \varphi(y)\,\kappa(x, \dd{y})\,\mu(\dd{x}).
  \]
\end{lemma}
\begin{proof}
  Suppose $\varphi : \Omega \to \Rnn$ is a simple function $\varphi = \sum_{i = 1}^n c_i\ch{U_i}$, notice that
  \begin{align*}
    \int_B \varphi\,\dd{\nu} 
    &= \sum_{i = 1}^n c_i \nu(U_i) \\
    &= \sum_{i = 1}^n c_i \int_A \int_{U_i} \kappa(x, \dd{y})\,\mu(\dd{x}) \\
    &= \int_A \int_{B} \varphi(y)\,\kappa(x, \dd{y})\,\mu(\dd{x}),
  \end{align*}
  where additivity and multiplicity follow from the same theorems as the ones in \cref{lem:append-hoare:continuation}. Now, consider a $(\Sigma_B, \mcal B\W)$-measurable function $\varphi : B \to \W$ approximated by the sequence of measurable simple functions $\set{\varphi_i : B \to \Rnn}_{i \in \N}$ via the simple function approximation theorem for $\W$ \cite[Theorem 2.89]{axler19}. By the monotone convergence theorem, we know
  \begin{align*}
    \int_B \varphi\,\dd{\nu}
    &= \int_B \lim_{i \to \infty} \varphi_i\,\dd{\nu} \\
    &= \lim_{i \to \infty} \int_B \varphi_i\,\dd{\nu} \\
    &= \lim_{i \to \infty} \int_A \int_B \varphi_i(y)\,\kappa(x, \dd{y})\,\mu(\dd{x}) \\
    &= \int_A \int_B \varphi(y)\,\kappa(x, \dd{y})\,\mu(\dd{x}).
  \end{align*}
  Hence, the desired property holds for all $(\Sigma_B, \mcal B\W)$-measurable functions.
\end{proof}

\begin{lemma}
  The following Hoare triple is sound:
  \[
    \inferrule[H-Let]
    {\vdash \set{P} M \set{X : \mcal A. Q} \\
      \vdash \forallrv X : \mcal A. \set{Q} N \set{Y : \mcal B. R}}
    {\vdash \set{P} \kr{let} X = M \kb{in} N \set{Y : \mcal B. R}} 
  \]
\end{lemma}
\begin{proof}
  Let $(\gamma, D, m)$ be a configuration such that $(\gamma, D, m) \vDash \set{P} M \set{X : \mcal A. Q}$ and $(\gamma, D, m) \vDash \forallrv X. \set{Q} N \set{Y : \mcal B. R}$. For all $m_{\pre} \in \mcal M$ with $(\gamma, D, m_{\pre}) \vDash P$,
  $m_{\fr} \in \mcal M$ with $(\mcal F_0, \mu_0) \deq m_{\pre} \bullet m_{\fr}$ defined,
  Borel measures $\mu^+_0 : \Sigma_\Omega \to [0, \infty]$ satisfying $\mu^+_{0}|_{\mcal F_0} = \mu_0$,
  and all probabilistic contexts $\Delta_{\ext}$ and all $D_{\ext} \in \msf{RV}\!\sem{\Delta_{\ext}}$.
  Assuming $$\int_{\Omega} \sem{\kr{let} X = M(\gamma) \kb{in} N(\gamma)}\!(D(\omega), \mcal B)\,\mu^+_0(\dd{\omega}) > 0,$$ then by definition of the semantics for $\kp{let}$, 
  \begin{align*}
    \int_\Omega \sem{\kr{let} X = M(\gamma) \kb{in} N(\gamma)}\!(D(\omega), \mcal B)\,\mu^+_0(\dd{\omega})
    = \int_\Omega \int_A \varphi(\omega, x)\sem{M(\gamma)}\!(D(\omega), \dd{x})\,\mu^+_0(\dd{\omega}) > 0,
  \end{align*}
  where $\varphi(\omega, x) \deq \sem{N(\gamma)}\!((D(\omega), x), \mcal B)$. By \cref{cor:append-hoare:posmeasposkern}, we know 
  \[
    \int_\Omega \sem{M(\gamma)}\!(D(\omega), \mcal A)\,\mu^+_0(\dd{\omega}) > 0. 
  \]
  Hence, there exists
  \begin{enumerate*}
    \item $X \in \msf{RV}(\mcal A)$, 
    \item $m_\post \in \mcal M$ such that $(\mcal F_1, \mu_1) \deq m_\post, m_\fr$ is defined, and
    \item Borel measure $\mu^+_1 : \Sigma_\Omega \to \W$ that extends $\mu_1$.
  \end{enumerate*}
  Moving to the premises of $\forallrv X : \mcal A. \set{Q} M \set{Y : \mcal B. R}$, we instantiate $X \in \msf{RV}(\mcal A)$ to the random variable obtained above, $m'_\pre \deq m_\post$, $m'_\fr \deq m_\fr$, $\mu'^+_0 \deq \mu^+_1$, $\Delta'_\ext \deq \Delta_\ext$ and $D'_\ext \deq D_\ext$. Notice that 
  $$\int_\Omega \sem{N(\gamma)}\!((D, X)(\omega), \mcal B)\,\mu^+_1(\dd{\omega}) = \int_\Omega \varphi(\omega, X(\omega))\,\mu_1^+(\dd{\omega}) > 0$$ 
  by instantiating $\Delta_\ext \deq \Omega$, we know, from the conclusion of $\set{P} M \set{X. Q}$, that the following identity holds for every $U \in \Sigma_\Omega \otimes \Sigma_A$:
  \begin{align*}
    \int_\Omega \int_A \ch{U}(\omega, x)\sem{M(\gamma)}\!(D(\omega), \dd{x})\,\mu_0^+(\dd{\omega})
    = \int_\Omega \ch{U}(\omega, X(\omega))\,\mu_1^+(\dd{\omega}).
  \end{align*}
  Notice that $(\omega, V) \mapsto \sem{M(\gamma)}\!(D(\omega), V)$ is a kernel from $\Omega$ to $\mcal A$. By \cref{lem:append-hoare:continuation}, we know
  \begin{align*}
    \int_\Omega \varphi(\omega, X(\omega))\,\mu_1^+(\dd{\omega})
    = \int_\Omega \int_A \varphi(\omega, x)\,\semo{M(\gamma)}(D(\omega), \dd{x})\,\mu_0^+(\dd{\omega}) > 0.
  \end{align*}

  Hence, there exists
  \begin{enumerate*}
    \item $Y \in \msf{RV}(\mcal B)$, 
    \item $m'_\post \in \mcal M$ such that $(\mcal F'_1, \mu'_1) \deq m'_\post \bullet m_\fr$ is defined, and
    \item a Borel measure $\mu'^+_1$ such that $\mu'^+_1|_{\mcal F'_1} = \mu'_1$
  \end{enumerate*}
  s.t. $(\gamma, (D, Y), m'_\post) \vDash R$, and for all $U \in \Sigma_{\sem{\Delta_\ext}} \otimes \Sigma_{\sem{\Delta}} \otimes \Sigma_A \otimes \Sigma_B$,
  \begin{align*}
    \int_\Omega \int_\mcal B \ch{U}(D_\ext, D, X, y) \sem{N(\gamma)}\!(D, X, \dd{y})\,\dd{\mu_1^+}
    = \int_\Omega \ch{U}(D_\ext, D, X, Y) \,\dd{\mu'^+_1}.
  \end{align*}
  With the same $Y, m'_\post, \mu'^+_1$, notice that
  \begin{align*}
    \int_\Omega &\int_B \ch{U}(D_\ext, D, y)\,\semo{\kr{let} X = M(\gamma) \kb{in} N(\gamma)}(D, \dd{y})\,\dd{\mu^+_0} \\
    &= \int_\Omega \int_B \ch{U}(D_\ext, D, y)\,\nu(D, \dd{y})\,\dd{\mu_0^+} \tag{semantics of $\kw{let}$} \\
    &= \int_\Omega \int_A \int_B \ch{A \times U}(x, D_\ext, y)\,\semo{N(\gamma)}(D, x, \dd{y})\,\semo{M(\gamma)}(D, \dd{x})\,\dd{\mu^+_0} \tag{\cref{lem:append-hoare:letintegral}} \\
    &= \int_\Omega \int_B \ch{A \times U}(X, D_\ext, y)\,\semo{N(\gamma)}(D, X, \dd{y})\,\dd{\mu_1^+} \tag{\cref{lem:append-hoare:continuation}} \\
    &= \int_\Omega \ch{A \times U}(X, D_\ext, D, Y)\,\dd{\mu'^+_1} \tag{$\set{Q} N \set{Y. R}$ holds} \\
    &= \int_\Omega \ch{U}(D_\ext(\omega), D(\omega), Y(\omega))\,\mu'^+_1(\dd{\omega}).
  \end{align*}
  where $\nu : \sem{\Delta} \times \Sigma_B \to \W$ is an $s$-finite kernel defined by
  \[
    \nu(\delta, E) \deq \int_A \semo{N(\gamma)}(\delta, x, E)\,\semo{M(\gamma)}(\delta, \dd{x}). \qedhere
  \]
\end{proof}

\begin{lemma}
  The following Hoare triple is sound:
  \[
    \inferrule[H-Frame]
      {\vdash \set{P} M \set{X : \mcal A. Q}}
      {\vdash \set{P * F} M \set{X : \mcal A. Q * F}}\ (X \notin \msf{fv}(F))
  \]
\end{lemma}
\begin{proof}
  Let $(\gamma, D, m) \vDash \set{P} M \set{X. Q}$ and $F$ be a proposition that does not contain $X$ as a free random variable. We need to show that $(\gamma, D, m) \vDash \set{P * F} M \set{X. Q * F}$.  
  For every $m_{\pre}$ such that $(\gamma, D, m_\pre) \vDash P * F$, we know, by definition of $\vDash$, there exists $m_{\pre 1}, m_{\pre 2}$ such that 
  \begin{align*}
    (\gamma, D, m_{\pre 1}) \vDash P,
    \qquad (\gamma, D, m_{\pre 2}) \vDash F\textnormal{, and}
    \qquad m_{\pre 1} \bullet m_{\pre 2} \sqsubseteq m_{\pre}.
  \end{align*}
  Now, suppose $m_\fr \in \mcal M$ such that $(\mcal F_0, \mu_0) \deq m_\pre \bullet m_\fr$ is defined. This means
  \[
    (\mcal F_0, \mu_0) = (m_{\pre 1} \bullet m_{\pre 2}) \bullet m_\fr = m_{\pre 1} \bullet (m_{\pre 2} \bullet m_\fr)
  \]
  Next, we instantiate the semantics of $(\gamma, D, m) \vDash \set{P} M \set{X. Q}$ with $m'_{\pre} \deq m_{\pre 1}$, $m'_\fr \deq m_{\pre 2} \bullet m_\fr$. By definition of the Hoare triple, we know that for every Borel measures $\mu_0^+ : \Sigma_\Omega \to \W$ satisfying $\mu_0^+|_{\mcal F_0}$ and $D_\ext \in \msf{RV}\!\sem{\Delta_{\ext}}$, there exists $X \in \msf{RV}(A)$, $m_\post \in \mcal M$ with $(\mcal F_1, \mu_1) \deq m_\post \bullet m'_\fr = m_\post \bullet (m_{\pre 2} \bullet m_\fr)$ defined, and Borel measure $\mu_1^+ : \Sigma_\Omega \to \W$ with $\mu^+_1|_{\mcal F_1} = \mu_1$ such that 
  \[ 
    (\gamma, (D, X), m_\post) \vDash Q.
  \] 
  Since $m_{\post} \bullet (m_{\pre 2} \bullet m_\fr) = (m_\post \bullet m_{\pre 2}) \bullet m_\fr$ is defined,  $m_\post \bullet m_{\pre 2}$ is also defined. Also, since $X$ is not a free variable of $F$, $(\gamma, D, m_{\pre 2}) \vDash F$ implies $(\gamma, (D, X), m_{\pre 2}) \vDash F$. By definition of $\vDash$, $(\gamma, (D, X), m_{\post}) \vDash Q$ and $(\gamma, (D, X), m_{\pre 2}) \vDash F$ implies $(\gamma, D, m_\post \bullet m_\fr) \vDash Q * F$.
\end{proof}

\begin{lemma}\label[lemma]{lem:append-hoare:cons}
  The following Hoare triple is sound:
  \[
    \inferrule[H-Cons]
      {P' \vdash P \\
      \vdash \set{P} M \set{X : \mcal A. Q} \\
      Q \vdash Q'}
      {\vdash \set{P'} M \set{X : \mcal A. Q'}}
  \]
\end{lemma}
\begin{proof}
  Assuming $P' \vdash P$, $(\gamma, D, m) \vDash \set{P} M \set{X. Q}$ and $Q \vdash Q'$. For every $m_{\pre}$ such that $(\gamma, D, m_\pre) \vDash P'$, we know $(\gamma, D, m_\pre) \vDash P$ by definition of $P' \vdash P$. This means for all $m_{\fr}$ with $(\mcal F_0, \mu_0) \deq m_\pre \bullet m_\fr$ defined, Borel measure $\mu_0^+ : \Sigma_\Omega \to \W$ satisfying $\mu_0^+|_{\mcal F_0} = \mu_0$ and random variables $D_\ext \in \msf{RV}\semo{\Delta_\ext}$. Assuming $\int_\Omega \semo{M(\gamma)}(D(\omega), A)\,\mu_0^+(\dd{\omega}) > 0$, then there exists 
  \begin{enumerate}
    \item $X \in \msf{RV}(A)$, 
    \item $m_\post \in \mcal M$ with $(\mcal F_1, \mu_1) \deq m_\post \bullet m_\fr$ defined, and 
    \item measure $\mu_1^+ : \Sigma_\Omega \to \W$ with $\mu^+_1|_{\mcal F_1} = \mu_1$ such that $(\gamma, (D, X), m_{\post}) \vDash Q$.
  \end{enumerate}
  It follows from (3), and the definition of $Q \vdash Q'$, that $(\gamma, (D, X), m_\post) \vDash Q'$.
\end{proof}

\newpage

\end{document}